\documentclass[twocolumn,twoside,trackchanges]{aastex701}

\usepackage{appendix}
\usepackage{tabularx}
\usepackage{amsmath, amssymb}
\usepackage{appendix}
\usepackage{hyperref}
\usepackage{booktabs}
\usepackage{longtable}

\begin{document}

\title{Tracing Radial Migration in the Outer Disk: A Comprehensive Analysis of the Old Open Cluster Berkeley 36}

\correspondingauthor{D. Bisht}
\email{devendrabisht297@gmail.com}  

\author[orcid=0000-0001-7940-3731, sname=Çınar, gname=Deniz Cennet]{Deniz Cennet Çınar}\thanks{E-mail: denizcennetcinar@gmail.com}
\affiliation{Programme of Astronomy and Space Sciences, Institute of Graduate Studies in Science, Istanbul University, Istanbul, 34116, Turkey}
\email{denizcennetcinar@gmail.com}

\author[orcid=0000-0002-8988-8434, sname=Bisht, gname=D.]{D. Bisht}
\affiliation{Indian Centre for Space Physics
466, Barakhola, Singabari Road, Netai Nagar, Kolkata, West Bengal, 700099}
\email{devendrabisht297@gmail.com}

\author[orcid=0000-0001-7359-3300, sname=Jiang, gname=D.]{Ing-Guey Jiang}\thanks{E-mail: jiang@phys.nthu.edu.tw}
\affiliation{Department of Physics and Institute of Astronomy, National Tsing-Hua University, Hsinchu 30013, Taiwan}
\email{jiang@phys.nthu.edu.tw}

\author[sname=Belwal, gname=K.]{K. Belwal}
\affiliation{Indian Centre for Space Physics 466, Barakhola, Singabari Road, Netai Nagar, Kolkata, West Bengal, 700099}
\email{kuldeepbelwal1997@gmail.com}

\author[0000-0003-3510-1509]{Sel\c{c}uk Bilir} \affiliation{Istanbul University, Faculty of Science, Department of Astronomy and Space Sciences, 34119, Beyaz\i t, Istanbul, Turkey} \email{sbilir@istanbul.edu.tr}

\begin{abstract}

We present a chemo-kinematical, structural, and photometric analysis of the old open cluster Berkeley~36 using Gaia~DR3 astrometry and photometry together with high-resolution spectroscopy from the Gaia-ESO Survey DR5.1. Applying a Gaussian Mixture Model to Gaia astrometry, we identify 946 high-probability cluster members. We derive a core radius of $2.63 \pm 0.21$ arcmin and a tidal radius of $14.84 \pm 0.47$ arcmin, indicating a moderately concentrated and dynamically relaxed system. By fixing the cluster metallicity to the spectroscopic value of $[\mathrm{Fe/H}] = -0.19 \pm 0.02$ dex and adopting an independently determined geometric distance of $4377 \pm 510$ pc, we minimize the classical age--reddening--metallicity degeneracy and derive a robust isochrone age of $6.8 \pm 0.5$ Gyr. Independent Ba-based chemical clocks yield a mean age of $7.75 \pm 1.94$ Gyr, supporting the isochrone solution.  The cluster exhibits a high main-sequence binary fraction of $48.0 \pm 1.7\%$ and hosts 83 blue straggler star candidates with an extended spatial distribution, contrary to classical mass segregation. Orbit integration shows that Berkeley~36 follows a nearly circular orbit in the outer Galactic disc ($R_{\rm GC}=11.42\pm0.44$ kpc). Accounting for the Galactic warp and disc flare increases the cluster's maximum vertical excursion by 86\% and the local disc scale height by 17\%, respectively. Comparison with its chemically inferred birth radius ($R_{\rm b}=6.71\pm0.66$ kpc) indicates an outward radial migration of approximately 5 kpc, predominantly driven by churning. These results establish Berkeley~36 as a benchmark for investigating radial migration, secular evolution, and the dynamical history of old Galactic open clusters.

\end{abstract}

\keywords{\uat{Open star clusters}{1160} --- \uat{Milky Way disk}{1050} --- \uat{Galaxy kinematics}{602} --- \uat{Blue straggler stars}{168} --- \uat{Stellar dynamics}{1596}}

\section{Introduction}
\label{sec:intro}

Open clusters (OCs) are among the most powerful tracers of the structure, kinematics, and chemical evolution of the Galactic disk. Because their member stars share a common origin, distance, initial chemical composition, and age, OCs provide nearly model-independent anchors for stellar evolutionary theory and serve as benchmarks for the age--metallicity relation \citep{Friel1995, lada2003, zinnecker2007, kruijssen2014}. Their spatial distribution across Galactocentric radii encodes the present-day metallicity gradient of the disk \citep[e.g.,][]{Friel2002, Jacobson2011, Netopil2016, Netopil2022, Joshi2024}, while their age spread, spanning from a few Myr to several Gyr, makes them uniquely suited for reconstructing the star-formation history and chemical enrichment of the thin disk over cosmological timescales \citep{Magrini2009, Donor2020}.

The launch of the Gaia mission \citep{GaiaCollaboration2016} has transformed star cluster studies. The high-precision astrometry provided by Gaia DR2, EDR3, and DR3 \citep{GaiaCollaboration2018, GaiaCollaboration2021, GaiaCollaboration2023} has enabled highly reliable membership determination from five-parameter astrometric solutions, greatly reducing the uncertainties of purely photometric methods. Gaia-based all-sky surveys have rediscovered poorly characterized clusters and significantly revised the fundamental parameters of hundreds of known clusters \citep{Cantat-Gaudin2018, Cantat-Gaudin2020, Hunt2024}. These homogeneous datasets now provide the basis for Galactic-scale studies of disk structure and evolution \citep[e.g.][]{Liu2019, Dias2021, Castro-Ginard2020, Plevne2026}. Determining accurate fundamental parameters is essential for exploiting the full potential of OCs in studies of stellar evolution and Galactic chemical evolution \citep{Friel1995, Netopil2016, Netopil2022, Otto2026}. However, despite the astrometric advances provided by Gaia, deriving reliable physical properties from photometry alone remains challenging for many OCs.

In traditional color--magnitude diagram (CMD) fitting, age, distance, and reddening are strongly correlated with metallicity \citep[e.g.,][]{Banks2020, Yontan2023a, Yontan2023b}. Variations in interstellar dust extinction and chemical composition can easily mimic one another along the main sequence (MS), meaning that entirely different parameter combinations can produce nearly identical CMD morphologies \citep[e.g.,][]{vonHippel2006, Bilir2010, Tasdemir23, Gokmen2023}. If all parameters are left free during isochrone fitting, this well-known degeneracy often leads to non-unique solutions and systematic errors in the derived cluster ages \citep[e.g.,][]{Yontan2015, Bostanci2015, Ak2016, Bossini2019, Cakmak2024}.

To overcome these photometric limitations and establish a physically consistent profile for Berkeley 36, it is crucial to constrain these variables using independent methods. The combination of Gaia astrometric membership and high-resolution spectroscopy from the Gaia-ESO Survey (GES; \citealt{Gilmore2022, Randich2022}) provides a powerful means of breaking the age--reddening--metallicity degeneracy. GES Data Release 5.1 (DR5.1) provides precise spectroscopic measurements of effective temperatures, surface gravities, and chemical abundances for individual cluster members \citep{Hourihane2023}. By securely anchoring the cluster's metallicity with GES data and establishing reliable distances via Gaia, the subsequent isochrone fitting is essentially reduced to a single free parameter: age. By combining independent astrometric and spectroscopic constraints, this approach substantially reduces the inherent degeneracies of photometric analyses and yields more robust cluster parameters \citep{Magrini2017, Viscasillas2022}.

Accurate cluster parameters not only improve our understanding of stellar evolution but also provide the foundation for investigating the dynamical evolution of the Galactic disk. Galactic disk evolution is shaped not only by star formation and chemical enrichment but also by dynamical processes that redistribute stars over time. Radial migration and orbital blurring mix stellar populations across the disk, complicating efforts to reconstruct the Milky Way’s formation history \citep{Sellwood2002, Roskar2008, Minchev2011}. Investigating these processes requires stellar tracers with accurately determined ages, distances, chemical compositions, and kinematics. OCs are ideal because their member stars share a common origin, allowing their fundamental parameters to be determined with high precision. Combined with Gaia astrometry, radial velocities provide full six-dimensional phase-space information, enabling the derivation of Galactic space velocities and orbital parameters \citep{Dias2002, Soubiran2018, Wu2009, Carrera2019, Tarricq2021, Yontan23c}. Orbit integration in realistic Galactic potentials further constrains birth radii, migration histories, and dynamical heating \citep{Allen1991, Irrgang2013, Bovy2015, Anders2017, Spina2021, Viscasillas2022}. Consequently, OCs are among the most powerful observational tracers for studying radial migration and blurring, providing key constraints on the formation and evolution of the Galactic disk.

Berkeley~36 ($\alpha_{2000} = 07^\mathrm{h}17^\mathrm{m}01^\mathrm{s}$, $\delta_{2000} = -13\degr08\arcmin48\arcsec$; $l= 227\degr.521, b=-0\degr.412$) is particularly well suited for such a study because it is an intermediate-to-old-age OC located toward the Galactic anticentre, where it serves as a valuable tracer of both the outer-disk metallicity gradient and Galactic dynamical evolution. Previous studies have been largely photometric in nature and limited in scope, providing only preliminary estimates of distance, reddening, and age without the benefit of spectroscopic metallicities or modern astrometric membership \citep[e.g.,][]{Hasegawa2004, Carraro2007, Tadross2011}. Despite recent advances in Gaia-based cluster studies, Berkeley~36 has not yet been investigated using a combined analysis of Gaia astrometry, GES spectroscopy, and TESS photometry, leaving its astrophysical and dynamical properties only partially constrained.

In this paper, we present a comprehensive analysis of Berkeley~36 by combining Gaia~DR3 astrometry, GES~DR5.1 spectroscopy, and TESS photometry. We determine the cluster’s structural and astrophysical parameters through CMD fitting constrained by spectroscopic metallicity, derive its full six-dimensional kinematics from Gaia astrometry and GES radial velocities, and investigate its Galactic orbit and radial migration history using a multi-component Galactic potential. We also examine the blue straggler and evolved star populations to assess the cluster’s dynamical evolutionary state. Together, these analyses provide a comprehensive picture of Berkeley~36, yielding new constraints on its evolutionary history and placing the cluster within the broader framework of Galactic disk formation and evolution.

\section{Data}\label{sec:data}
The Gaia space observatory \citep{GaiaCollaboration2016}, conceived and operated under the European Space Agency (ESA), has fundamentally transformed our capacity to construct three-dimensional maps of the Milky Way. Gaia~DR3 \citep{GaiaCollaboration2021} supplies astrometric and photometric measurements for approximately 1.46 billion stellar sources, constituting an unparalleled database for investigations of cluster membership, stellar kinematics, and Galactic structure. The catalog provides, for each source, celestial coordinates ($\alpha$, $\delta$), trigonometric parallaxes ($\varpi$), two-dimensional proper motion vectors ($\mu_\alpha \cos\delta$, $\mu_\delta$), and broadband photometry in the $G$, $G_{\rm BP}$, and $G_{\rm RP}$ passbands \citep{Riello2021}. This combination of high-precision positional measurements and uniform photometric calibration renders Gaia~DR3 particularly well suited for membership determination and CMD studies. We extracted all sources from the Gaia~DR3 archive lying within a circular aperture of radius $r=21$~arcmin centered on the nominal coordinates of Berkeley~36. The query construction and the applied data-quality criteria are documented in Appendix~\ref{app:data_quality}.

Elemental abundance measurements were drawn from GES~DR5.1 \citep{Hourihane2023}, which corresponds to the survey's sixth internal data release (iDR6). The spectroscopic data used in this work were obtained from the ESO Science Archive Facility \citep{ESO_GES}. The GES program acquired high-resolution spectra for more than $10^5$ stars with the VLT/FLAMES multi-object spectrograph \citep{Pasquini2002} between December~2011 and January~2018. Observations employed both the UVES arm ($R \approx 47{,}000$; gratings U520 and U580, spanning approximately 4140--6840~\AA) and the GIRAFFE arm ($R \approx 17{,}000$--$31{,}000$). Stellar atmospheric parameters and individual elemental abundances were derived in a self-consistent manner across all working groups and subsequently placed onto a homogeneous scale through the WG15 homogenization framework described by \citet{Hourihane2023}. Cluster members identified through our Gaia~DR3 membership analysis were cross-matched with the GES catalog using the VizieR XMatch service \citep{Boch2012}, adopting a positional tolerance of $1\arcsec$. This initial cross-match yielded 219 common sources. Following the quality-control recommendations of GES~DR5.1, nine stars flagged by the NIA and SRP quality indicators were excluded, as these correspond to sources with too few spectral lines for reliable abundance determinations\footnote{\url{https://www.eso.org/rm/api/v1/public/releaseDescriptions/191}.}. The resulting sample, therefore, comprised 210 stars.

The signal-to-noise ratios (SNRs) of the corresponding spectra span a wide range, from approximately 4 to 130. To determine an appropriate quality threshold, we examined three SNR limits ($\mathrm{SNR} \geq 20$, 25, and 30). Increasing the threshold from 20 to 30 reduces the number of stars with abundance measurements from 106 to 48, while the median metallicity changes only from $[\mathrm{Fe/H}] = -0.19$ to $-0.18$ dex and the median radial velocity varies by less than $\approx 1~\mathrm{km\,s^{-1}}$ (62.25 to $61.21~\mathrm{km\,s^{-1}}$). Since these variations are negligible compared to the substantial reduction in sample size, we adopted $\mathrm{SNR} \geq 20$, thereby maximizing the number of spectroscopic cluster members without introducing a significant systematic bias in the cluster's bulk chemical or kinematic properties. Finally, only stars with available atmospheric parameters and chemical abundance measurements were retained for the subsequent analysis. This resulted in a final spectroscopic sample of 106 stars. 

The available atmospheric parameters include the effective temperature ($T_{\rm eff}$), surface gravity ($\log g$), and metallicity ($[\mathrm{Fe/H}]$), together with abundance measurements for a wide range of chemical elements. In the present analysis, we focus on five $s$-process elements, four $\alpha$-elements, and the odd-$Z$ element Na. Elemental abundances are expressed in the standard logarithmic scale, $A(\mathrm{El}) = 12 + \log\left[n(\mathrm{El})/n(\mathrm{H})\right]$, while abundances normalised to the solar composition are given by $[\mathrm{El}/\mathrm{H}] = \log\left[n(\mathrm{El})/n(\mathrm{H})\right] - \log\left[n(\mathrm{El})/n(\mathrm{H})\right]_{\odot}$.

As a preparatory step, a kinematic pre-selection was applied by retaining only sources whose parallaxes and proper motions lie within $1\sigma$ of the cluster mean values reported in \citet{Hunt2024}. This criterion corresponds to the ranges $-1.864 < \mu_{\alpha}\cos\delta < -1.538$ mas yr$^{-1}$, $0.806 < \mu_{\delta} < 1.114$ mas yr$^{-1}$, and $0.102 < \varpi < 0.346$ mas. This pre-selection substantially reduces field-star contamination before the formal membership analysis. The resulting filtered samples serve as input to the membership procedure detailed in Section~\ref{sec:membership} and are subsequently used for isochrone fitting, Galactic orbit integration, chemical abundance comparison, and dynamical evolution.

The Transiting Exoplanet Survey Satellite \citep{ricker2015transiting} photometric data used in this work were obtained from the Mikulski Archive for Space Telescopes (MAST) at the Space Telescope Science Institute. The specific TESS observations analysed in this study, including
the light curves of probable Berkeley 36 members and the foreground eclipsing binary candidates, are available via
\dataset[doi:10.17909/531f-ks67]{https://doi.org/10.17909/531f-ks67} \citep{Belwal2026TESS}.

\section{Membership Analysis}\label{sec:membership}

A robust separation of true cluster stars from the surrounding Galactic disk population is a prerequisite for all subsequent steps in this study. While published Gaia DR3-based membership lists for Berkeley~36 already exist \citep{Cantat-Gaudin2020, Hunt2024}, we construct our own catalog so that the sample used throughout this paper is internally uniform and treated with a single, well-defined method. Membership was assigned through a Gaussian Mixture Model \citep[GMM;][]{dempster1977maximum,mclachlan2000finite} fitted to the Gaia DR3 proper motions and trigonometric parallaxes of stars projected toward Berkeley~36. In this approach, the observed astrometric distribution is decomposed into two Gaussian components, one tracing the cluster and one tracing the field, and each star is assigned a probability of belonging to the cluster component based on its position in proper-motion/parallax space. Such a probabilistic treatment is particularly well suited to lines of sight where the cluster and disk populations partially overlap kinematically, and it is now routinely applied in OC studies \citep{agarwal2021ml, belwal2025unveiling, belwal2026time, Bisht2026, Bisht2026b}.

\begin{figure*}
\centering
\includegraphics[width=0.95\linewidth]{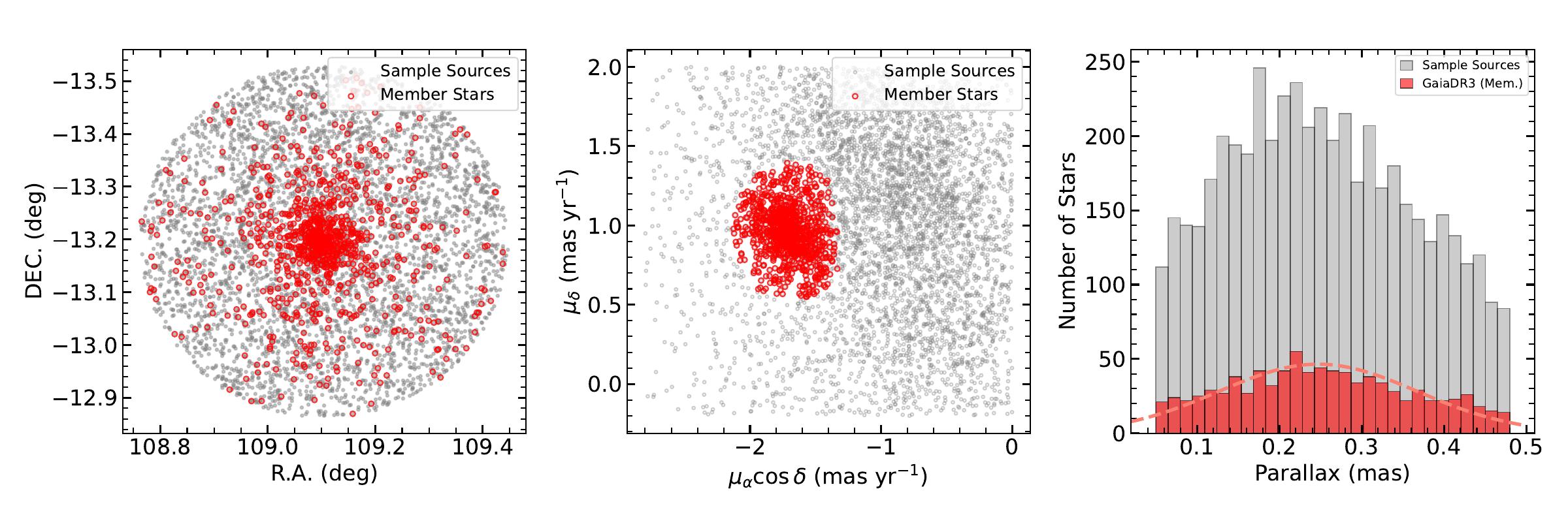}
\caption{Positional, kinematic, and astrometric properties of Berkeley~36. Left: sky distribution of the surveyed stars in equatorial coordinates ($\alpha$, $\delta$). Middle: proper-motion vector-point diagram showing the clear offset between the cluster and the field population. Right: trigonometric parallax distribution of the same stars. Grey points and grey bars mark all stars within the searched field, while red circles and red bars indicate the stars retained as probable cluster members from the Gaia DR3 astrometry.}
\label{fig:spatial_kinematic_dist}
\end{figure*}

Working from the Gaia DR3 sources within the adopted search radius, we retained only stars with a full five-parameter astrometric solution and complete photometry across the three Gaia passbands. We further imposed a positive trigonometric parallax ($\varpi > 0$), an upper limit on the proper-motion uncertainty ($\epsilon_\mu$), and $\mathrm{RUWE} \leq 1.4$ \citep{Lindegren2021} to remove sources with unreliable single-star astrometric fits, likely unresolved binaries, or otherwise flawed measurements. Before running the GMM, an approximate cluster locus in ($\mu_\alpha \cos\delta$, $\mu_\delta$, $\varpi$) space was located with a $k$-Nearest Neighbours search \citep[kNN;][]{cover1967nearest}, which was used to initialise the fit and limit early contamination from field stars. The astrometric quantities were standardised, and a two-component GMM was then optimised using the Expectation-Maximisation algorithm, yielding a membership probability $P$ for each star. Stars with $P \geq P_{\rm min}$ were retained as probable members, with the threshold chosen to keep the sample as complete as possible while limiting field contamination for Berkeley~36.

With this procedure, we obtain $N_{\rm mem} = 946$ probable members of Berkeley~36 with $P \geq 0.7$. The resulting mean proper motion is consistent with previous catalogues \citep{Cantat-Gaudin2020, Hunt2024}. A comparison with the membership catalogue of \citet{Hunt2024} shows substantial agreement between the two samples: of the 678 candidate members identified by \citet{Hunt2024}, 331 have $P \geq 0.7$, all of which are included in our final sample. In addition, 587 stars are common to both catalogues, while the remaining 91 stars identified by \citet{Hunt2024} do not meet our adopted membership-probability threshold of $P=0.7$. Our final catalogue contains 946 members, indicating that our selection recovers the high-probability members identified by \citet{Hunt2024} while also identifying additional candidate members. As shown in Figure~\ref{fig:spatial_kinematic_dist}, the selected members trace a well-defined spatial and kinematic overdensity, while their parallaxes show a relatively concentrated distribution consistent with the cluster population. Together, these features support the identification of Berkeley~36 as a physically associated stellar group rather than a chance alignment of field stars.

The cluster center was located by finding the position of maximum stellar density on the sky, using the equatorial coordinates of the member stars from Gaia DR3. Separate one-dimensional density profiles were built along $\alpha$ and $\delta$, and each was fitted with a Gaussian function; the peak of each fit was taken as the corresponding coordinate of the density maximum. The resulting central coordinates are $\alpha = 07^{\text{h}}16^{\text{m}}23^{\text{s}}.86$, $\delta = -13^\circ11^\prime51^{\prime\prime}\!.35$, equivalent to Galactic coordinates $l = 227^\circ\!.496$ and $b = -0^\circ\!.569$ for Berkeley~36. As in similar analyses, the probability cut was set to $P_{\rm min}$=0.7, representing a compromise between sample size and purity. We also verified that small changes in this threshold do not significantly affect the derived cluster properties.

The mean astrometric solution for Berkeley~36 was computed from the Gaia DR3 measurements of the adopted member stars. We find a mean proper motion of $\langle\mu_{\alpha}\cos\delta, \mu_{\delta}\rangle = (-1.701 \pm 0.163,\ 0.960 \pm 0.154)$~mas~yr$^{-1}$, while a Gaussian fit to the trigonometric parallax distribution gives $\varpi = 0.224 \pm 0.122$~mas, corresponding to a parallax-based distance of $d_{\varpi} = 4.46 \pm 2.43$~kpc. These astrometric parameters are consistent with earlier determinations \citep[e.g.,][]{Cantat-Gaudin2020, Hunt2024}; the full set of derived quantities is listed in Table~\ref{tab:summary}.

\section{Structural Parameters Determination}
\label{sec:rdp}
To characterize the cluster's internal structure, we analyzed the radial stellar density distributions of high-probability cluster members identified through our Gaia astrometric analysis. The radial density profile (RDP) was constructed using an equal-area annular binning scheme, ensuring uniform sampling of the stellar density field and comparable Poisson uncertainties across bins, with each annulus containing at least 20 member stars ($N_i \geq 20$).

The surface density $\rho(r_i)$ at projected distance $r_i$ from the cluster center was obtained as $\rho(r_i) = N_i / A_i$, where $N_i$ is the star count and $A_i$ the annular area. We fitted the observed profiles with the empirical surface-density model of \citet{King62}:
\begin{equation}
\rho(r) = \rho_0 \left[ \frac{1}{\sqrt{1 + (r/r_{\rm c})^2}} - \frac{1}{\sqrt{1 + (r_{\rm t}/r_{\rm c})^2}} \right]^2 + \rho_{\rm bg},
\label{eq:king_model}
\end{equation}
where $\rho_0$ is the central density, $r_{\rm c}$ is the core radius (where $\rho = 0.5\,\rho_0$), $r_{\rm t}$ is the tidal radius, and $\rho_{\rm bg}$ is the residual background density. Although $\rho_{\rm bg} \approx 0$ is expected for a member-only sample, it was retained as a free parameter to account for residual contamination. Best-fit parameters were estimated by minimizing the negative log-likelihood:
\begin{equation}
\ln \mathcal{L} = -\frac{1}{2} \sum_i \left( \frac{\rho_i - \rho_{i,{\rm model}}}{\sigma_{\rho_i}} \right)^2,
\label{eq:log_likelihood}
\end{equation}
where $\sigma_{\rho_i}$ is the Poisson uncertainty in each annulus.

\begin{figure}
    \centering
    \includegraphics[width=1\linewidth]{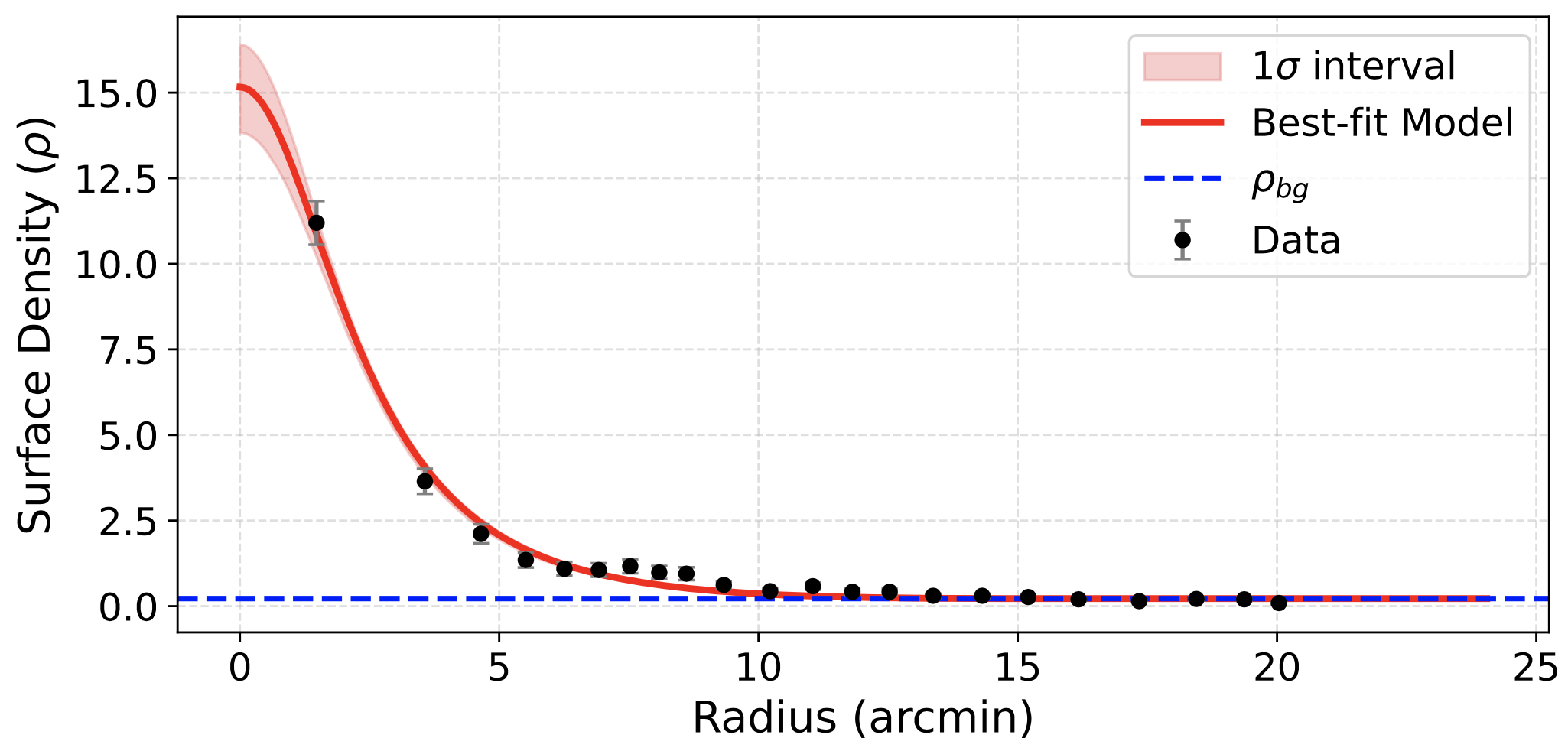}\\
    \includegraphics[width=1\linewidth]{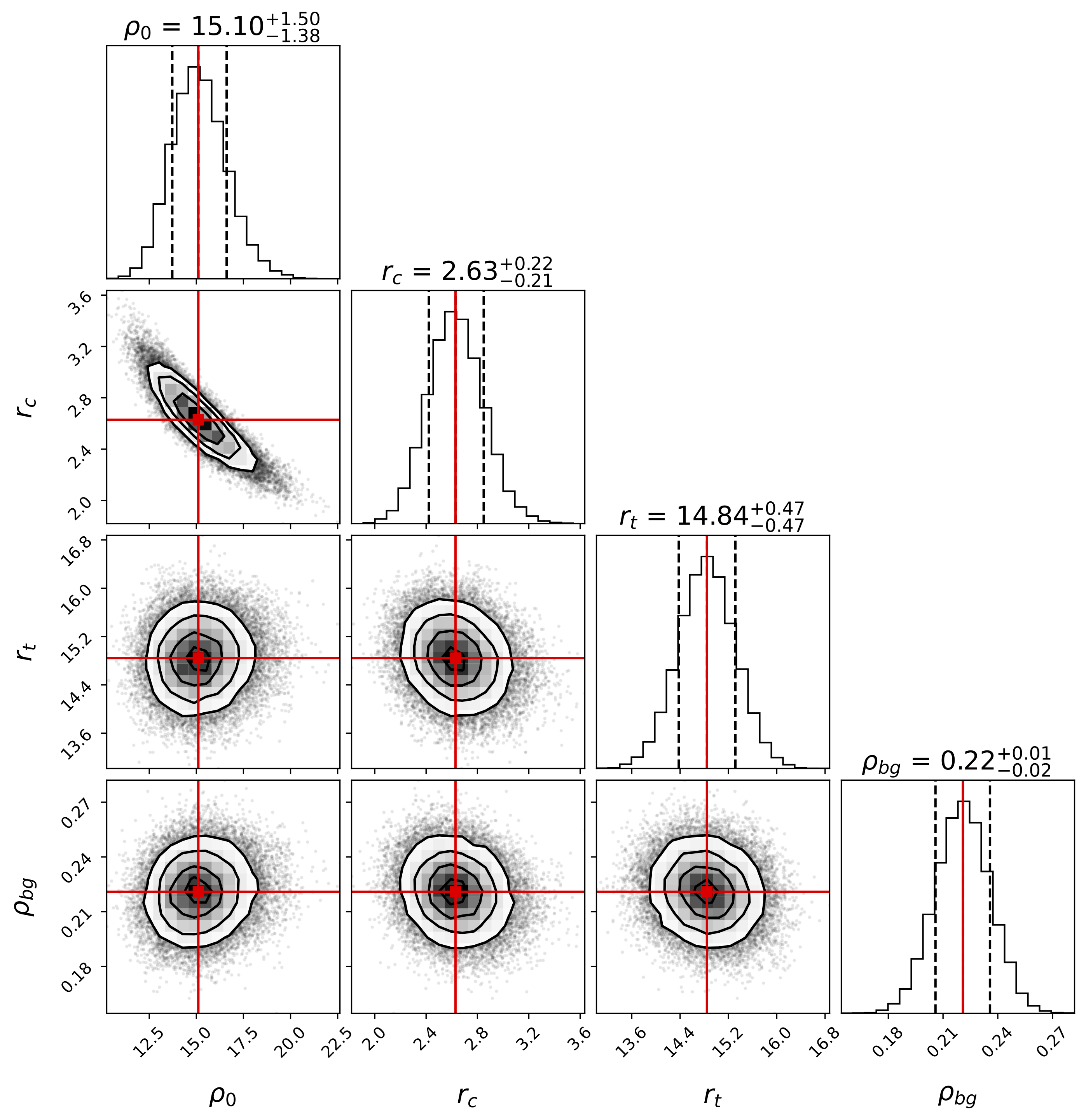}
\caption{RDP and MCMC posterior distributions for Berkeley~36. \textit{Upper panel:} Observed surface density ($\rho$) as a function of radial distance from the cluster center (black points with Poisson error bars). The solid red curve is the best-fitting \citet{King62} model, with the shaded region indicating the $1\sigma$ confidence interval. The horizontal blue dashed line marks the background density level ($\rho_{\rm bg}$). \textit{Lower panels:} Corner plot of the posterior distributions for the fitted King model parameters ($\rho_0$, $r_{\rm c}$, $r_{\rm t}$, $\rho_{\rm bg}$), with median values (solid red lines) and $1\sigma$ boundaries (dashed lines).}
    \label{fig:RDP_fits}
\end{figure}

Parameter optimization was performed using the MCMC ensemble sampler \texttt{emcee} \citep{emcee} with 100 walkers, 5,000 steps, and a 500-step burn-in, adopting broad uniform priors on all parameters. Chain convergence was confirmed via the \citet{gelman1992} $\hat{R}$ diagnostic ($\hat{R} < 1.1$). The resulting best-fit model and $1\sigma$ confidence interval are shown in the upper panel of Figure~\ref{fig:RDP_fits}, while the posterior distributions of the fitted parameters are shown in the lower-corner plot. The narrow posterior distributions demonstrate that the structural parameters are tightly constrained despite the cluster's large distance. The structural parameters are summarised in Table~\ref{tab:summary}.

For Berkeley~36, the MCMC analysis yields a central density $\rho_0 = 15.10^{+1.50}_{-1.38}$, a core radius $r_{\rm c} = 2.63^{+0.22}_{-0.21}$ arcmin, a tidal radius $r_{\rm t} = 14.84^{+0.47}_{-0.47}$ arcmin, and a background density $\rho_{\rm bg} = 0.22^{+0.01}_{-0.01}$. All posterior distributions are well constrained and approximately unimodal, and the low correlation among $r_{\rm t}$, $r_{\rm c}$, and $\rho_{\rm bg}$ indicates that the fit is robust, with the only appreciable degeneracy occurring between $\rho_0$ and $r_{\rm c}$, as expected from the functional form of the King profile.

Following \citet{King62}, the concentration parameter $C = \log(r_{\rm t}/r_{\rm c})$ indicates a cluster's physical compactness and dynamical state. For Berkeley~36, we obtain $C \approx 0.75$ ($r_{\rm t}/r_{\rm c} \approx 5.6$), a value comparable to those typically found for moderately concentrated, dynamically relaxed OCs, and consistent with an evolved system that has undergone significant two-body relaxation.

\section{Color--Magnitude Diagram and Fundamental Parameters}\label{sec:fundamental}

\begin{table*}
\centering
\footnotesize
\renewcommand{\arraystretch}{1}
\caption{Derived physical and kinematic parameters of Berkeley~36.}
\label{tab:summary}
\begin{tabular}{llr}
\hline\hline
Parameter & Symbol & This study \\
\hline
\multicolumn{3}{c}{\textit{Astrometric Parameters}} \\
\hline
Right ascension                 & $\alpha$ (hh:mm:ss)  & 07:16:23.86  \\
Declination                     & $\delta$ (dd:mm:ss)  & $-$13:11:51.35 \\
Galactic longitude              & $l$ (degree)       & 227.496 \\
Galactic latitude               & $b$ (degree)       & $-$0.569 \\
Mean proper motion (RA)         & $\langle\mu_{\alpha}\cos\delta\rangle$ (mas yr$^{-1}$) & $-1.701 \pm 0.163$ \\
Mean proper motion (Dec)        & $\langle\mu_{\delta}\rangle$ (mas yr$^{-1}$)        & $0.960 \pm 0.154$ \\
Mean trigonometric parallax     & $\langle\varpi\rangle$ (mas)                & $0.224 \pm 0.122$ \\
Mean parallax distance          & $\langle d_{\varpi}\rangle$ (kpc)       & $4.46 \pm 2.43$ \\
\hline
\multicolumn{3}{c}{\textit{Structural Parameters}} \\
\hline
Central surface density         & $\rho_0$ (stars arcmin$^{-2}$)    & $15.10^{+1.50}_{-1.38}$ \\
Background density              & $\rho_{\rm bg}$ (stars arcmin$^{-2}$) & $0.22^{+0.01}_{-0.01}$ \\
Core radius                     & $r_{\rm c}$ (arcmin)       & $2.63^{+0.22}_{-0.21}$   \\
Tidal radius                    & $r_{\rm t}$ (arcmin)       & $14.84^{+0.47}_{-0.47}$  \\
Concentration parameter         & $C$                        &  $0.75^{+0.05}_{-0.05}$ \\
\hline
\multicolumn{3}{c}{\textit{Fundamental Parameters}} \\
\hline
Geometric distance        & $\langle d_{\rm geo}\rangle$ (pc) & $4377 \pm 510$ \\
True distance modulus           & $(m-M)_0$ (mag)                     & $13.206 \pm 0.253$ \\
Apparent distance modulus       & $(m-M)$ (mag)                       & $14.800 \pm 0.379$ \\
$V$-band extinction             & $\langle A_{\rm V} \rangle$ (mag)   & $1.594 \pm 0.283$ \\
Color excess ($UBV$)            & $E(B-V)$ (mag)                      & $0.514 \pm 0.091$ \\
Color excess (Gaia)             & $E(G_{\rm BP}-G_{\rm RP})$ (mag)    & $0.725 \pm 0.129$ \\
$G$-band extinction           & $\langle A_{\rm G}\rangle$ (mag)    & $1.350 \pm 0.045$ \\
Iron abundance                  & $[\rm{Fe/H}]$ (dex)                 & $-0.19 \pm 0.02$ \\
Metallicity                     & $Z$                                 & $0.01001 \pm 0.0004$ \\
Age (isochrone)                 & $t_{\rm iso}$ (Gyr)                 & $6.8 \pm 0.5$ \\
Age (chemical clock)              & $t_{\rm [Ba/X]}$ (Gyr)             & $7.75 \pm 1.94$\\
\hline
\multicolumn{3}{c}{\textit{Galactic Orbital Parameters}} \\
\hline
Radial velocity            & $\langle V_{\rm rad,OC}\rangle$ (km s$^{-1}$)        & $62.80 \pm 0.04$ \\
Perigalactic radius             & $R_{\rm peri}$ (kpc)              & $11.369 \pm 0.437$ \\
Apogalactic radius              & $R_{\rm apo}$ (kpc)               & $12.214 \pm 1.139$ \\
Mean orbital radius             & $R_{\rm m}$ (kpc)                 & $11.791 \pm 0.610$ \\
Orbital eccentricity            & $e$                               & $0.036 \pm 0.027$ \\
Maximum vertical height         & $Z_{\rm max}$ (kpc)               & $0.380 \pm 0.023$ \\
Present Galactocentric radius   & $R_{\rm GC}$ (kpc)                & $11.422 \pm 0.437$ \\
Guiding radius                  & $R_{\rm g}$ (kpc)                 & $11.751 \pm 0.772$ \\
Chemical birth radius           & $R_{\rm b}$ (kpc)                 & $6.71\pm 0.66$ \\
Orbital period                  & $T_{\rm p}$ (Myr)                 & $345 \pm 26$ \\
\hline
\multicolumn{3}{c}{\textit{Space Velocities}} \\
\hline
Velocity component & $U_{\rm LSR}$ (km s$^{-1}$) & $-58.56 \pm 3.79$ \\
Velocity component & $V_{\rm LSR}$ (km s$^{-1}$) & $-8.93 \pm 3.40$ \\
Velocity component & $W_{\rm LSR}$ (km s$^{-1}$) & $-15.98 \pm 2.03$ \\
Space velocity & $S_{\rm LSR}$ (km s$^{-1}$) & $61.36 \pm 5.48$ \\
\hline
\hline
\end{tabular}
\end{table*}

Accurate fundamental parameters such as reddening, distance, metallicity, and age are prerequisites for interpreting the formation history and dynamical state of OCs and for tracing the chemical and structural evolution of the Galactic disk \citep{Friel1995, Netopil2016}. However, their derivation from CMD analysis is non-trivial. Clusters at low Galactic latitudes suffer from substantial interstellar extinction and spatially variable reddening \citep{Burki1975, Carraro2017}, while the OC population itself spans a broad metallicity range, with cluster-to-cluster differences reaching $\sim$0.25~dex \citep{Cinar2024, Tasedemir2026}. Compounding these difficulties, reddening, distance modulus, metallicity, and age are mutually correlated in isochrone space \citep{vonHippel2006, Bilir2006a, Bilir2010, Andreuzzi2011, Akbulut2021, Koc2022, Yontan2026}
, so that unconstrained simultaneous fitting can yield degenerate or physically inconsistent solutions \citep{Bilir2016, Bostanci2018, Yontan2019, Yontan2022, Tanik2025, Karagoz25}. To mitigate these degeneracies in our analysis of Berkeley~36, we adopted a sequential parameter-fixation strategy. Line-of-sight extinction was estimated on a star-by-star basis from the three-dimensional dust maps of \citet{Green2019}, providing a spatially resolved treatment of reddening across the cluster field. Individual stellar distances were drawn from the Bayesian photogeometric catalog of \citet{BailerJones2021}, which combines Gaia trigonometric parallaxes with photometric priors to yield reliable distance estimates for Berkeley~36. Metallicity was anchored to the value derived from high-resolution GES spectroscopy. With reddening, distance, and metallicity independently constrained, the CMD fit reduces to a single free parameter, cluster age, thereby yielding more robust, uniquely determined results than a fully free isochrone fit.

\subsection{Geometric Distance}\label{sec:photogeometric_distance}
One of the key advantages of the Gaia mission is the ability to determine stellar distances directly through trigonometric parallax measurements. However, the accuracy of this method decreases significantly beyond approximately 2 kpc from the Sun, where increasing relative uncertainties in parallax measurements and reduced signal-to-noise ratios limit its reliability \citep{Plevne2020, Doner2023}. In this regime, directly inverting observed parallaxes can yield biased and physically inconsistent distance estimates, necessitating statistical corrections for robust applications. Given that Berkeley~36 is located at a distance of approximately 3.5 kpc, using raw parallax values for individual cluster members is inappropriate. Instead, we adopt the Bayesian geometric distance estimates of \citet{BailerJones2021}, which incorporate parallax measurements within a probabilistic framework. This method also accounts for prior assumptions about the spatial distribution of stars in the Galaxy, thereby providing more stable and reliable distance estimates, particularly for distant and low-precision parallax sources. The adoption of these Bayesian distances plays a crucial role in reducing parameter degeneracies in the subsequent cluster analysis.

In this context, individual distances for the cluster members were adopted from the catalog of \citet{BailerJones2021}, who derived Bayesian distance estimates for approximately 1.47 billion Gaia EDR3 sources. This catalog combines trigonometric parallax measurements with a direction-dependent prior based on a three-dimensional model of the Milky Way. It provides two types of distance estimates for each source: the purely geometric distance ($d_{\rm geo}$), which relies predominantly on trigonometric parallax, and the photogeometric distance ($d_{\rm pgeo}$), which additionally incorporates photometric information such as stellar magnitudes and color indices. In this study, we adopt $d_{\rm geo}$ because it relies solely on trigonometric parallax and the astrometric prior, without being influenced by photometric information that may introduce additional systematics for cluster members with uncertain or blended photometry. For each source, \citet{BailerJones2021} additionally provides the 16th and 84th percentiles of the posterior distance distribution, which we adopt as the lower and upper $1\sigma$ uncertainty bounds on the individual distance estimates.

Because the membership catalogue does not include Gaia EDR3 source identifiers, the cluster members were matched with the catalogue of \citet{BailerJones2021} using their sky coordinates. For this purpose, a cone search around the cluster field was performed using the VizieR service, and the sources were cross-matched within a 1$^{\prime\prime}$ radius, consistent with Gaia's astrometric precision. Only confirmed cluster members were retained in the final sample. The median geometric distance of the cluster members, together with the 16th and 84th percentiles of the distribution, was $d_{\rm geo} = 4377\pm510$~pc. This result confirms Berkeley~36 as a distant OC located well beyond the solar neighbourhood.

\subsection{Interstellar Extinction}

\begin{figure}
    \centering
    \includegraphics[width=0.94\linewidth]{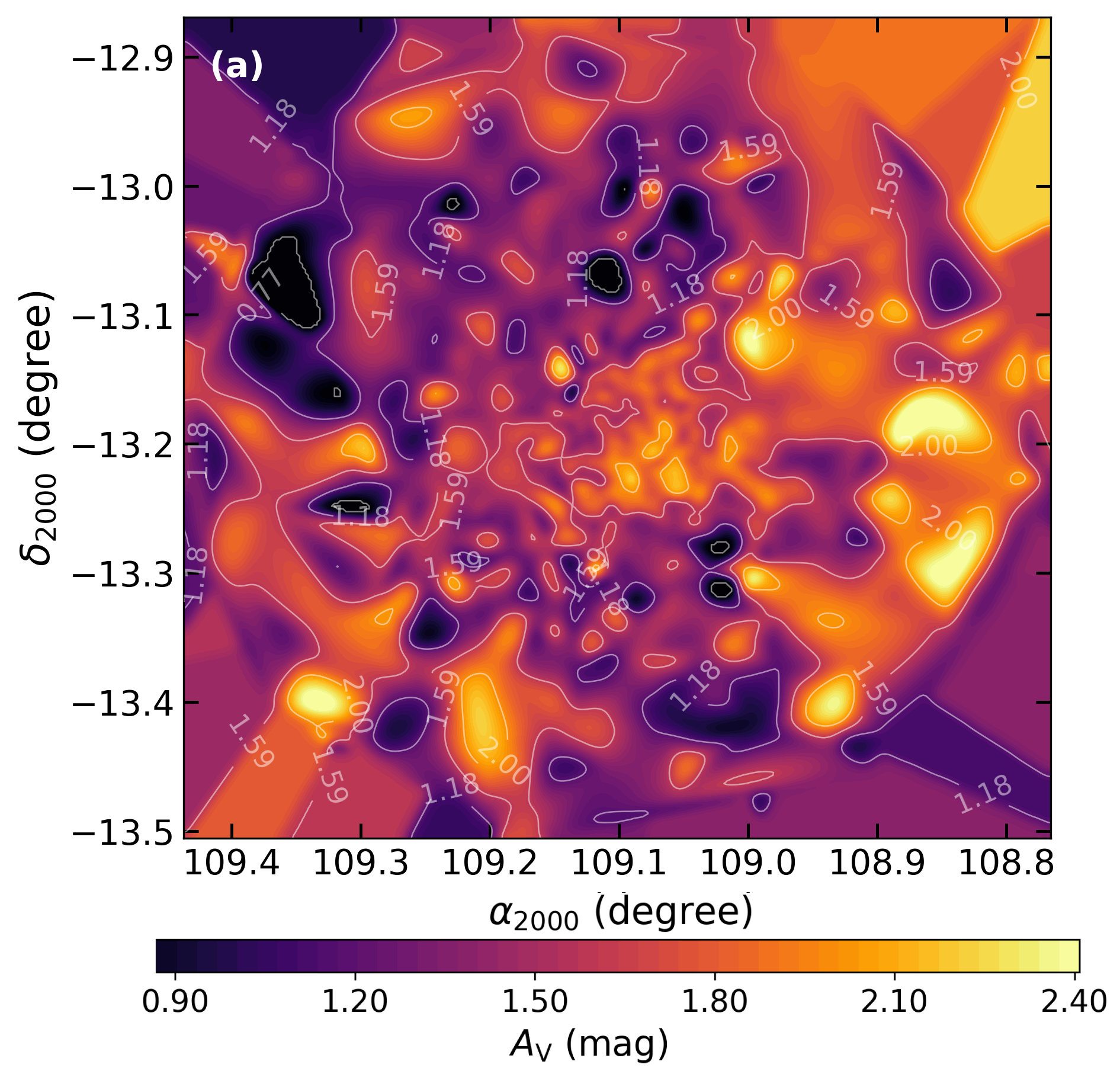}
    \includegraphics[width=0.85\linewidth]{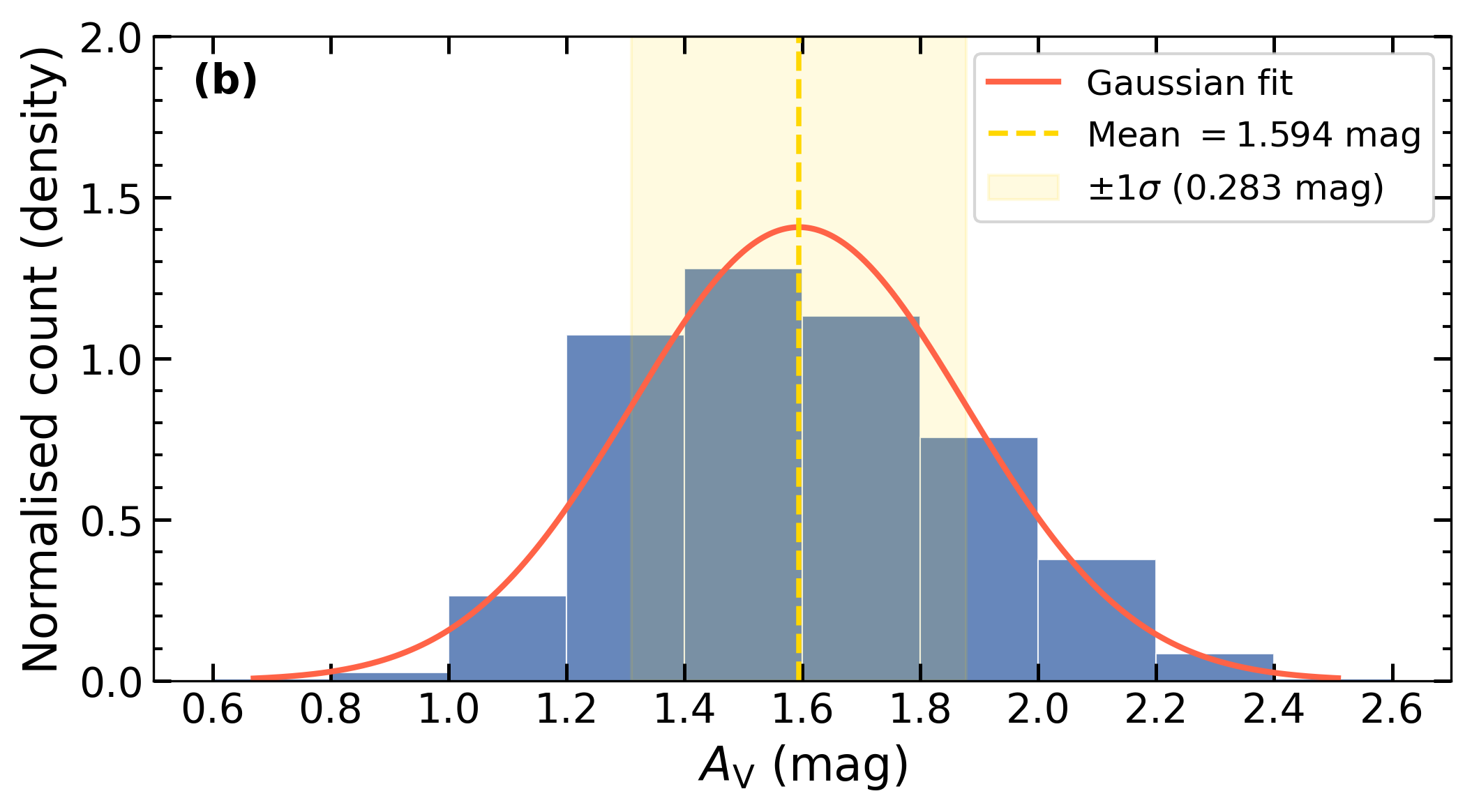}
    \caption{Interstellar extinction properties of the Berkeley~36 member stars. (a) Spatial distribution of $A_{\rm V}$ across the cluster field, reconstructed from the {\sc Bayestar19} dust model \citep{Green2019}; the contours and color scale trace the extinction level. (b) Distribution of individual $A_{\rm V}$ values, with the red curve showing the Gaussian fit, the vertical dashed line marking the mean, and the shaded band indicating the $\pm1\sigma$ range.}
    \label{fig:av_map}
\end{figure}

Accurate OC parameters are strongly hampered by interstellar reddening, which is tightly coupled to both distance and age when fitting isochrones. Properly accounting for interstellar extinction is therefore a prerequisite for breaking the degeneracies that otherwise arise among the photometric parameters. Rather than assuming a single, cluster-wide extinction value, we exploit a three-dimensional, distance-resolved reddening approach in which line-of-sight extinction is estimated individually for each member star. This is achieved by combining the precise photogeometric distances obtained in Section~\ref{sec:photogeometric_distance} with a Galactic dust model, yielding a spatially resolved treatment of reddening across the cluster field.

We derived the line-of-sight reddening of the cluster members from the three-dimensional {\sc Bayestar19} dust map of \citet{Green2019}, which is calibrated following \citet{Schlafly2011} and combines Pan-STARRS~1 and 2MASS photometry to yield distance-dependent extinction estimates that are broadly compatible with Gaia astrometry. For each member, the reddening was interpolated at the star's sky position and geometric distance ($d_{\rm geo}$); to suppress the effect of small-scale dust inhomogeneities, we adopted the median of the interpolated posterior as the representative value for that star. Since {\sc Bayestar19} reports reddening in the Gaia photometric system, we converted the native $E(G_{\rm BP}-G_{\rm RP})$ values to the Johnson–Cousins system via $E(G_{\rm BP}-G_{\rm RP}) = 1.41\times E(B-V)$ and $A_{\rm G} = 0.83627 \times A_{\rm V}$ \citep{Canbay2023}, and obtained the visual extinction from $A_{\rm V} = 3.1\times E(B-V)$ \citep{Cardelli1989}.

Figure~\ref{fig:av_map}a shows the extinction field toward Berkeley~36, constructed by interpolating the individual $A_{\rm V}$ estimates of member stars onto a regular spatial grid. The resulting map reveals clear evidence of differential reddening: extinction is not uniformly distributed across the field but instead traces the underlying dust structure, with localized patches reaching $A_{\rm V} \sim 2.4$~mag while the bulk of the cluster area shows comparatively lower values around the cluster median. The corresponding histogram of individual member extinctions (Figure~\ref{fig:av_map}b) is well approximated by a single Gaussian component, with no evidence for multiple reddening populations or strong asymmetry, consistent with the closeness of the mean ($1.594$~mag) and median ($1.571$~mag) values of the sample. This near-symmetric behavior points to a foreground dust screen that varies smoothly on the cluster's angular scale rather than being dominated by compact, high-density clumps. 

Among the 946 member stars with reliable dust-map estimates, the mean reddening is $\langle E(B-V) \rangle = 0.514 \pm 0.091$~mag, corresponding to a mean $V$-band extinction of $\langle A_{\rm V} \rangle = 1.594 \pm 0.283$~mag. The equivalent extinction in the Gaia photometric system is $E(G_{\rm BP}-G_{\rm RP}) = 0.725 \pm 0.129$~mag and $A_{\rm G} = 1.350 \pm 0.045$~mag. The substantially higher reddening derived for Berkeley~36, compared to typical nearby OCs, reflects its location in a heavily obscured region of the Galactic disk.

\subsection{Metallicity}

Spectroscopic [Fe/H] and [$\alpha$/Fe] abundances for Berkeley~36 were taken from the GES DR5.1 catalog \citep{Hourihane2023}, which offers a solid basis for both isochrone selection and chemical characterization \citep{Carrera2011}. Applying an SNR $\geq 20$ quality cut to the high-probability cluster members yielded a final sample of 106 stars with 319 individual spectra, covering $13.8<G~{\rm (mag)}\leq 18.5$ and concentrated mainly along the MS and MSTO regions of the CMD; further details of the sample selection are provided in Section~\ref{sec:data}.

\begin{figure}
    \centering
    \includegraphics[width=0.95\linewidth]{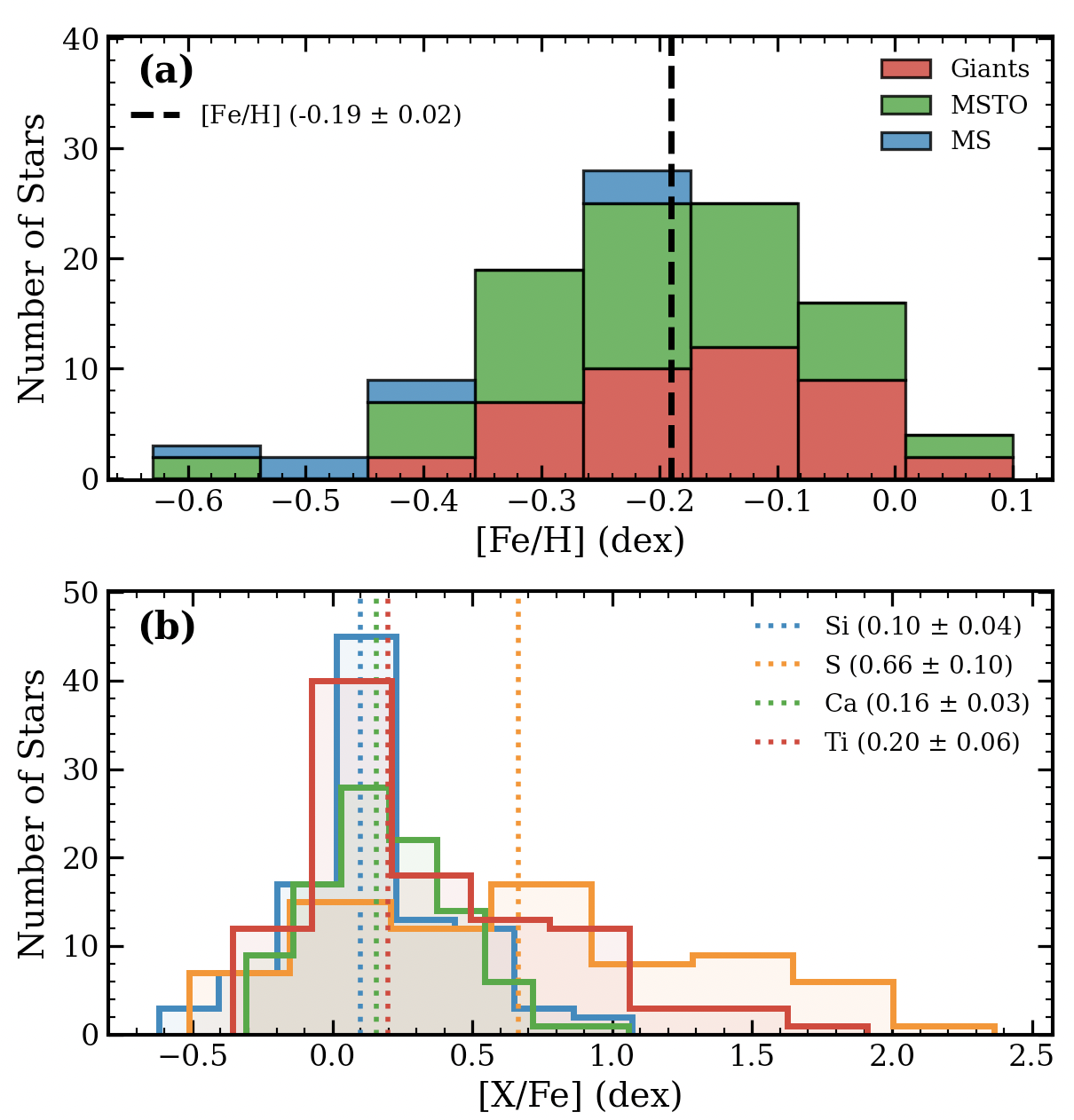}
    \caption{Chemical abundance distributions of Berkeley~36 member stars with SNR $\geq20$ from GES data. Panel (a): [Fe/H] distribution, with bin colors indicating stellar evolutionary stage. Panel (b): abundance distributions of Si, S, Ca, and Ti. Dotted vertical lines mark the median of each distribution, with uncertainties given as the standard error of the median.}
    \label{fig:element_histo}
\end{figure}

As shown in Figure~\ref{fig:element_histo}a, the stars with measured iron abundances are distributed predominantly along the giant branch, with a smaller number located around the MSTO and MS regions. The individual [Fe/H] measurements obtained from the Gaia-ESO Survey spectra for the 106 stars span roughly $-0.7<{\rm [Fe/H]~(dex)}\leq +0.2$ dex. This relatively broad observed range should not be interpreted as the intrinsic metallicity dispersion of Berkeley~36, as the sample includes stars at different evolutionary stages and individual spectroscopic measurements naturally carry different uncertainties. Open clusters are generally expected to exhibit relatively small intrinsic metallicity dispersions \citep[e.g.,][]{Bovy2016, Sinha2024}; therefore, the observed spread is more appropriately regarded as reflecting the combined effects of measurement uncertainties and the heterogeneous evolutionary stages represented in the spectroscopic sample. The distribution is centered around $\langle \rm [Fe/H] \rangle = -0.19 \pm 0.02$ dex, which we adopt as the representative metallicity of Berkeley~36.

Not every $\alpha$ element has uniform spectral-line coverage in the literature, so the analysis was restricted to O, Mg, Si, S, Ca, and Ti. All 106 stars in our spectroscopic sample have reliable [Fe/H] measurements, whereas $\alpha$-element measurements are not available for every star. Consequently, the number of stars with available measurements varies among the individual elements: 6, 6, 102, 75, 98, and 102 for O, Mg, Si, S, Ca, and Ti, respectively. The resulting median ratios are [O/Fe] $=0.23\pm0.03$, [Mg/Fe] $=0.03\pm0.05$, [Si/Fe] $= 0.10 \pm 0.04$, [S/Fe] $= 0.66 \pm 0.10$, [Ca/Fe] $= 0.16 \pm 0.03$, and [Ti/Fe] $= 0.20 \pm 0.06$ dex, corresponding to a weighted mean [$\alpha$/Fe] = $0.17 \pm 0.02$ dex (Figure~\ref{fig:element_histo}b). We note that the O and Mg ratios rest on only six member stars each and should therefore be regarded as tentative, whereas the Si, Ca, and Ti abundances are based on considerably larger samples and provide more reliable estimates of the $\alpha$-element abundance pattern. The unusually broad [S/Fe] distribution should be treated with caution. The enhanced O, Ca, and Ti abundances indicate that the chemical enrichment of Berkeley~36 was primarily driven by core-collapse (Type~II) supernovae, which dominate the production of $\alpha$-elements in the early stages of Galactic chemical evolution \citep{McWilliam1997, Kobayashi2006, Nomoto2013}. The varying enhancement levels among the individual $\alpha$-elements likely reflect the metallicity- and progenitor-mass-dependent nucleosynthetic yields of massive stars, as predicted by theoretical supernova models \citep{Woosley1995, Kobayashi2006, Kobayashi2020}. We do not interpret the sulfur abundance as evidence for a distinct enrichment pattern because of the unusually large observed scatter in [S/Fe].

The comparatively large [S/Fe] $= 0.66 \pm 0.10$ dex value, based on 75 member stars, should be treated with caution. The [S/Fe] distribution exhibits an unusually large scatter, extending over nearly 3 dex, which is substantially larger than the intrinsic abundance dispersions typically expected for open clusters \citep[e.g.,][]{Bovy2016, Sinha2024}. We therefore do not consider the sulfur measurements sufficiently reliable for drawing conclusions about the chemical properties or enrichment history of Berkeley~36.

To place this measurement on the scale required by the stellar evolution models \citep[e.g.,][]{Alzhrani2025b, Alzhrani2025,  Cinar2024, Cinar2025, Cinar2026, Elsanhoury2025, Elsanhoury2026}, the mean iron abundance $\langle {\rm [Fe/H]} \rangle = -0.19 \pm 0.02$ dex was converted to a heavy-element mass fraction using the calibrated relation \citep[e.g.,][]{Tasdemir2025, Bilir2026, Bisht2026b, Bisht2026c, Canbay2026b}:
\begin{equation}
Z_{\rm x} = \frac{Z}{0.7515 - 2.78 \times Z}
\end{equation}
\begin{equation}
\text{[Fe/H]} = \log (Z_{\rm x}) - \log \left( \frac{Z_{\odot}}{1 - 0.248 - 2.78 \times Z_{\odot}} \right),
\end{equation}
with $Z_{\odot}=0.0152$, giving $Z = 0.01001 \pm 0.0004$, which is adopted in the subsequent isochrone fitting.

\begin{figure*}
    \centering
    \includegraphics[width=1\linewidth]{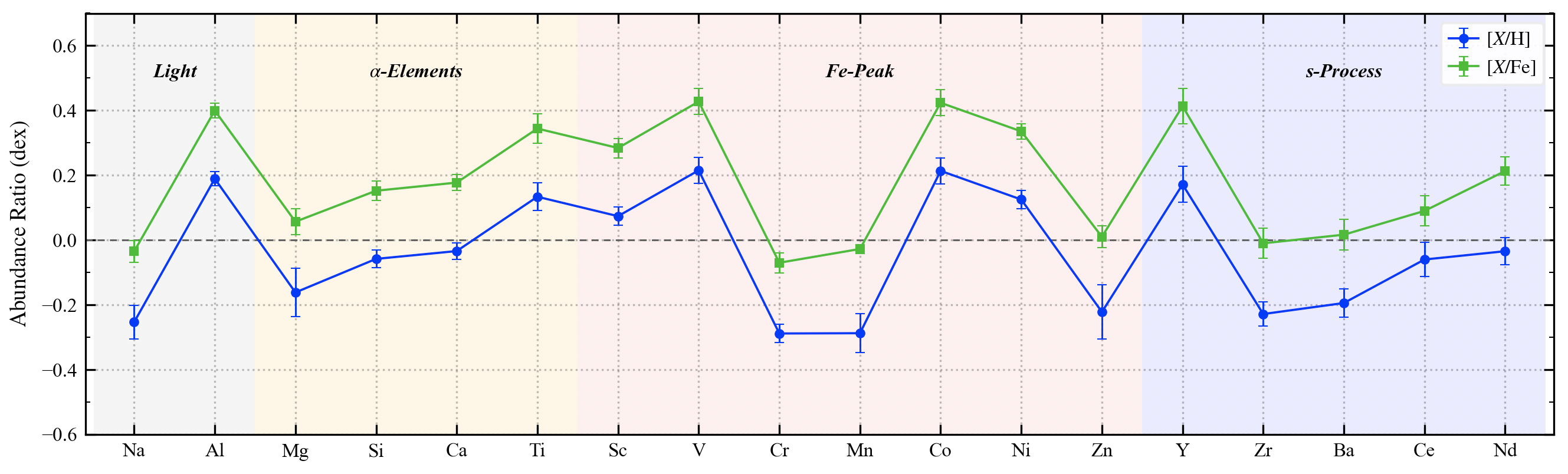}
    \caption{Elemental abundance pattern of Berkeley~36 derived from member stars with high-resolution spectroscopy. Blue circles show [X/H] and green squares show [X/Fe] ratios (in dex) for each element, with error bars representing the associated uncertainties. Elements are grouped into four categories, indicated by the shaded background regions and labels: light elements (Na, Al), $\alpha$-elements (Mg, Si, Ca, Ti), Fe-peak elements (Sc, V, Cr, Mn, Co, Ni, Zn), and s-process elements (Y, Zr, Ba, Ce, Nd). The dashed horizontal line marks the solar abundance ratio.}

    \label{fig:abundance_pattern}
\end{figure*}

Figure~\ref{fig:abundance_pattern} presents the elemental abundance pattern of Berkeley~36, showing both [X/H] and [X/Fe] ratios for a set of light, $\alpha$-, Fe-peak, and s-process elements. The light elements Na and Al display distinct behavior: while [Na/H] is slightly sub-solar, [Al/Fe] shows a clear enhancement of $\sim+0.4$ dex, consistent with the well-known Na-Al abundance trends observed in evolved cluster giants. Among the $\alpha$-elements (Mg, Si, Ca, Ti), the [X/Fe] ratios are all mildly enhanced relative to solar, ranging from $\sim+0.05$ to $+0.35$ dex, a pattern typical of an old, moderately metal-poor OC whose stars formed before significant enrichment by Type Ia supernovae. The Fe-peak elements exhibit more scatter: Sc, V, Co, and Ni are all enhanced in [X/Fe] by $+0.3$ to $+0.4$ dex, whereas Cr and Mn remain close to the solar value, and Zn is essentially solar within the uncertainties. This dichotomy likely reflects the differing nucleosynthetic origins of these species, since Sc, V, Co, and Ni are produced primarily in core-collapse supernovae (Type~II), while Cr and Mn are more closely tied to Type Ia events. Finally, the s-process elements show a mixed picture: Y is enhanced by $\sim+0.4$ dex in [Y/Fe], while Zr, Ba, Ce, and Nd remain close to the solar ratio, suggesting only a modest contribution from asymptotic giant branch (AGB) nucleosynthesis to the present-day chemical inventory of the cluster. Overall, the abundance pattern of Berkeley~36 is broadly consistent with that expected for an old Galactic disk OC, exhibiting generally enhanced $\alpha$-element abundances and no evidence for anomalous s-process enrichment \citep[e.g.,][]{Bragaglia2008, Magrini2017, Bragaglia2018}.

\subsection{Isochrone Fitting}

\begin{figure*}[!ht]
\centering
\includegraphics[width=0.45\linewidth]{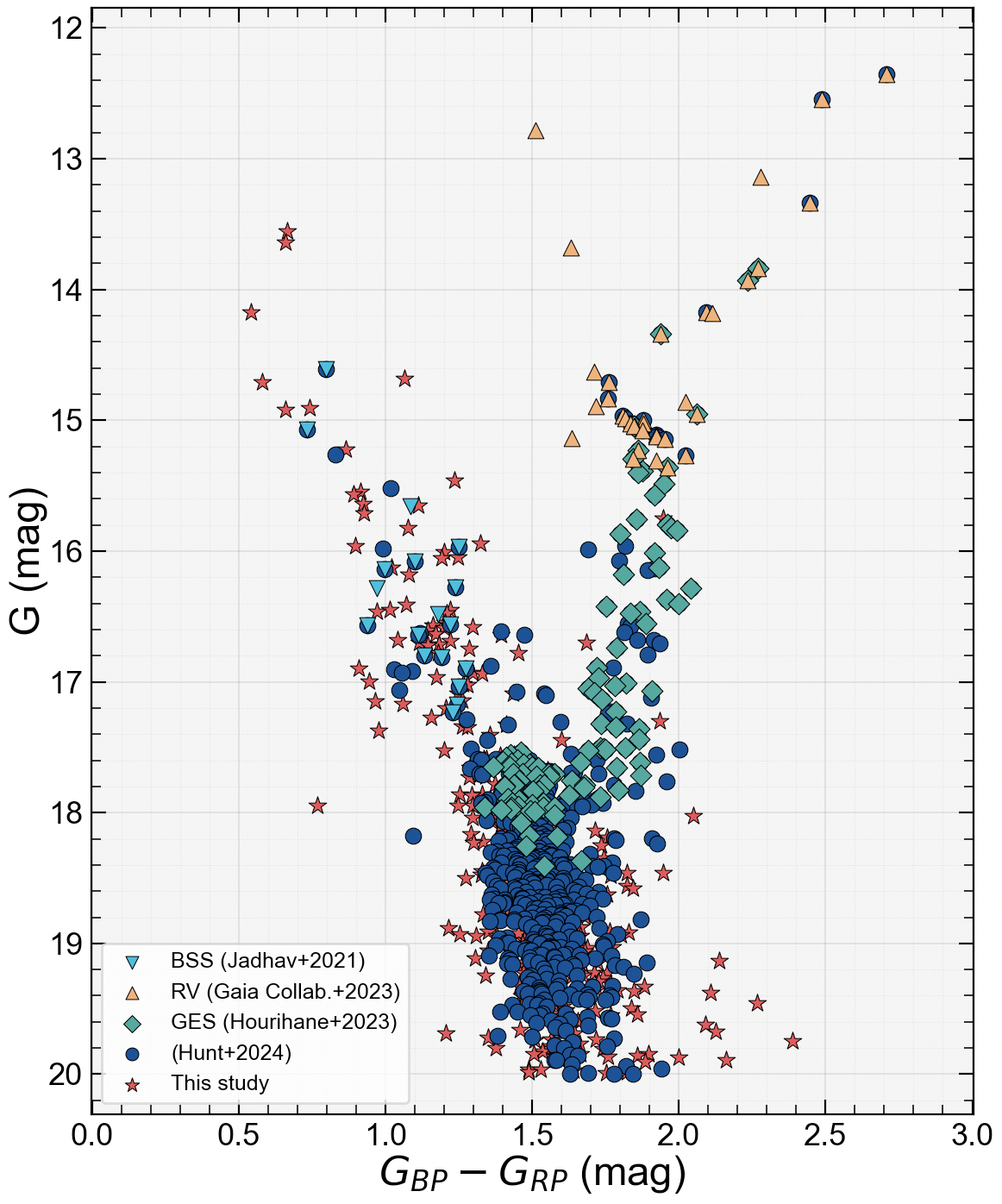}
\includegraphics[width=0.47\linewidth]{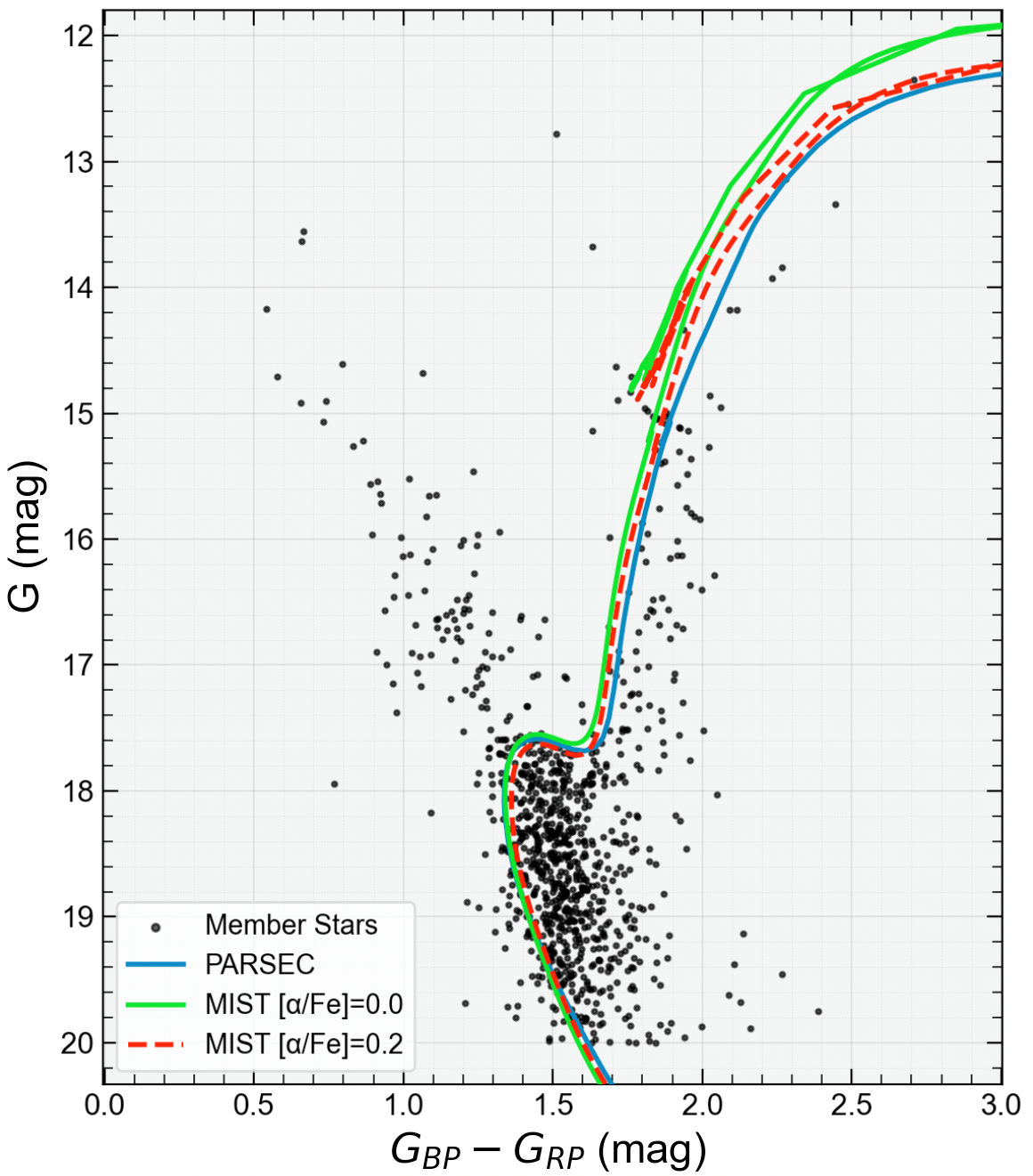}

\caption{Members are color-coded according to the cross-matched literature catalogs, as indicated in the legend (left panel). CMDs of the high-probability cluster members. In the right panel, cluster members are shown as black points. The solid curves represent the best-fitting theoretical PARSEC (blue) and MIST (green and red) isochrones.}
\label{fig:CMDs}
\end{figure*}

CMDs remain one of the most powerful diagnostics for recovering the fundamental parameters of star clusters, since the loci traced by genuine members can be directly compared with theoretical stellar evolutionary tracks \citep[e.g.,][]{Carrera2011}. The left panel of Figure~\ref{fig:CMDs} shows the $G$ versus $G_{\rm BP}-G_{\rm RP}$ diagram of Berkeley~36, built exclusively from the high-probability members recovered in our astrometric membership analysis. The resulting sequence is well populated from the MS up through the MSTO and continues into the RGB and red clump region. As a consistency check, we compared our astrometrically selected members against independent catalogs from the literature, namely the list of relevant catalogs used for the Berkeley 36 cross-match, \citep[e.g,][]{Jadhav2021bss, Hourihane2023, Hunt2024}. The sequences obtained from our own selection agree closely with these external samples across the full magnitude range probed, supporting the reliability of the adopted membership criteria.

Before the isochrone comparison, the basic cluster parameters entering the CMD analysis were fixed using independent methods: metallicity from spectroscopy, and reddening and distance from photometric/astrometric data. For Berkeley~36 we adopted a heliocentric distance of $\langle d \rangle = 4377\pm510$~pc, a Gaia $G$-band extinction of $\langle A_{\rm G}\rangle = 1.350 \pm 0.045 $~mag, and a metallicity of $\langle [{\rm Fe/H}]\rangle = -0.19\pm 0.02$~dex. With these quantities held fixed, the cluster's age was then constrained by matching the observed CMD morphology to theoretical isochrones.

We tested the observed sequences against two independent sets of theoretical isochrones, PARSEC CMD 3.9\footnote{\url{https://stev.oapd.inaf.it/cgi-bin/cmd}} \citep{Bressan2012} and MIST v2.5\footnote{\url{https://mist.science/interp_isos.html}} \citep{Choi2016, Dotter2016}, both evaluated at the spectroscopic metallicity quoted above. For the MIST grid we assumed a rotation rate of $v/v_{\rm crit}=0$ and considered $\alpha$-enhancement values of $[\alpha/{\rm Fe}]=0.0 \text{ and/or } 0.2$~dex. %As in the case of Berkeley~32 \citep[e.g.,][]{ThisPaper2026}, the two isochrone families produce very similar fits to the observed CMD (Figure~\ref{fig:CMDs}), showing that the recovered parameters do not depend strongly on the choice of evolutionary models. 
The MIST turn-off falls at slightly bluer colors relative to PARSEC, a difference attributable to small physical differences between the two model sets (solar calibration, rotational mixing, convective overshoot); comparable offsets have been reported for other clusters \citep{Bastian2025}. Changing the $\alpha$-element abundance in the MIST grid affects the fitted CMD, indicating that the age solution is not sensitive to the assumed $\alpha$-enhancement over the range tested. The best-fitting isochrones give an age of $6.8 \pm 0.5$~Gyr for Berkeley~36, with the MS, MSTO, and RGB simultaneously reproduced within the quoted uncertainties by both model grids. %This age is compared against independent age estimates in Section~\ref{sec:Ymg_age}.

\subsection{Binary Fraction}\label{sec:binaryfraction}

The fraction of binary systems in an OC carries useful information about its dynamical history, internal kinematics, and long-term stability \citep[e.g.][]{vonHippel2002, Sollima2007}. Because unresolved binaries can shift and broaden the main sequence in a CMD, ignoring them can lead to biased estimates of a cluster's stellar population \citep{Milone2012}. Following the photometric approach described by \citet{Milone2012} and \citet{Donada2023}, we examined the binary content of Berkeley~36.

Main-sequence stars were selected from the turnoff magnitude ($G_{\rm MSTO} = 18$ mag) down to four magnitudes fainter. This selection yielded a sample of 337 MS stars. A fourth-degree polynomial was fitted to the median ridge line of this sample to represent the single-star sequence. To reduce the impact of differential reddening, all magnitudes and color indices were first corrected for extinction, using the cluster's distance modulus and mean reddening (Table~\ref{tab:summary}), so that the fit was performed in the de-reddened $(G_{\rm BP}-G_{\rm RP})_0$ versus $M_{\rm G}$ CMD.

The magnitude difference between an unresolved binary and a single star of the same mass was estimated from $\Delta m = -2.5\log_{10}(1 + q^{\alpha})$ with the mass--luminosity index fixed at $\alpha = 3.5$, appropriate for main-sequence stars \citep{Eker2015, Eker2018, Eker2024}, and $q = M_2/M_1$ the mass ratio of the two components. For $q = 0.5$, this yields $\Delta m \approx -0.107$ mag, which we adopted as the minimum magnitude offset for classifying a star as a binary candidate, that is, any star lying more than 0.107 mag above the single-star ridge line.

\begin{figure*}
    \centering
    \includegraphics[width=1\linewidth]{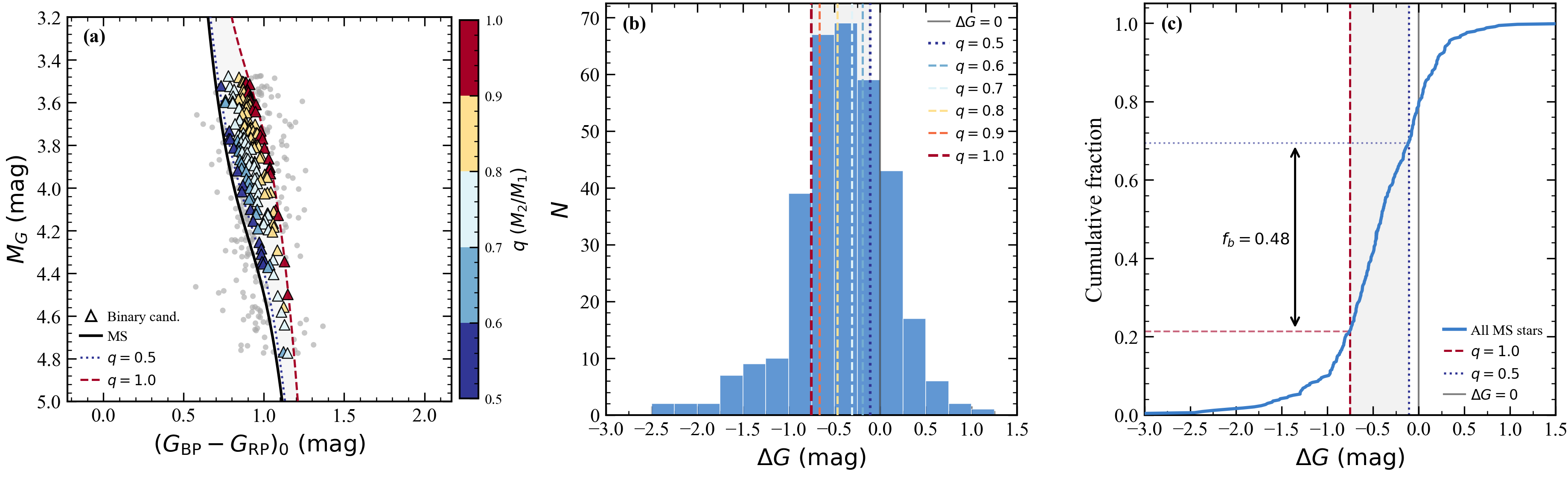}
    \caption{Binary fraction analysis of Berkeley~36. (a) CMD of the MS stars. The solid line shows the fitted single-star sequence; the dashed and dotted lines mark the lower and upper limits of the binary selection region, respectively. Triangle symbols indicate binary candidates. (b) Distribution of $\Delta G$ residuals. The shaded region denotes the binary selection zone between the lower and upper limits. (c) Cumulative distribution of $\Delta G$. The binary fraction $f_b$ is indicated by the vertical arrow spanning the two selection limits.}
    \label{fig:berk36_binary}
\end{figure*}

Figure~\ref{fig:berk36_binary}a shows the resulting CMD, with the fitted single-star locus and the two selection boundaries overplotted. The bulk of the sample follows the main-sequence fit closely, but a second, fainter-in-offset sequence running parallel and above it is clearly visible, as expected for a population of unresolved binaries; the scatter widens toward fainter magnitudes as photometric errors grow. Figure~\ref{fig:berk36_binary}b shows the corresponding $\Delta G$ histogram, spanning $-3.075$ to $2.328$ mag with a median of $-0.314$ mag; most stars cluster near the single-star locus, but a distinct secondary excess appears beyond the adopted $\Delta m$ threshold. The cumulative $\Delta G$ distribution in Figure~\ref{fig:berk36_binary}c marks the region between the two boundaries used to isolate binary candidates.

Out of 337 MS stars, 165 fall within the binary-candidate region, corresponding to a binary fraction of $f_{\rm b} = 0.480 \pm 0.017$ (68\% confidence interval: $0.463, 0.497$). This is consistent, within uncertainties, with the binary fraction we derived for Berkeley~32 and with values reported for other old OCs analyzed with comparable techniques \citep[e.g.,][]{Milone2012, Donada2023, Cinar2026b}.

Splitting the binary candidates by mass ratio, we find 15 systems with $0.5 \leq q \leq 0.6$, 30 with $0.6 < q \leq 0.7$, 49 with $0.7 < q \leq 0.8$, 57 with $0.8 < q \leq 0.9$, and 14 with $0.9 < q \leq 1$. As with Berkeley~32, the distribution is not flat: it rises steadily from low $q$ to a clear maximum in the $0.8$--$0.9$ bin, with the next-highest bin ($0.7$--$0.8$) also well populated, before dropping sharply near $q \sim 1$. The concentration near $q \sim 0.8$--$0.9$ is consistent with the pile-up expected for detached binaries, while the broader shoulder extending down to $q \sim 0.7$ may include a mix of semi-detached, contact, or otherwise interacting systems rather than a single well-defined population \citep[e.g.,][]{Bilir2005, Demircan2006, Eker2006, Ibanoglu2006, Eker2014}. Spectroscopic confirmation would be needed to disentangle these sub-populations. Overall, the mass-ratio distribution in Berkeley~36 mirrors that found for Berkeley~32, suggesting that similar dynamical processes shape the binary population in both clusters despite differences in age or metallicity. As before, since our method loses sensitivity below $q \approx 0.5$, the quoted binary fraction should be taken as a lower limit on the cluster's true binary content.

\subsection{Blue and Yellow Straggler Stars}\label{sec:BSS_YSS}

We also investigated the Blue Straggler Star (BSS) and Yellow Straggler Star (YSS) populations of Berkeley~36 using cluster members with membership probabilities greater than 70\%. As expected for an old and dynamically evolved OC, Berkeley~36 exhibits a noticeable BSS population \citep{Linck2026}.

BSSs occupy a well-defined region of the CMD, lying blueward and brighter than the cluster MSTO but still consistent with core hydrogen burning, since they are understood to be the products of mass transfer, mergers, or stellar collisions within binary systems \citep{McCrea1964, Leonard1989, Andronov2006}. Following the approach adopted by \citet{Jadhav2021bss} and \citet{Linck2026}, we defined the boundaries of the BSS region in the CMD using the zero-age main sequence (ZAMS) and the terminal-age main sequence (TAMS), computed from PARSEC isochrones at the cluster age, metallicity, distance modulus, and mean extinction listed in Table~\ref{tab:summary}. The ZAMS marks the blue (bluest, unevolved) edge of the region, corresponding to the locus of stars that have just begun hydrogen burning, while the TAMS marks its red edge, corresponding to the reddest colors a star can reach while still on the main sequence before evolving toward the subgiant branch. Together, the ZAMS and TAMS define the full width in color that a genuine MS star can occupy at any given magnitude; a star located blueward of this band, and brighter than the MSTO, cannot be explained by single-star evolution and is therefore flagged as a BSS candidate. We additionally overlaid the equal-mass binary sequence, obtained by shifting the isochrone brighter by $-0.753$ mag (for $q=1$) to help distinguish genuine BSSs from unresolved near-equal-mass binaries of normal MS stars that can mimic a similar CMD position.

YSS candidates were identified following the same methodology, as stars lying just redward of the TAMS boundary and above the base of the giant branch -- a region populated by BSSs that have started to evolve off the main sequence and cool toward the giant branch while retaining an anomalously high luminosity for their color, as discussed by \citet{Linck2026}.

To identify previously reported BSS candidates, we cross-matched our member sample with the catalog of \citet{Jadhav2021bss}, recovering 18 stars in common. Applying the CMD criteria described above, we identified 10 YSS candidates. In addition to the literature-confirmed sources, we identified 65 new photometric BSS candidates from their positions in the CMD, as shown in Figure~\ref{fig:BSS}, bringing the total BSS sample to 83 stars. Although the membership probabilities are generally high, spectroscopic follow-up observations would be useful to assess potential contamination from field stars and unresolved binaries. The full list of BSS and YSS candidates is presented in Appendix{~\ref{sec:appendix_bss_b36}.

\begin{figure}
    \centering
    \includegraphics[width=1\linewidth]{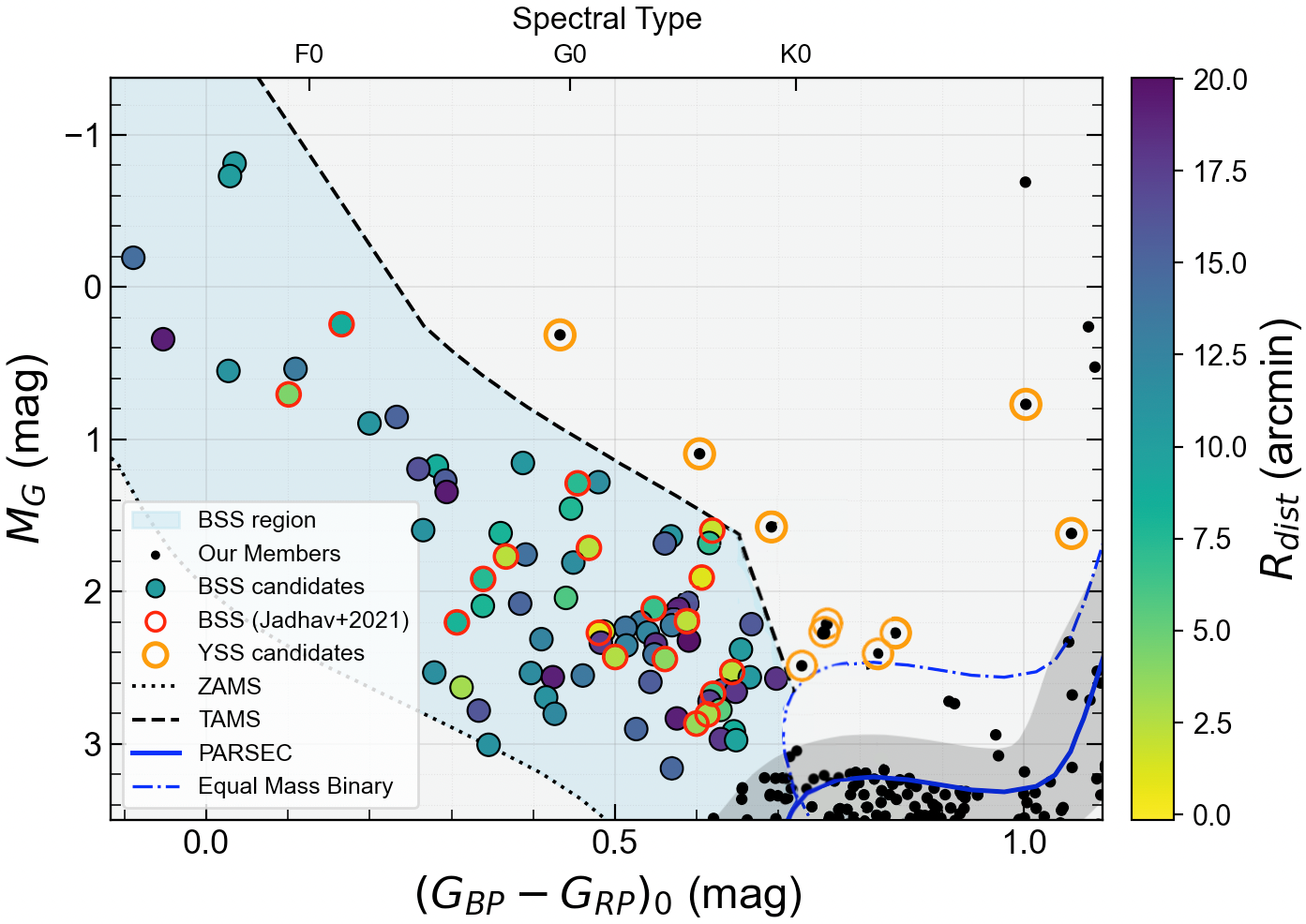}
    \caption{CMD of the Berkeley 36 in the $M_G$ versus $(G_{\rm BP}-G_{\rm RP})_0$ plane. The spectral types adopted from \citet{Pecaut2013} are shown along the upper axis. The stars are color-coded according to their projected distance from the cluster centre, as indicated by the color scale on the right-hand side of the figure. Member stars are displayed as black dots. The solid blue line represents the best-fit PARSEC isochrone, while the dashed and dotted lines indicate the derived TAMS and the ZAMS, respectively. The blue shaded regions delineate the search areas for BSS candidates. Open orange symbols represent the identified YSS candidates, and blue inverted triangles highlight BSS following the classification by \citet{Jadhav2021bss}. The grey-shaded region around the best-fit isochrone represents the uncertainty in the derived cluster age.}
    \label{fig:BSS}
\end{figure}

\subsection{Mass Segregation}\label{sec:relax_b36}

Mass segregation is one of the key diagnostics of the dynamical evolution of stellar clusters. It arises from two-body relaxation, through which kinetic energy is redistributed among cluster members, driving the system toward partial energy equipartition. As a result, more massive stars progressively lose kinetic energy and migrate toward the cluster center, while lower-mass stars gain kinetic energy and are displaced to larger radii \citep{PortegiesZwart2010, Dib2018}.

To investigate this effect in Berkeley~36, cluster members were divided into four mutually exclusive classes based on their position in the CMD: supergiants ($N = 10$), giants ($N = 81$), MS stars ($N = 719$), and blue straggler stars (BSS, $N = 86$), which trace different evolutionary stages and, to first order, different stellar masses. The spatial distribution of each population was examined using the radial cumulative distribution functions (RCDFs) shown in Figure~\ref{fig:rcdf_berk36}.

The giant population shows the steepest RCDF, rising faster than the MS distribution at all radii and indicating a stronger central concentration. The supergiant population follows a broadly similar, though noisier, trend consistent with its small sample size, remaining close to or slightly above the MS curve out to intermediate radii. In contrast, the BSS population is markedly less centrally concentrated than either the MS or giant stars, with its RCDF lying below the other three curves across essentially the entire radial range.

Two-sample Kolmogorov--Smirnov (KS) tests were applied to each pair of populations to assess the statistical significance of these differences. The supergiant--MS comparison gives $p_{\rm KS} = 0.550$, and the supergiant--giant comparison gives $p_{\rm KS} = 0.237$, neither indicating a significant difference, consistent with the limited number of supergiants. The giant--MS comparison yields $p_{\rm KS} = 0.071$, a marginal difference suggestive of mild central concentration among the more evolved, higher-mass stars. By contrast, the BSS population differs significantly from both the giant ($p_{\rm KS} < 0.001$) and MS ($p_{\rm KS} < 0.001$) distributions, confirming that its spatial distribution is statistically distinct from the rest of the cluster population.

These results indicate that the giant population is somewhat more centrally concentrated than the MS stars, consistent with expectations from mass segregation, whereas the BSS population shows the opposite behavior, being significantly more extended than both the giant and MS distributions. This is a notable departure from the classical mass-segregation picture, since BSS are typically expected to behave dynamically as relatively massive objects and to be centrally concentrated as a result of collisional or binary-mediated formation channels \citep[e.g.,][]{Bailyn1995, Ferraro2026}. However, observations of several old OCs have shown that BSS do not always exhibit a strong central concentration, and instead suggest that primordial binaries and mass-transfer processes constitute the dominant formation channel \citep{Mathieu2009, Geller2011, Geller2012}. In addition, numerical simulations by \citet{Mapelli2004, Mapelli2006} demonstrated that BSS formed through binary evolution can remain preferentially distributed in the outer regions of clusters, in contrast to collisionally formed BSS, which are expected to be more centrally concentrated. Similar conclusions have also been reached by population synthesis and Monte Carlo cluster simulations, which predict that the radial distribution of BSS is highly sensitive to the relative contributions of binary evolution and dynamical interactions \citep{Hypki2017}. The extended spatial distribution of BSS in Berkeley~36 is therefore consistent with these observational and theoretical studies and may instead point toward a predominantly primordial or outer-region formation origin for at least part of this population, rather than one purely governed by dynamical mass segregation.

\begin{figure}
    \centering
    \includegraphics[width=0.97\linewidth]{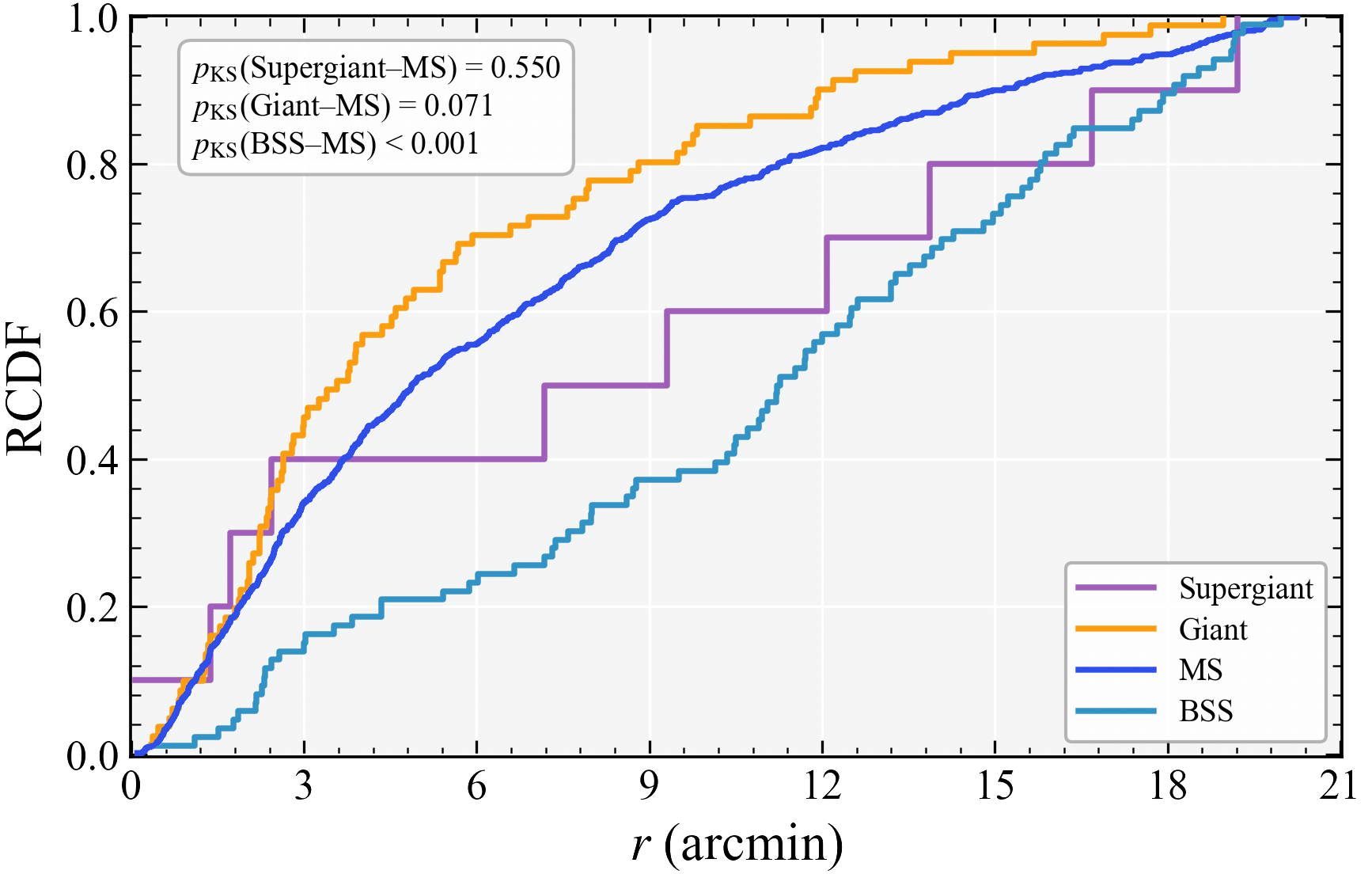}
    \caption{RCDFs of supergiant, giant, MS, and BSS populations in Berkeley~36. The radial distance corresponds to the projected angular separation from the cluster center in arcminutes. Populations were selected according to their $G$-band magnitude ranges and membership probabilities.}
    \label{fig:rcdf_berk36}
\end{figure}

\section{Dynamical Orbital Parameters}\label{sec:orbit}

\subsection{Radial Velocity Determination}\label{sec:rv}

\begin{figure}
    \centering
    \includegraphics[width=1\linewidth]{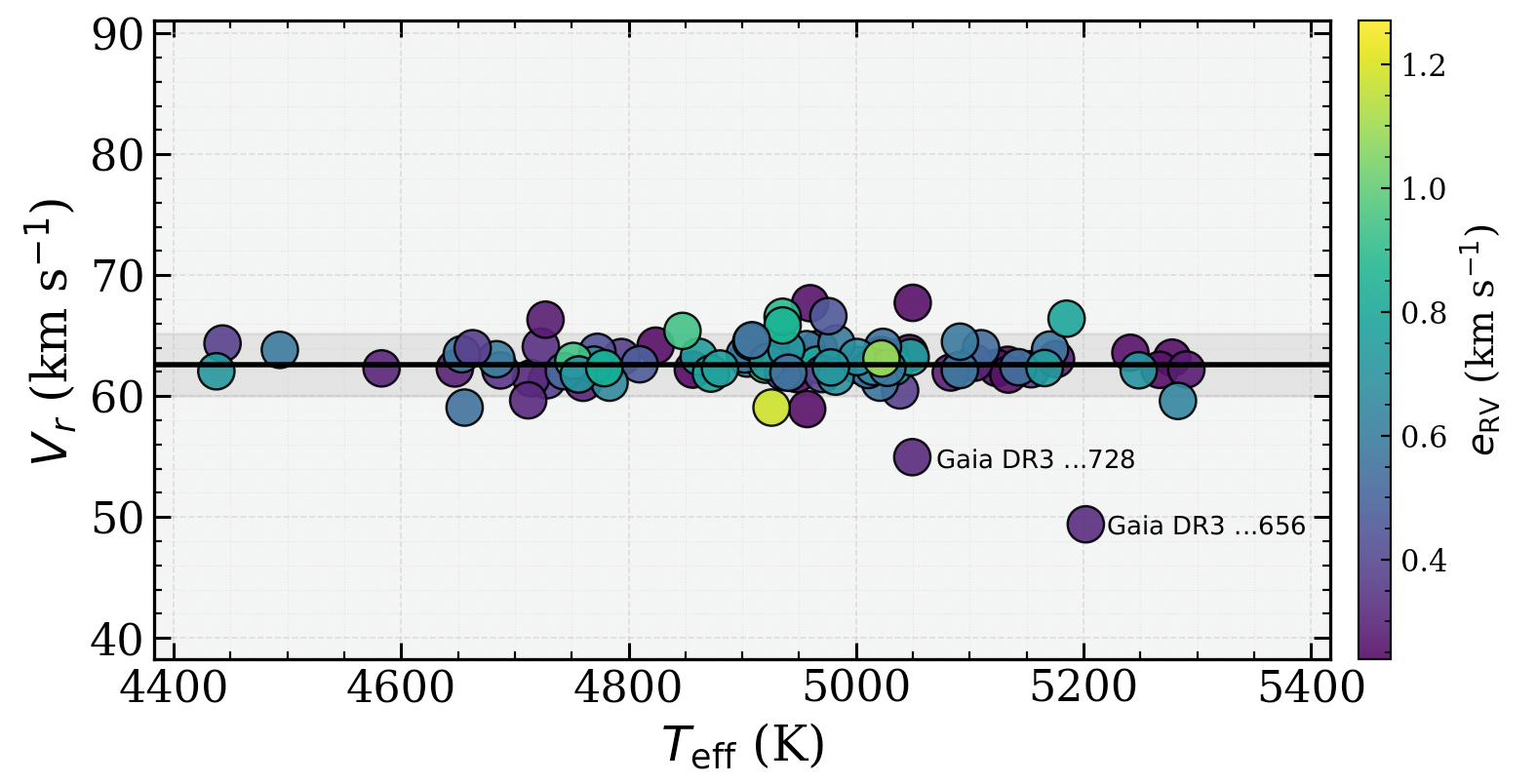}
    \caption{Radial velocity of the confirmed Berkeley~36 members plotted against effective temperature, with individual points color-coded by their radial-velocity uncertainty ($e_{\rm RV}$; km~s$^{-1}$). The horizontal solid line marks the adopted systemic velocity of the cluster, and the shaded band shows the associated $1\sigma$ dispersion about that value.}
    \label{fig:radial_vel_be36}
\end{figure}

The systemic radial velocity of Berkeley~36 was derived from spectroscopic measurements in GES DR5.1 \citep{Hourihane2023}, as described in Section~\ref{sec:data}, and was selected over other available surveys owing to its higher spectral resolution. Following the weighting scheme of \citet{Soubiran2018} and \citet{Carrera2022}, individual member velocities were combined into an inverse-variance-weighted mean, with each star assigned a weight $w_i = 1/\sigma_{v_{\rm rad}, i}^2$, from which we obtain both the cluster's systemic velocity and its internal dispersion. Based on $N = 106$ confirmed members, we derive $\langle V_{\rm rad, OC}\rangle = 62.8 \pm 0.04$~km~s$^{-1}$ (standard error of the mean), with an internal velocity dispersion of $\sigma_{V_{\rm rad, OC}} = 2.58$~km~s$^{-1}$.

Figure~\ref{fig:radial_vel_be36} illustrates the distribution of individual stellar velocities as a function of effective temperature ($T_{\rm eff}$) over the observed range (4400-5300 K). Most stars scatter tightly around the adopted mean, confirming that the sample forms a kinematically coherent group. Two stars, Gaia DR3~3032953420220233728 and Gaia DR3~3032952148909926656, lie beyond the nominal $3\sigma$ interval, yet their astrometric quality and CMD placement remain consistent with cluster membership. Their [Fe/H] values of $-0.26$ and $-0.13$ dex, respectively, are also broadly consistent with the cluster mean of $-0.19\pm0.02$ dex, providing no strong evidence that they are field interlopers.} We therefore interpret their discrepant velocities as the signature of orbital motion in unresolved binary systems rather than non-membership. Excluding these two stars from the weighted mean shifts $\langle V_{\rm rad, OC}\rangle$ by only 0.3 km~s$^{-1}$, well within the quoted uncertainty, so their inclusion does not affect the derived systemic velocity or the kinematic conclusions drawn from it.

\subsection{Cluster Orbit Analysis}

To reconstruct the dynamical history of Berkeley~36, we integrated its orbit through a model Galactic potential using the \texttt{galpy} package \citep{Bovy2015}. Such an analysis constrains how far the cluster has migrated radially and vertically since its formation, and how strongly it has been perturbed by the disk potential over its lifetime. Our default choice of potential is the axisymmetric \textsc{MWPotential2014}, built from a Miyamoto--Nagai disk \citep{Miyamoto1975}, an NFW dark-matter halo \citep{Navarro1996}, and a power-law bulge with an exponential cutoff \citep{Bovy2013}, so that
\begin{equation}
\Phi_{\rm total}(R, z) = \Phi_{\rm bulge}(r) + \Phi_{\rm disk}(R,z) + \Phi_{\rm halo}(r),
\end{equation}
with $r = \sqrt{R^2 + z^2}$ the spherical Galactocentric radius. As a consistency check, we repeated the integration with the \textsc{McMillan2017} potential \citep{McMillan2017}, an independently calibrated mass model with updated disk and halo parameters; the two potentials produced orbital solutions that agree within their respective uncertainties. We therefore report results based on \textsc{MWPotential2014}, following common practice in recent open-cluster orbit studies \citep[e.g.,][]{Cantat-Gaudin20, Tarricq2021, Donor2020}.

Because an axisymmetric potential cannot capture perturbations from the Galactic bar and spiral arms, we also tested the orbit's sensitivity to these non-axisymmetric components. The bar was represented with \texttt{DehnenBarPotential}, i.e., a rotating quadrupolar term of the form
\begin{equation}
\Phi_{\rm bar}(R, \phi, t) = A(t)\cos\left[2(\phi - \Omega_{\rm bar} t)\right] f(R),
\end{equation}
where $\Omega_{\rm bar}$ is the pattern speed of the bar and $A(t)$ its (time-dependent) amplitude, and spiral structure was added through a steady-state spiral perturbation (\texttt{SpiralArmsPotential}). Standard literature values were adopted for the bar pattern speed ($\Omega_{\rm bar}\sim 40$~km~s$^{-1}$~kpc$^{-1}$), the bar scale length ($R_{\rm bar}\sim3.5$~kpc), and a moderate perturbation amplitude, consistent with previous dynamical modeling of the Galaxy.

The six phase-space coordinates required for the integration, sky position ($\alpha$, $\delta$), heliocentric distance ($d$), proper motion ($\mu_\alpha\cos\delta$, $\mu_\delta$), and the systemic radial velocity from Section~\ref{sec:rv}, were propagated forward and backward in time using a 1~Myr timestep, over a baseline set by the cluster's estimated age of $\sim$6.8~Gyr. We assumed a local circular speed of $V_{\rm rot} = 220$~km~s$^{-1}$ and a vertical solar offset of $Z_0 = 25\pm5$~pc \citep{Bovy2012}, and converted to Galactocentric coordinates following the geometric prescription of \citet{Tuncel2019}, with $R_0 = 8.2\pm0.1$~kpc. We also compute the cluster's guiding radius, the radius of the circular orbit that shares the cluster's specific angular momentum, as a tracer of radial mixing and possible migration \citep[e.g.,][]{Binney2008, Schoenrich2009}.

\begin{figure}
    \centering
    \includegraphics[width=0.85\linewidth]{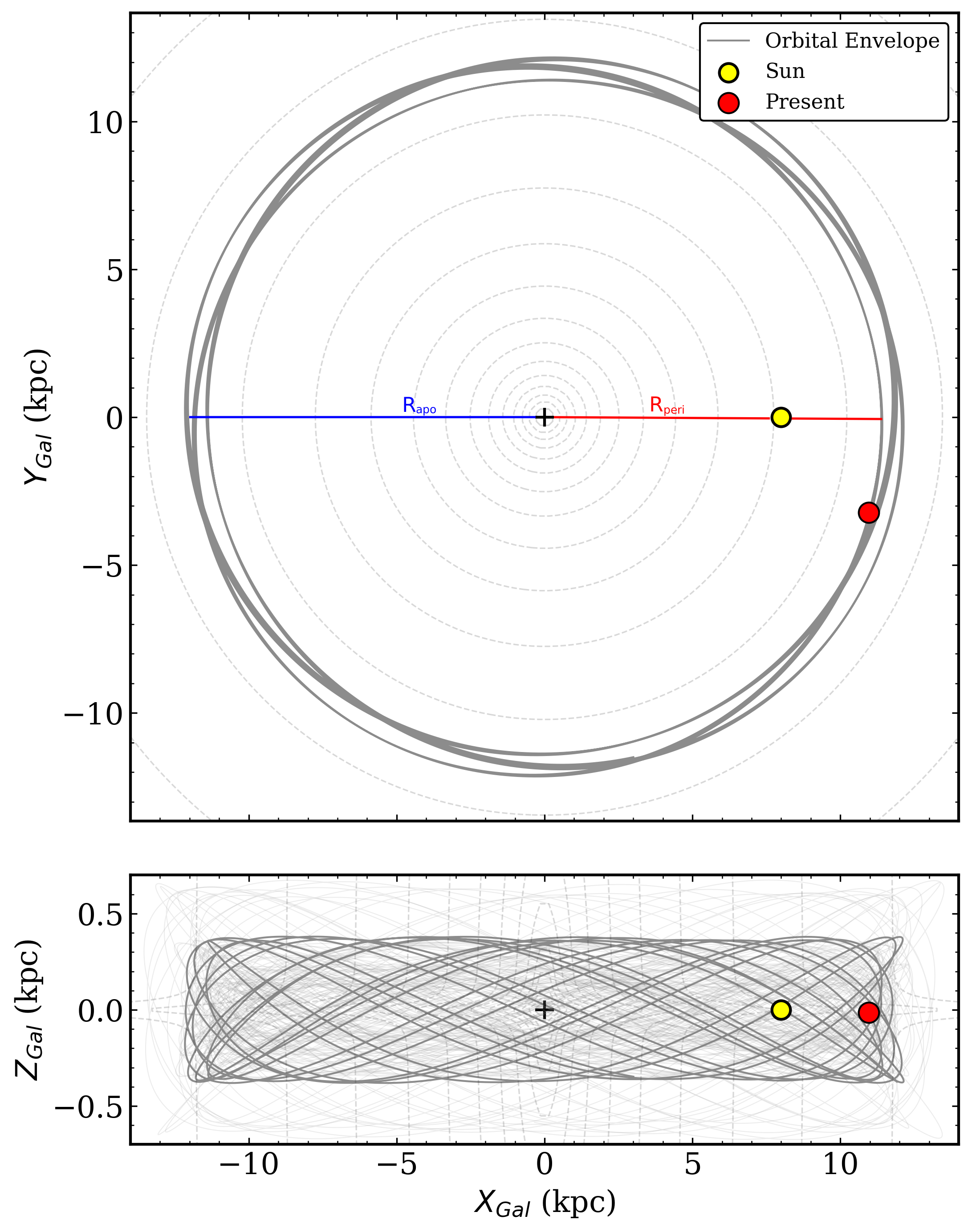}
    \caption{Reconstructed Galactic orbit of Berkeley~36. \textit{Top:} projection onto the $X_{\mathrm{Gal}}$--$Y_{\mathrm{Gal}}$ plane. \textit{Bottom:} projection onto the $X_{\mathrm{Gal}}$--$Z_{\mathrm{Gal}}$ plane; thin gray curves trace the family of orbits obtained from the Monte Carlo resampling of the observational uncertainties. The cluster's present-day position is marked with a filled red circle, the Sun's position with a filled yellow circle, and the background gray shading traces the Galactic mass density.}
    \label{fig:orbit_be36}
\end{figure}

To propagate observational errors into the orbital solution, we generated an ensemble of Monte Carlo realizations of the input phase-space vector and re-integrated the orbit for each realization. From the resulting distribution we derive the perigalactic distance ($R_{\rm peri}$), apogalactic distance ($R_{\rm apo}$), mean orbital radius ($R_{\rm m}$), orbital eccentricity ($e$), maximum vertical excursion ($Z_{\rm max}$), guiding radius ($R_{\rm g}$), and orbital period, listed in Table~\ref{tab:summary}. Berkeley~36 is found to follow a nearly circular orbit, with $e = 0.036$, $R_{\rm peri} = 11.369$~kpc and $R_{\rm apo} = 12.214$~kpc (Figure~\ref{fig:orbit_be36}), giving a mean orbital radius of $R_{\rm m} \simeq 11.79$~kpc and a guiding radius of $R_{\rm g} = 11.751$~kpc -- both well beyond the solar circle, placing the cluster firmly in the outer Galactic disk. The current Galactocentric distance is $R_{\rm GC} = 11.422$~kpc, closely matching $R_{\rm m}$ and $R_{\rm g}$, suggesting that the cluster is currently near its time-averaged orbital radius rather than caught in transit between extremes. The modest vertical amplitude, $Z_{\rm max} = 380$~pc, together with the low eccentricity, points to dynamically quiet, disk-like kinematics rather than a history of strong scattering. Given its outer-disk location and near-circular motion, Berkeley~36 appears to have experienced comparatively little radial migration relative to more eccentric OCs, though we caution that any inference about its birth radius from long-term backward integration remains subject to the adopted Galactic potential and to the (still uncertain) treatment of bar and spiral perturbations, and should be read as indicative rather than definitive.

The Galactic space velocity components ($U$, $V$, $W$) of Berkeley~36 were computed using the \textsc{galpy} package, based on the cluster's mean proper motion, distance, and radial velocity. These raw velocities were then corrected to the Local Standard of Rest (LSR) using the solar motion values of \citet{Coskunoglu2011}, namely $(U_{\odot}, V_{\odot}, W_{\odot}) = (8.83\pm0.24, 14.19\pm0.34, 6.57\pm0.21)$~km~s$^{-1}$. This correction was applied by adding the solar motion components to the original velocities, yielding the LSR-corrected values, which are also listed in Table~\ref{tab:summary}. Finally, the total space velocity of the cluster was calculated from the LSR-corrected components as
\begin{equation}
    S_{\rm LSR} = \sqrt{U_{\rm LSR}^2 + V_{\rm LSR}^2 + W_{\rm LSR}^2},
\end{equation}
with the associated uncertainty obtained through standard error propagation. The resulting space velocity of Berkeley~36 is $S_{\rm LSR} = 61.36 \pm 5.48$~km~s$^{-1}$. The obtained value falls within the velocity range typical of young thin-disc stars \citep{Leggett1992, Nissen2004}.

\subsection{Galactic Warp and Flare Corrections}

Berkeley~36 lies at a present-day Galactocentric distance of $R_{\rm GC}=11.422$~kpc, beyond the solar circle, where both the Galactic warp and the flare of the thin disc become significant. The vertical displacement produced by the warp is modeled as
\begin{equation}
Z_{\rm warp}(R,\phi)=\gamma_{\rm w}\,[R-R_{\rm warp}]\,\sin(\phi-\phi_{\rm w}),
\end{equation}
with a warp onset radius $R_{\rm warp}=8.4$~kpc and line-of-nodes angle $\phi_{\rm w}=5^{\circ}$ \citep{Derriere2001}, consistent with Besançon-type Galaxy models \citep{Robin12} and 2MASS/Gaia-based studies placing the warp onset near $8$--$9$~kpc \citep{Lopez2002,Reyle2009,Chrobakova2022}. The amplitude $\gamma_{\rm w}=0.09$~kpc$^{-1}$ was chosen to match literature vertical displacements at $R_{\rm GC}\approx9$--$12$~kpc. Orbits were then corrected via $Z_{\rm cor}=Z-Z_{\rm warp}(R,\phi)$.

The disc flare is described by a radius-dependent scale height,
\begin{equation}
h_{\rm z}(R)=h_{\rm z,0}\left[1+\gamma_{\rm f}(R-R_0)\right],
\end{equation}
with $h_{\rm z,0}=300$~pc, and $\gamma_{\rm f}=0.05$~kpc$^{-1}$, following 2MASS, SDSS, and Gaia-based measurements of outer-disc thickening \citep{Momany2006, Cabrera-Lavers2007, Bilir2006b, Bilir2008, Lopez2014}.

Integrating the orbit in the \texttt{MWPotential2014} potential gives a maximum vertical amplitude of $Z_{\rm max}=381$~pc, a low eccentricity of $e=0.036$, and a mean Galactocentric radius of $R_{\rm m}=11.79$~kpc. Once the warp correction is applied, the vertical amplitude rises to $Z_{\rm max, wf}=709$~pc, an 86\% increase, showing that the warp substantially modulates the cluster's vertical excursion at this large Galactocentric distance. The cluster remains within 100~pc of the local warp surface for only 25\% of its orbital period ($W_{\rm mf}$), and the median $|Z|/h_{\rm z}$ along the orbit is 0.75, compared with just 0.04 at its present-day position, indicating that Berkeley~36 currently sits close to the flared local disc plane despite reaching much greater heights elsewhere in its orbit. At its current position, the effective scale height is $h_{\rm z}=351$~pc, 17\% larger than the local value, consistent with the expected outward thickening of the outer thin disk.

\section{Variability Search in the Berkeley 36 Field}

We inspected the available TESS observations of the high-probability members of Berkeley 36 to search for photometric variability \citep{Belwal2026TESS}. TESS light curves were available for eight probable cluster members. However, all of these stars exhibit very high contamination fractions ($\sim0.89$--$0.97$), indicating severe source blending caused by the large TESS pixel scale in the crowded cluster field. Consequently, the extracted light curves are not suitable for reliable variability analysis, and no robust periodic variability could be established. We therefore do not identify any confirmed variable members of Berkeley~36 from the currently available TESS observations.

\begin{figure*}[!ht]
\centering
\includegraphics[width=0.95\linewidth]{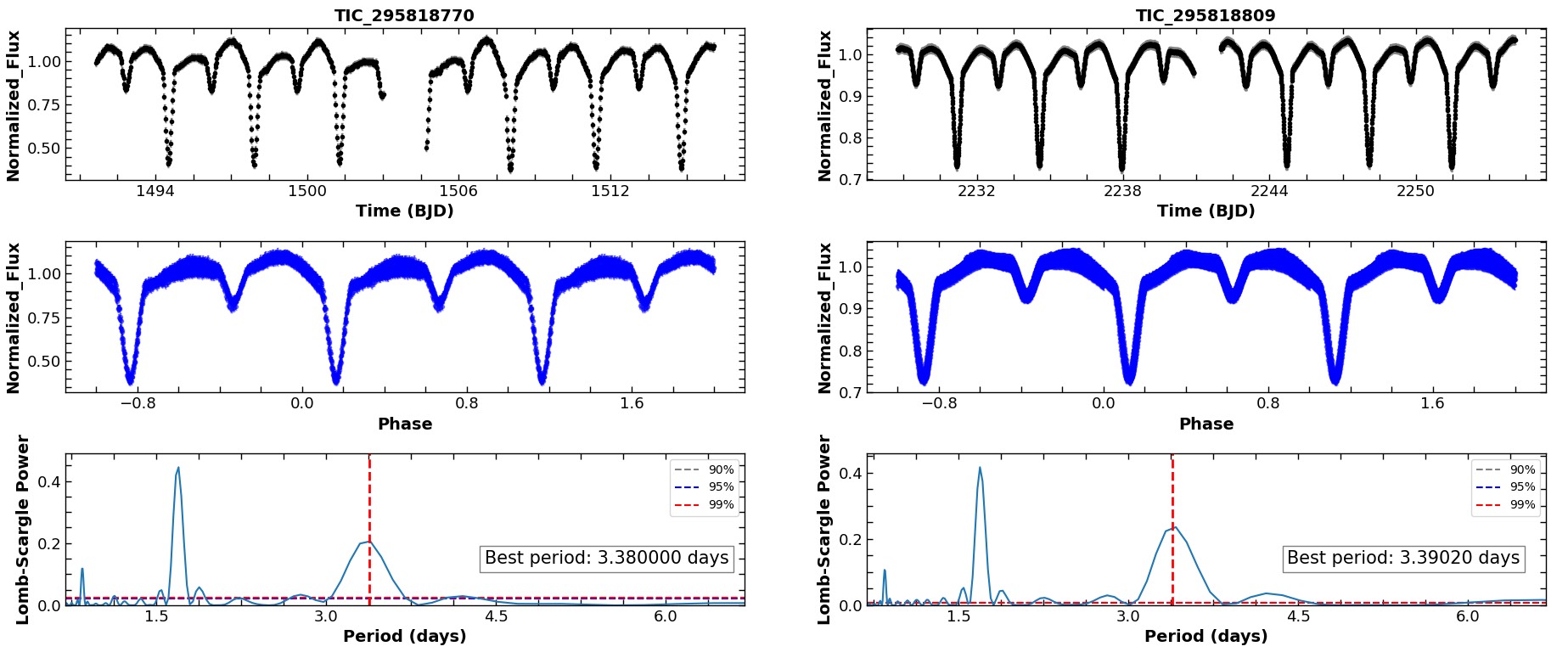}
\caption{TESS light curves of the two foreground eclipsing binaries identified in the field of Berkeley~36. For each source, the upper panel shows the detrended TESS light curve, the middle panel presents the phase-folded light curve using the derived orbital period, and the lower panel displays the corresponding Lomb--Scargle periodogram. The adopted orbital periods are $3.3800$~d for TIC~295818770 and $3.3902$~d for TIC~295818809. The morphology of the phased light curves is characteristic of detached Algol-type (EA) eclipsing binaries. Both systems are foreground field stars.}
\label{fig:lcs}
\end{figure*}

During the inspection of the surrounding TESS field, we identified two stars displaying well-defined eclipsing variability, namely TIC~295818770 and TIC~295818809 (Figure~\ref{fig:lcs}). Cross-matching with our Gaia DR3 membership catalogue shows that both objects have negligible membership probabilities and are therefore unrelated foreground field stars. Both sources are also classified as detached eclipsing binaries (EA type) in the Gaia DR3 variability catalogue, confirming their eclipsing nature.

The orbital periods derived from the Lomb--Scargle periodogram are $P = 3.3800 \pm 0.08$~d for TIC~295818770 and $P = 3.3902 \pm 0.07$~d for TIC~295818809. Their phased light curves exhibit deep primary eclipses and shallower secondary eclipses, characteristic of detached Algol-type eclipsing binary systems. To estimate their basic stellar parameters, we modelled their spectral energy distributions using the available multi-band photometry. The resulting effective temperatures are approximately $6846$~K and $7594$~K for TIC~295818770 and TIC~295818809, respectively, consistent with late-F and early-A type stars. The derived distances are approximately $1.74$~kpc for both systems, significantly smaller than the distance of Berkeley~36. Together with their Gaia astrometry and low membership probabilities, these results confirm that both eclipsing binaries are foreground field stars projected toward the cluster.

\section{Discussion}\label{sec:discussion}

\subsection{Comparison with Previous Studies}

Table~\ref{tab:lit_comp_be36} shows the parameter estimates that have been published for Berkeley~36 over the past decade and a half, split into a pre-Gaia group and a post-Gaia group using 2018 as the dividing line. As with most OCs in this part of the disk, the early determinations are noticeably scattered, especially for distance and age, since they relied on ground-based photometry and proper motions of limited precision. Once Gaia astrometry became available, the reported distances tightened considerably, clustering mostly between about 4 and 5~kpc, and the extinction values settled into a narrower band around $A_{\rm V}\sim1.4$--2.0~mag.

Our own results place the cluster at $\langle d_{\rm geo}\rangle=4377\pm510$~pc with $\langle A_V\rangle=1.594\pm0.283$~mag, both of which sit comfortably inside the range defined by the post-Gaia literature. The metallicity we derive, $[\mathrm{Fe/H}]=-0.19\pm0.02$~dex, is likewise close to the handful of spectroscopic estimates available (e.g.\ \citealt{Spina2022}; \citealt{Dias2021}), which cluster between $-0.03$ and $-0.32$~dex. The age we obtain from isochrone fitting, $6.8\pm0.5$~Gyr, is however substantially older than most of the values quoted in the table, which mostly fall in the 1--4.5~Gyr range; only \citet{Otto2026} and \citet{Cantat-Gaudin2020} report ages approaching 6.8~Gyr. This discrepancy is discussed further in Section~\ref{sec:fundamental}, where we examine how membership selection and isochrone set choice affect the derived age.

\begin{table}
\caption{Published astrophysical parameters for Berkeley~36. The horizontal line separates pre-Gaia studies; dashes indicate unreported parameters.}
\label{tab:lit_comp_be36}
\renewcommand{\arraystretch}{1}
\centering
\begin{tabular}{cccccc}
\hline\hline
Year & $d$ & $A_{\rm V}$ & Age & $[\mathrm{Fe/H}]$ & Ref. \\
 & (kpc) & (mag) & (Myr) & (dex) & \\
\hline
2026 & 4.38 & 1.59 & 6800 & $-0.19$ & [01]   \\        
2026 & --   & --   & 6761 & $-0.16$ & [02] \\
2024 & 3.56 & 1.63 & 1820 & $-0.03$ & [03] \\
2023 & 3.88 & 1.95 & 1042 & --      & [04] \\
2023 & 4.57 & 1.99 & 4467 & $-0.32$ & [05] \\
2022 & --   & --   & --   & $-0.26$ & [06] \\
2021 & 4.98 & 1.80 & 3412 & $-0.03$ & [07] \\
2020 & 4.37 & 1.42 & 6761 & --      & [08] \\
2020 & 4.09 & 1.87 & 1175 & --      & [09] \\
\hline
2014 & 6.14 & 1.26 & 3162 & --      & [10] \\
2013 & 5.04 & 1.84 & 2512 & --      & [11] \\
2011 & 7.54 & 1.42 & 2239 & --      & [12] \\
\hline
\hline
\end{tabular}
\begin{minipage}{\linewidth}
\footnotesize
\tablenotetext{}{
\textbf{References:}
[01]~This study,
[02]~\citet{Otto2026},
[03]~\citet{Cavallo2024},
[04]~\citet{Hunt2024},
[05]~\citet{Angelo2023},
[06]~\citet{Spina2022},
[07]~\citet{Dias2021},
[08]~\citet{Cantat-Gaudin2020},
[09]~\citet{Kounkel2020},
[10]~\citet{Dias2014},
[11]~\citet{Kharchenko2013},
[12]~\citet{Bukowiecki2011}.}
\end{minipage}
\end{table}
\label{subsec:be36_chemical_clocks}

As an independent check on the isochrone age of Berkeley~36, we derived stellar ages from the age--chemical-clock relations of \citet{Viscasillas2022}, who calibrated multivariate relations of the form $[s/\alpha] = m_1 \cdot \mathrm{Age} + m_2 \cdot R_{\rm GC} + m_3 \cdot \mathrm{[Fe/H]} + c$ using 62 OCs observed by the Gaia-ESO survey. Since Berkeley~36 lies at $R_{\rm GC} = 11.422$~kpc, well within the outer-disc regime ($R_{\rm GC} > 9$~kpc), we adopted the outer-disc coefficients from their Table~A.3, inverted to express age directly as a function of the abundance ratio, $ R _ {\ rm GC} $, and [Fe/H]:3, inverted to express age directly as a function of the abundance ratio, $R_{\rm GC}$, and [Fe/H]:
\begin{equation}
\mathrm{Age} = c' + m_1' \, [s/\alpha] + m_2' \, R_{\rm GC} + m_3' \, \mathrm{[Fe/H]}.
\end{equation}
We applied this relation to the four barium-based clocks available in our sample, [Ba/Al], [Ba/Mg], [Ba/Si], and [Ba/Ca], since ratios involving Ba show the tightest correlation with age among the indicators tested by \citet{Viscasillas2022}. The exclusion of [S/Fe] from the $\alpha$-element abundance discussion does not affect the chemical-clock analysis, since none of the four adopted barium-based clocks ([Ba/Al], [Ba/Mg], [Ba/Si], and [Ba/Ca]) involves sulfur. Therefore, the same ten stars and the same abundance ratios were used to derive the chemical-clock age.

Before computing ages, the individual stellar abundances were processed using the same normalization scheme as in \citet{Viscasillas2022}. Elemental abundances with a reported error $\geq 0.1$~dex were discarded to avoid propagating low-quality measurements into the age relations. Each star was then classified as a giant ($\log g \leq 3.5$) or a dwarf
($\log g > 3.5$), and the raw [X/H] abundances were placed on a common solar scale by subtracting the corresponding M67 giant or dwarf reference abundances (Table~1 of \citealt{Viscasillas2022}), which removes the small but non-negligible offset between dwarf and giant abundance scales noted by the authors. Finally, for each element, the normalized abundances across the sample were cleaned using the interquartile range (IQR) criterion, rejecting values outside $[Q_1 - 1.5\,\mathrm{IQR},\, Q_3 + 1.5\,\mathrm{IQR}]$, consistent
with the outlier-rejection procedure used to build the original calibration sample.

Ages were computed on a star-by-star basis for the ten Berkeley~36 member stars with usable Ba and $\alpha$-element abundances, yielding up to four individual age estimates per star. The individual chemical clocks based on the [Ba/Al], [Ba/Mg], [Ba/Si], and [Ba/Ca] abundance ratios yield mean ages of $7.52\pm1.54$, $6.17\pm0.76$, $8.28\pm2.08$, and $5.72\pm1.62$~Gyr, respectively, demonstrating an overall consistency among the independent calibrations despite the different sample sizes available for each clock. For each star, we adopted the mean of the available clock ages and its standard deviation as an internal precision estimate. The final cluster age was then derived by averaging these star-by-star mean ages, rather than by averaging the mean ages of the individual chemical clocks. Averaging over all stars in the sample, we obtain a mean cluster age of $\langle t \rangle \simeq 7.75 \pm 1.94$~Gyr, in reasonable agreement with the isochrone age of $6.8\pm0.5$~Gyr. The relatively large star-to-star scatter is not unexpected: \citet{Viscasillas2022} report a precision of about $0.9$~Gyr for the [Ba/Al] relation in the outer disc when applied to cluster-averaged abundances, but individual member stars in their own sample show appreciably larger dispersion (their Figure~9), reflecting both measurement uncertainties in the neutron-capture-element abundances and the intrinsic non-uniqueness of any single age--chemical-clock relation (see their Section~7). Given that Berkeley~36 is located in the outer disc, where radial migration is expected to play a comparatively minor role \citep{Viscasillas2022}, we consider the chemical-clock age estimate to be a genuine, largely unbiased confirmation of the isochrone-based age, rather than an estimate confused by migrated interlopers.

\subsection{Chemical Birth Radius and Radial Migration}

The chemical birth radius ($R_{\rm b}$) of Berkeley~36 was estimated following the chemo-dynamical approach of \citet{Minchev2018}, as adopted in subsequent studies \citep[e.g.,][]{Lu2024, Ratcliffe2025, Ratcliffe2026, Cinar2026c}. In this context, $R_{\rm b}$ is inferred from the cluster's age and metallicity under the assumption of a time-dependent Galactic radial metallicity gradient. Using $\rm [Fe/H]=-0.19\pm0.02$~dex and an age of $\approx6.8\pm0.5$~Gyr, we obtain $R_{\rm b}=6.71\pm 0.66$~kpc, smaller than the solar radius ($R_\odot\approx8.2$~kpc), suggesting an inner-disk origin. This estimate depends on the adopted chemo-dynamical model and should be regarded as approximate rather than a unique solution.

Comparing $R_{\rm b}$ with the cluster's present-day guiding radius ($R_{\rm g}=11.751$~kpc) and Galactocentric radius ($R_{\rm GC}=11.422$~kpc) yields $|R_{\rm g}-R_{\rm b}|=5.04$~kpc and $|R_{\rm GC}-R_{\rm b}|=4.71$~kpc, while the offset between $R_{\rm g}$ and $R_{\rm GC}$ itself is only $|R_{\rm g}-R_{\rm GC}|=0.33$~kpc. Following the usual interpretation, the difference between the guiding radius and the birth radius, $R_{\rm g}-R_{\rm b}$, is commonly interpreted as an estimate of the cumulative effect of churning, i.e., the long-term change in guiding-center radius driven by angular momentum exchange with spiral arms and the Galactic bar \citep{Sellwood2002,Anders2017}. Recent cosmological simulations show that spiral-arm interactions can drive significant outward migration of open clusters through churning without substantial kinematic heating \citep{Wiggins2025}. In contrast, the difference between the present-day Galactocentric radius and the birth radius, $R_{\rm GC}-R_{\rm b}$, reflects the combined effects of churning and epicyclic motion (blurring), whereas the offset $R_{\rm GC}-R_{\rm g}$ represents the contribution from epicyclic motion alone, arising from orbital eccentricity \citep{Binney2008,Sellwood2002}. For Berkeley~36, the blurring contribution is small, consistent with its low orbital eccentricity ($e=0.036$). Therefore, the inferred radial migration is expected to be dominated by churning, with only a minor contribution from epicyclic blurring. Since $R_{\rm g}>R_{\rm b}$, Berkeley~36 has migrated outward from an inner-disk birth site to its current position well beyond the solar circle.

\begin{figure*}
    \centering
    \includegraphics[width=0.65\linewidth]{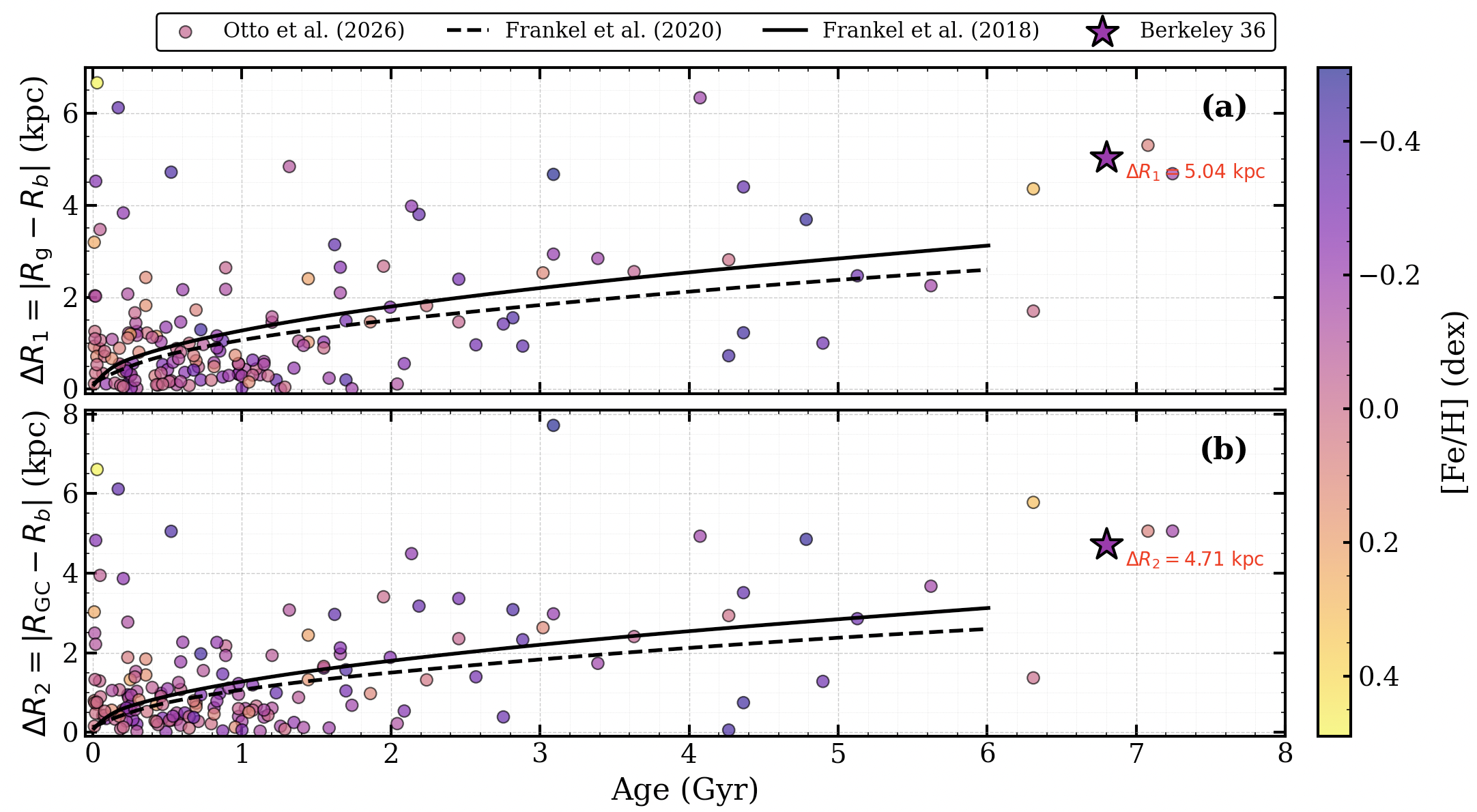}
    \caption{Distribution of OCs in the $|R_{\rm g} - R_{\rm b}|\times t$ (panel a) and $|R_{\rm GC}- R_{\rm b}|\times t$ (panel b) planes. Colors indicate the cluster metallicity ([Fe/H]). The circular symbols correspond to the OC sample of \citet{Otto2026}, while Berkeley~36 is marked with a star symbol. The solid and dashed black curves show the empirical age-dependent limits reported by \citet{Frankel_2018} and \citet{Frankel2020}, respectively.}
    \label{fig:rguide_rgc}
\end{figure*}
 
Figure~\ref{fig:rguide_rgc} places Berkeley~36 in the age--radius-deviation plane together with the OC sample of \citet{Otto2026}. At an age of $\sim6.8$~Gyr, the cluster's $|R_{\rm g}-R_{\rm b}|=5.04$~kpc (panel a) and $|R_{\rm GC}-R_{\rm b}|=4.71$~kpc (panel b) lie clearly above the empirical secular-heating envelopes of \citet{Frankel_2018} and \citet{Frankel2020}, which reach only $\sim$3~kpc by this age. This places Berkeley~36 among the small number of old clusters in the comparison sample with the largest inferred migration distances, alongside a handful of \citet{Otto2026} clusters at similar ages ($\sim$7--7.3~Gyr) showing comparably large offsets ($\sim$5.2--5.3~kpc). Berkeley~36 is therefore not an isolated outlier but part of a small population of old OCs whose inferred radial migration exceeds that expected from standard secular evolution involving both churning and blurring, suggesting that stronger or more episodic angular-momentum exchange, for example transient spiral structure or bar-resonance scattering, may have acted on this cluster's orbit, in addition to or instead of steady-state secular diffusion. As with the extreme values seen among the youngest clusters in the sample ($t<0.5$~Gyr, $|R_{\rm g}-R_{\rm b}|$ up to $\sim6.5$~kpc), part of the offset could also reflect uncertainties in the adopted birth-radius model, the metallicity gradient, or the orbital parameters, and this caveat should be kept in mind when interpreting Berkeley~36's position above the envelope.
 
Normalizing the migration distance by age gives an effective rate of $|R_{\rm g}-R_{\rm b}|/t\approx0.74$~kpc~Gyr$^{-1}$ (and $\approx0.69$~kpc~Gyr$^{-1}$ using $R_{\rm GC}$ instead of $R_{\rm g}$), somewhat below the $1.0\pm0.6~\rm kpc~Gyr^{-1}$ inferred here for the older \citet{Otto2026} clusters with $|\Delta R|>1$~kpc, and below the $1.5\pm0.5~\rm kpc~Gyr^{-1}$ reported by \citet{Chen2020}. This indicates that, while individually large, Berkeley~36's total migration distance is still broadly consistent with several Gyr of sustained radial mixing, rather than requiring an implausibly fast migration episode, even though its position above the Frankel envelopes suggests it has migrated further than the population's typical secular-heating expectation for its age.
 
Berkeley~36's inner-disk birth radius ($R_{\rm b}=6.71$~kpc) combined with its mildly sub-solar metallicity ($\rm[Fe/H]=-0.19$~dex) is broadly consistent with the negative radial metallicity gradient expected from inside-out disk formation \citep[e.g.,][]{Chiappini1997, Minchev2014}, although its metallicity is only moderately depleted relative to solar for its inferred birth radius, which may reflect intrinsic scatter in the adopted gradient or the cluster's subsequent migration history. Taken together, the large outward displacement from $R_{\rm b}$ to $R_{\rm GC}$, the dominance of churning over blurring, and the cluster's old age ($\sim$6.8~Gyr) support a scenario in which Berkeley~36 may have formed in the inner disk and subsequently migrated outward through secular radial migration, making it a useful benchmark for probing strong-migration outliers relative to the general old-cluster population.
 
\section{Summary and Conclusion}
\label{sec:conclusion}
 
In this study, we have presented a comprehensive, multi-wavelength characterization of the OC cluster Berkeley~36, combining Gaia~DR3 astrometry and photometry with high-resolution spectroscopy from GES~DR5.1 in a single, internally consistent analysis. By anchoring the cluster's metallicity and distance with independent astrometric and spectroscopic constraints, we reduced the classical age--reddening--metallicity degeneracy to a single free parameter, allowing us to derive a self-consistent set of structural, chemical, kinematic, and dynamical properties for Berkeley~36. Our principal results can be summarized as follows:
 
\begin{itemize}
    \item Using a GMM applied to Gaia~DR3 proper motions and parallaxes, we identified $N_{\rm mem} = 946$ probable members of Berkeley~36 at $P \geq 0.7$, centred at $\alpha = 07^{\text{h}}16^{\text{m}}23^{\text{s}}$.86, $\delta = -13^\circ11^\prime51^{\prime\prime}\!.35$, with a mean proper motion of $(-1.701 \pm 0.163,\ 0.960 \pm 0.154)$~mas~yr$^{-1}$ and a parallax-based distance of $d_{\varpi} = 4.46 \pm 2.43$~kpc.
 
    \item The radial density profile, fitted with a \citet{King62} model via MCMC, yields a core radius $r_{\rm c} = 2.63^{+0.22}_{-0.21}$~arcmin and a tidal radius $r_{\rm t} = 14.84^{+0.47}_{-0.47}$~arcmin, corresponding to a concentration parameter $C \approx 0.75$, characteristic of a moderately concentrated, dynamically relaxed system.
 
    \item Fixing the metallicity from GES spectroscopy ($\langle[\mathrm{Fe/H}]\rangle = -0.19 \pm 0.02$~dex) and the distance and reddening from photometric/astrometric data ($\langle d\rangle = 4377 \pm 510$~pc, $\langle A_{\rm G}\rangle = 1.350 \pm 0.045$~mag), isochrone fitting with both PARSEC and MIST models gives a consistent cluster age of $6.8 \pm 0.5$~Gyr.

    \item Independent Ba-based chemical clocks, combining the [Ba/Al], [Ba/Mg], [Ba/Si], and [Ba/Ca] abundance ratios, yield a mean cluster age of $\langle t \rangle \simeq 7.75 \pm 1.94$~Gyr from star-by-star estimates, in good agreement with the isochrone age of $6.8 \pm 0.5$~Gyr. 

    \item A photometric analysis of the main sequence reveals a binary fraction of $f_{\rm b} = 0.480 \pm 0.017$, with the mass-ratio distribution peaking at $q \sim 0.8$--$0.9$, mirroring the pattern previously found for Berkeley~32.
 
    \item We identified 83 BSS candidates (18 previously known and 65 newly detected) and 10 YSS candidates. Unexpectedly, the BSS population is significantly less centrally concentrated than the giant and main-sequence populations, in contrast to the classical mass-segregation picture, suggesting a substantial primordial or outer-region contribution to this population.
 
    \item The cluster's systemic radial velocity, derived from $N=106$ GES members, is $\langle V_{\rm rad, OC}\rangle = 62.80 \pm 0.04$~km~s$^{-1}$, with an internal dispersion of $\sigma_{V_{\rm rad, OC}} = 2.58$~km~s$^{-1}$, confirming that the sample forms a kinematically coherent group.
 
    \item Orbit integration in the \textsc{MWPotential2014} potential shows that Berkeley~36 follows a nearly circular orbit ($e = 0.036\pm0.027$) in the outer disc, with a guiding radius $R_{\rm g} = 11.751\pm 0.772$~kpc and a current Galactocentric distance $R_{\rm GC} = 11.422\pm 0.437$~kpc. Accounting for the Galactic warp increases the maximum vertical excursion by 86\%, while the Galactic disc flare increases the local scale height by 17\%, highlighting the importance of both effects at large Galactocentric radii.
    
    \item The chemical birth radius, $R_{\rm b} = 6.71\pm 0.66$~kpc, lies well inside the cluster's present-day guiding radius, implying an outward migration of $\sim$5~kpc dominated by churning rather than blurring. This places Berkeley~36 among the old OCs with the largest inferred migration distances relative to its age, exceeding the secular-heating expectations of \citet{Frankel_2018} and \citet{Frankel2020}.
 
    \item A search for photometric variability in the TESS field of Berkeley~36 did not yield any confirmed cluster-member variables, owing to severe blending in the crowded field; the two eclipsing binaries detected nearby (TIC~295818770 and TIC~295818809) were found to be unrelated foreground field stars.
\end{itemize}
 
These results depict Berkeley~36 as an old, dynamically evolved OC that may have formed in the inner Galactic disc and subsequently migrated outward to its present location beyond the solar circle, primarily through churning rather than orbital blurring. Its large migration distance relative to its age, together with its centrally concentrated giant population and unusually extended BSS distribution, makes Berkeley~36 a valuable benchmark for studies of secular disc evolution and cluster dynamical history. Further progress on this system would benefit from spectroscopic follow-up of the BSS candidates to assess field contamination and binarity, deeper or higher-cadence photometric monitoring capable of overcoming the blending limitations of TESS in this crowded field, and the inclusion of Berkeley~36 in future homogeneous, Galaxy-wide compilations of migration histories as additional spectroscopic and astrometric data become available.

\begin{acknowledgments}
We sincerely thank the referee and the Data Editor for their constructive comments and valuable suggestions, which have helped us improve the quality, clarity, and overall presentation of this manuscript. Ing-Guey Jiang acknowledges support from the National Science and Technology Council (NSTC), Taiwan, under grants NSTC 113-2112-M-007-030 and NSTC 114-2112-M-007-029. This study is a part of the PhD Thesis of Deniz Cennet Çınar. This work uses data from the European Space Agency's Gaia mission, processed by the Gaia Data Processing and Analysis Consortium (DPAC). Based on data obtained from the ESO Science Archive Facility with DOI: https://doi.org/10.18727/archive/25
\end{acknowledgments}

\vspace{5mm}
\facilities{Gaia \citep{GaiaCollaboration2016, GaiaCollaboration2023}}, Gaia-ESO Survey DR5.1 \citep{GES2012}, Pan-STARRS1 \citep{Chambers2016}, TESS \citep{Ricker2015}, 2MASS \citep{skrutskie2006two}

\software{\texttt{astropy} \citep{astropy13, astropy18, astropy22},
\texttt{Matplotlib} \citep{Hunter07},
\texttt{NumPy} \citep{Harris20},
\texttt{SciPy} \citep{Virtanen20},
\texttt{scikit-learn} \citep{scikit-learn},
\texttt{emcee} \citep{emcee},
\texttt{galpy} \citep{Bovy2015}}

\bibliography{references}{}

\begin{thebibliography}{}
\expandafter\ifx\csname natexlab\endcsname\relax\def\natexlab#1{#1}\fi
\providecommand{\url}[1]{\href{#1}{#1}}
\providecommand{\dodoi}[1]{doi:~\href{http://doi.org/#1}{\nolinkurl{#1}}}
\providecommand{\doeprint}[1]{\href{http://ascl.net/#1}{\nolinkurl{http://ascl.net/#1}}}
\providecommand{\doarXiv}[1]{\href{https://arxiv.org/abs/#1}{\nolinkurl{https://arxiv.org/abs/#1}}}

% type= article
\bibitem[{M. Agarwal {et~al.}(2021)Agarwal, Rao, Vaidya, \& Bhattacharya}]{agarwal2021ml}
Agarwal, M., Rao, K.~K., Vaidya, K., \& Bhattacharya, S. 2021, \bibinfo{title}{ML-MOC: machine learning (kNN and GMM) based membership determination for open clusters,} Monthly Notices of the Royal Astronomical Society, 502, 2582

% type= article
\bibitem[{T. {Ak} {et~al.}(2016){Ak}, {Bostanc{\i}}, {Yontan}, {Bilir}, {G{\"u}ver}, {Ak}, {{\"U}rg{\"u}p}, \& {Paunzen}}]{Ak2016}
{Ak}, T., {Bostanc{\i}}, Z.~F., {Yontan}, T., {et~al.} 2016, \bibinfo{title}{{CCD <inline-formula id=``IEq1''><mml:math><mml:mi mathvariant=``italic''>UBV</mml:mi></mml:math></inline-formula> photometry of the open cluster NGC 6819},} \apss, 361, 126, \dodoi{10.1007/s10509-016-2707-2}

% type= article
\bibitem[{B. {Akbulut} {et~al.}(2021){Akbulut}, {Ak}, {Yontan}, {Bilir}, {Ak}, {Banks}, {Kaan Ulgen}, \& {Paunzen}}]{Akbulut2021}
{Akbulut}, B., {Ak}, S., {Yontan}, T., {et~al.} 2021, \bibinfo{title}{{A study of the Czernik 2 and NGC 7654 open clusters using CCD UBV photometric and Gaia EDR3 data},} \apss, 366, 68, \dodoi{10.1007/s10509-021-03975-x}

% type= article
\bibitem[{C. {Allen} \& A. {Santillan}(1991){Allen} \& {Santillan}}]{Allen1991}
{Allen}, C., \& {Santillan}, A. 1991, \bibinfo{title}{{An improved model of the galactic mass distribution for orbit computations.},} \rmxaa, 22, 255

% type= article
\bibitem[{A.~Y. {Alzhrani} {et~al.}(2025{\natexlab{a}}){Alzhrani}, {HarooN}, {Elsanhoury}, \& {{\c{C}}{\i}nar}}]{Alzhrani2025b}
{Alzhrani}, A.~Y., {HarooN}, A.~A., {Elsanhoury}, W.~H., \& {{\c{C}}{\i}nar}, D.~C. 2025{\natexlab{a}}, \bibinfo{title}{{Enhancing SED-based astrometric, photometric, and kinematic studies of SAI 72 and SAI 75 using Gaia DR3},} Journal of Astrophysics and Astronomy, 46, 50, \dodoi{10.1007/s12036-025-10076-6}

% type= article
\bibitem[{A.~Y. {Alzhrani} {et~al.}(2025{\natexlab{b}}){Alzhrani}, {Haroon}, {Elsanhoury}, \& {{\c{C}}{\i}nar}}]{Alzhrani2025}
{Alzhrani}, A.~Y., {Haroon}, A.~A., {Elsanhoury}, W.~H., \& {{\c{C}}{\i}nar}, D.~C. 2025{\natexlab{b}}, \bibinfo{title}{{In-depth analysis of photometric and kinematic characteristics of SAI 16, SAI 81 and SAI 86 open clusters utilizing Gaia DR3},} Journal of Astrophysics and Astronomy, 46, 58, \dodoi{10.1007/s12036-025-10083-7}

% type= article
\bibitem[{F. {Anders} {et~al.}(2017){Anders}, {Chiappini}, {Rodrigues}, \& {others}}]{Anders2017}
{Anders}, F., {Chiappini}, C., {Rodrigues}, T.~S., \& {others}. 2017, \bibinfo{title}{{Galactic archaeology with asteroseismology and spectroscopy: Red giants observed by CoRoT and APOGEE},} \aap, 597, A30, \dodoi{10.1051/0004-6361/201527204}

% type= article
\bibitem[{G. {Andreuzzi} {et~al.}(2011){Andreuzzi}, {Bragaglia}, {Tosi}, \& {Marconi}}]{Andreuzzi2011}
{Andreuzzi}, G., {Bragaglia}, A., {Tosi}, M., \& {Marconi}, G. 2011, \bibinfo{title}{{Old open clusters and the Galactic metallicity gradient: Berkeley 20, Berkeley 66 and Tombaugh 2},} \mnras, 412, 1265, \dodoi{10.1111/j.1365-2966.2010.17986.x}

% type= article
\bibitem[{N. {Andronov} {et~al.}(2006){Andronov}, {Pinsonneault}, \& {Terndrup}}]{Andronov2006}
{Andronov}, N., {Pinsonneault}, M.~H., \& {Terndrup}, D.~M. 2006, \bibinfo{title}{{Mergers of Close Primordial Binaries},} \apj, 646, 1160, \dodoi{10.1086/505127}

% type= article
\bibitem[{M.~S. {Angelo} {et~al.}(2023){Angelo}, {Santos}, {Maia}, \& {Corradi}}]{Angelo2023}
{Angelo}, M.~S., {Santos}, Jr., J.~F.~C., {Maia}, F.~F.~S., \& {Corradi}, W.~J.~B. 2023, \bibinfo{title}{{Enlightening the dynamical evolution of Galactic open clusters: an approach using Gaia DR3 and analytical descriptions},} \mnras, 522, 956, \dodoi{10.1093/mnras/stad1038}

% type= article
\bibitem[{ {Astropy Collaboration} {et~al.}(2013){Astropy Collaboration}, {Robitaille}, {Tollerud}, {Greenfield}, {Droettboom}, {Bray}, {Aldcroft}, {Davis}, {Ginsburg}, {Price-Whelan}, {Kerzendorf}, {Conley}, {Crighton}, {Barbary}, {Muna}, {Ferguson}, {Grollier}, {Parikh}, {Nair}, {Unther}, {Deil}, {Woillez}, {Conseil}, {Kramer}, {Turner}, {Singer}, {Fox}, {Weaver}, {Zabalza}, {Edwards}, {Azalee Bostroem}, {Burke}, {Casey}, {Crawford}, {Dencheva}, {Ely}, {Jenness}, {Labrie}, {Lim}, {Pierfederici}, {Pontzen}, {Ptak}, {Refsdal}, {Servillat}, \& {Streicher}}]{astropy13}
{Astropy Collaboration}, {Robitaille}, T.~P., {Tollerud}, E.~J., {et~al.} 2013, \bibinfo{title}{{Astropy: A community Python package for astronomy},} \aap, 558, A33, \dodoi{10.1051/0004-6361/201322068}

% type= article
\bibitem[{ {Astropy Collaboration} {et~al.}(2018){Astropy Collaboration}, {Price-Whelan}, {Sip{\H{o}}cz}, {G{\"u}nther}, {Lim}, {Crawford}, {Conseil}, {Shupe}, {Craig}, {Dencheva}, {Ginsburg}, {VanderPlas}, {Bradley}, {P{\'e}rez-Su{\'a}rez}, {de Val-Borro}, {Aldcroft}, {Cruz}, {Robitaille}, {Tollerud}, {Ardelean}, {Babej}, {Bach}, {Bachetti}, {Bakanov}, {Bamford}, {Barentsen}, {Barmby}, {Baumbach}, {Berry}, {Biscani}, {Boquien}, {Bostroem}, {Bouma}, {Brammer}, {Bray}, {Breytenbach}, {Buddelmeijer}, {Burke}, {Calderone}, {Cano Rodr{\'\i}guez}, {Cara}, {Cardoso}, {Cheedella}, {Copin}, {Corrales}, {Crichton}, {D'Avella}, {Deil}, {Depagne}, {Dietrich}, {Donath}, {Droettboom}, {Earl}, {Erben}, {Fabbro}, {Ferreira}, {Finethy}, {Fox}, {Garrison}, {Gibbons}, {Goldstein}, {Gommers}, {Greco}, {Greenfield}, {Groener}, {Grollier}, {Hagen}, {Hirst}, {Homeier}, {Horton}, {Hosseinzadeh}, {Hu}, {Hunkeler}, {Ivezi{\'c}}, {Jain}, {Jenness}, {Kanarek}, {Kendrew}, {Kern}, {Kerzendorf}, {Khvalko}, {King}, {Kirkby}, {Kulkarni},
  {Kumar}, {Lee}, {Lenz}, {Littlefair}, {Ma}, {Macleod}, {Mastropietro}, {McCully}, {Montagnac}, {Morris}, {Mueller}, {Mumford}, {Muna}, {Murphy}, {Nelson}, {Nguyen}, {Ninan}, {N{\"o}the}, {Ogaz}, {Oh}, {Parejko}, {Parley}, {Pascual}, {Patil}, {Patil}, {Plunkett}, {Prochaska}, {Rastogi}, {Reddy Janga}, {Sabater}, {Sakurikar}, {Seifert}, {Sherbert}, {Sherwood-Taylor}, {Shih}, {Sick}, {Silbiger}, {Singanamalla}, {Singer}, {Sladen}, {Sooley}, {Sornarajah}, {Streicher}, {Teuben}, {Thomas}, {Tremblay}, {Turner}, {Terr{\'o}n}, {van Kerkwijk}, {de la Vega}, {Watkins}, {Weaver}, {Whitmore}, {Woillez}, {Zabalza}, \& {Astropy Contributors}}]{astropy18}
{Astropy Collaboration}, {Price-Whelan}, A.~M., {Sip{\H{o}}cz}, B.~M., {et~al.} 2018, \bibinfo{title}{{The Astropy Project: Building an Open-science Project and Status of the v2.0 Core Package},} \aj, 156, 123, \dodoi{10.3847/1538-3881/aabc4f}

% type= article
\bibitem[{ {Astropy Collaboration} {et~al.}(2022){Astropy Collaboration}, {Price-Whelan}, {Lim}, {Earl}, {Starkman}, {Bradley}, {Shupe}, {Patil}, {Corrales}, {Brasseur}, {N{\"o}the}, {Donath}, {Tollerud}, {Morris}, {Ginsburg}, {Vaher}, {Weaver}, {Tocknell}, {Jamieson}, {van Kerkwijk}, {Robitaille}, {Merry}, {Bachetti}, {G{\"u}nther}, {Aldcroft}, {Alvarado-Montes}, {Archibald}, {B{\'o}di}, {Bapat}, {Barentsen}, {Baz{\'a}n}, {Biswas}, {Boquien}, {Burke}, {Cara}, {Cara}, {Conroy}, {Conseil}, {Craig}, {Cross}, {Cruz}, {D'Eugenio}, {Dencheva}, {Devillepoix}, {Dietrich}, {Eigenbrot}, {Erben}, {Ferreira}, {Foreman-Mackey}, {Fox}, {Freij}, {Garg}, {Geda}, {Glattly}, {Gondhalekar}, {Gordon}, {Grant}, {Greenfield}, {Groener}, {Guest}, {Gurovich}, {Handberg}, {Hart}, {Hatfield-Dodds}, {Homeier}, {Hosseinzadeh}, {Jenness}, {Jones}, {Joseph}, {Kalmbach}, {Karamehmetoglu}, {Ka{\l}uszy{\'n}ski}, {Kelley}, {Kern}, {Kerzendorf}, {Koch}, {Kulumani}, {Lee}, {Ly}, {Ma}, {MacBride}, {Maljaars}, {Muna}, {Murphy}, {Norman},
  {O'Steen}, {Oman}, {Pacifici}, {Pascual}, {Pascual-Granado}, {Patil}, {Perren}, {Pickering}, {Rastogi}, {Roulston}, {Ryan}, {Rykoff}, {Sabater}, {Sakurikar}, {Salgado}, {Sanghi}, {Saunders}, {Savchenko}, {Schwardt}, {Seifert-Eckert}, {Shih}, {Jain}, {Shukla}, {Sick}, {Simpson}, {Singanamalla}, {Singer}, {Singhal}, {Sinha}, {Sip{\H{o}}cz}, {Spitler}, {Stansby}, {Streicher}, {{\v{S}}umak}, {Swinbank}, {Taranu}, {Tewary}, {Tremblay}, {de Val-Borro}, {Van Kooten}, {Vasovi{\'c}}, {Verma}, {de Miranda Cardoso}, {Williams}, {Wilson}, {Winkel}, {Wood-Vasey}, {Xue}, {Yoachim}, {Zhang}, {Zonca}, \& {Astropy Project Contributors}}]{astropy22}
{Astropy Collaboration}, {Price-Whelan}, A.~M., {Lim}, P.~L., {et~al.} 2022, \bibinfo{title}{{The Astropy Project: Sustaining and Growing a Community-oriented Open-source Project and the Latest Major Release (v5.0) of the Core Package},} \apj, 935, 167, \dodoi{10.3847/1538-4357/ac7c74}

% type= article
\bibitem[{C.~A.~L. {Bailer-Jones} {et~al.}(2021){Bailer-Jones}, {Rybizki}, {Fouesneau}, {Demleitner}, \& {Andrae}}]{BailerJones2021}
{Bailer-Jones}, C.~A.~L., {Rybizki}, J., {Fouesneau}, M., {Demleitner}, M., \& {Andrae}, R. 2021, \bibinfo{title}{{Estimating Distances from Parallaxes. V. Geometric and Photogeometric Distances to 1.47 Billion Stars in Gaia Early Data Release 3},} \aj, 161, 147, \dodoi{10.3847/1538-3881/abd806}

% type= article
\bibitem[{C.~D. {Bailyn}(1995){Bailyn}}]{Bailyn1995}
{Bailyn}, C.~D. 1995, \bibinfo{title}{{Blue Stragglers and Other Stellar Anomalies:Implications for the Dynamics of Globular Clusters},} \araa, 33, 133, \dodoi{10.1146/annurev.aa.33.090195.001025}

% type= article
\bibitem[{T. {Banks} {et~al.}(2020){Banks}, {Yontan}, {Bilir}, \& {Canbay}}]{Banks2020}
{Banks}, T., {Yontan}, T., {Bilir}, S., \& {Canbay}, R. 2020, \bibinfo{title}{{Vilnius photometry and Gaia astrometry of Melotte 105},} Journal of Astrophysics and Astronomy, 41, 6, \dodoi{10.1007/s12036-020-9621-2}

% type= article
\bibitem[{N. {Bastian} {et~al.}(2025){Bastian}, {Kamann}, {Niederhofer}, \& {Saracino}}]{Bastian2025}
{Bastian}, N., {Kamann}, S., {Niederhofer}, F., \& {Saracino}, S. 2025, \bibinfo{title}{{Testing the role of merging binaries in the formation of the split main sequence in young clusters},} \aap, 700, A241, \dodoi{10.1051/0004-6361/202555369}

% type= misc
\bibitem[{K. {Belwal}(2026){Belwal}}]{Belwal2026TESS}
{Belwal}, K. 2026, TESS observations of Berkeley 36, Mikulski Archive for Space Telescopes (MAST), \dodoi{10.17909/531f-ks67}

% type= article
\bibitem[{K. Belwal {et~al.}(2026)Belwal, Bisht, Jiang, Bisht, Raj, Chakrabarti, \& Bhowmick}]{belwal2026time}
Belwal, K., Bisht, D., Jiang, I.-G., {et~al.} 2026, \bibinfo{title}{Time-series Photometric Detection and Physical Characterization of Variable Stars in Four Intermediate-to Old-age Galactic Open Clusters,} The Astronomical Journal, 171, 288

% type= article
\bibitem[{K. Belwal {et~al.}(2025)Belwal, Bisht, Jiang, Yadav, Raj, Rangwal, Dattatrey, Bisht, \& Durgapal}]{belwal2025unveiling}
Belwal, K., Bisht, D., Jiang, I.-G., {et~al.} 2025, \bibinfo{title}{Unveiling dynamics and variability in open clusters: insights from a comprehensive analysis of six galactic clusters,} Monthly Notices of the Royal Astronomical Society, 544, 988

% type= article
\bibitem[{S. {Bilir} {et~al.}(2008){Bilir}, {Cabrera-Lavers}, {Karaali}, {Ak}, {Yaz}, \& {L{\'o}pez-Corredoira}}]{Bilir2008}
{Bilir}, S., {Cabrera-Lavers}, A., {Karaali}, S., {et~al.} 2008, \bibinfo{title}{{Estimation of Galactic Model Parameters in High Latitudes with SDSS},} \pasa, 25, 69, \dodoi{10.1071/AS07026}

% type= article
\bibitem[{S. {Bilir} {et~al.}(2006{\natexlab{a}}){Bilir}, {G{\"u}ver}, \& {Aslan}}]{Bilir2006a}
{Bilir}, S., {G{\"u}ver}, T., \& {Aslan}, M. 2006{\natexlab{a}}, \bibinfo{title}{Separation of dwarf and giant stars with {ROTSE-III}d,} Astronomische Nachrichten, 327, 693, \dodoi{10.1002/asna.200510614}

% type= article
\bibitem[{S. {Bilir} {et~al.}(2010){Bilir}, {G{\"u}ver}, {Khamitov}, {Ak}, {Ak}, {Co{\c{s}}kuno{\u{g}}lu}, {Paunzen}, \& {Yaz}}]{Bilir2010}
{Bilir}, S., {G{\"u}ver}, T., {Khamitov}, I., {et~al.} 2010, \bibinfo{title}{{CCD BV and 2MASS photometric study of the open cluster NGC 1513},} \apss, 326, 139, \dodoi{10.1007/s10509-009-0233-1}

% type= article
\bibitem[{S. {Bilir} {et~al.}(2006{\natexlab{b}}){Bilir}, {Karaali}, {G{\"u}ver}, {Karata{\textcommabelow s}}, \& {Ak}}]{Bilir2006b}
{Bilir}, S., {Karaali}, S., {G{\"u}ver}, T., {Karata{\textcommabelow s}}, Y., \& {Ak}, S.~G. 2006{\natexlab{b}}, \bibinfo{title}{{Galactic model parameters for field giants separated from field dwarfs by their 2MASS and V apparent magnitudes},} Astronomische Nachrichten, 327, 72, \dodoi{10.1002/asna.200510480}

% type= article
\bibitem[{S. {Bilir} {et~al.}(2005){Bilir}, {Karata{\textcommabelow s}}, {Demircan}, \& {Eker}}]{Bilir2005}
{Bilir}, S., {Karata{\textcommabelow s}}, Y., {Demircan}, O., \& {Eker}, Z. 2005, \bibinfo{title}{{Kinematics of W Ursae Majoris type binaries and evidence of the two types of formation},} \mnras, 357, 497, \dodoi{10.1111/j.1365-2966.2005.08609.x}

% type= article
\bibitem[{S. {Bilir} {et~al.}(2026){Bilir}, {Ta{\textcommabelow s}demir}, {Erayd{\i}n}, {{\c{C}}{\i}nar}, {Alfonso}, \& {Canbay}}]{Bilir2026}
{Bilir}, S., {Ta{\textcommabelow s}demir}, S., {Erayd{\i}n}, E., {et~al.} 2026, \bibinfo{title}{{Gaia DR3 Analysis of Four Open Clusters Toward the Galactic Anticenter},} Research in Astronomy and Astrophysics, 26, 085002, \dodoi{10.1088/1674-4527/ae587d}

% type= article
\bibitem[{S. {Bilir} {et~al.}(2016){Bilir}, {Bostanc{\i}}, {Yontan}, {G{\"u}ver}, {Bak{\i}{\c{s}}}, {Ak}, {Ak}, {Paunzen}, \& {Eker}}]{Bilir2016}
{Bilir}, S., {Bostanc{\i}}, Z.~F., {Yontan}, T., {et~al.} 2016, \bibinfo{title}{{CCD UBV photometry and kinematics of the open cluster NGC 225},} Advances in Space Research, 58, 1900, \dodoi{10.1016/j.asr.2016.06.039}

% type= book
\bibitem[{J. {Binney} \& S. {Tremaine}(2008){Binney} \& {Tremaine}}]{Binney2008}
{Binney}, J., \& {Tremaine}, S. 2008, {Galactic Dynamics: Second Edition}

% type= article
\bibitem[{D. {Bisht} {et~al.}(2026{\natexlab{a}}){Bisht}, {Jiang}, {Elsanhoury}, {Belwal}, {{\c{C}}inar}, {Raj}, {Biswas}, {Dattatrey}, {Rangwal}, {Sariya}, {Bisht}, \& {Durgapal}}]{Bisht2026}
{Bisht}, D., {Jiang}, I.-G., {Elsanhoury}, W.~H., {et~al.} 2026{\natexlab{a}}, \bibinfo{title}{{Dynamical and Photometric Analysis of NGC 146 and King 14: Evidence for a Comoving, Unbound Cluster Pair},} \aj, 171, 72, \dodoi{10.3847/1538-3881/ae285b}

% type= article
\bibitem[{D. {Bisht} {et~al.}(2026{\natexlab{b}}){Bisht}, {Jiang}, {Belwal}, {{\c{C}}inar}, {Dattatrey}, {Rangwal}, {Raj}, {Biswas}, {Bisht}, \& {Durgapal}}]{Bisht2026b}
{Bisht}, D., {Jiang}, I.-G., {Belwal}, K., {et~al.} 2026{\natexlab{b}}, \bibinfo{title}{{Multiwavelength Study of Blue Straggler Stars in Tombaugh 2: Evidence for Binary Mass Transfer and Constraints on Cluster Dynamical State},} \apj, 1003, 182, \dodoi{10.3847/1538-4357/ae6330}

% type= article
\bibitem[{D. {Bisht} {et~al.}(2026{\natexlab{c}}){Bisht}, {Jiang}, {Belwal}, {{\c{C}}{\i}nar}, {Durgapal}, {Biswas}, {Raj}, {Rangwal}, {Bisht}, \& {Manu}}]{Bisht2026c}
{Bisht}, D., {Jiang}, I.-G., {Belwal}, K., {et~al.} 2026{\natexlab{c}}, \bibinfo{title}{{Blue Straggler Stars in Berkeley 18: A Multiwavelength Study of Their Physical Properties and Dynamical Evolution},} \mnras, \dodoi{10.1093/mnras/stag1068}

% type= inproceedings
\bibitem[{T. {Boch} {et~al.}(2012){Boch}, {Pineau}, \& {Derriere}}]{Boch2012}
{Boch}, T., {Pineau}, F., \& {Derriere}, S. 2012, \bibinfo{title}{{The CDS Cross-Match Service},} in Astronomical Society of the Pacific Conference Series, Vol. 461, Astronomical Data Analysis Software and Systems XXI, ed. P.~{Ballester}, D.~{Egret}, \& N.~P.~F. {Lorente}, 291

% type= article
\bibitem[{D. {Bossini} {et~al.}(2019){Bossini}, {Vallenari}, {Bragaglia}, {Cantat-Gaudin}, {Sordo}, {Balaguer-N{\'u}{\~n}ez}, {Jordi}, {Moitinho}, {Soubiran}, {Casamiquela}, {Carrera}, \& {Heiter}}]{Bossini2019}
{Bossini}, D., {Vallenari}, A., {Bragaglia}, A., {et~al.} 2019, \bibinfo{title}{{Age determination for 269 Gaia DR2 open clusters},} \aap, 623, A108, \dodoi{10.1051/0004-6361/201834693}

% type= article
\bibitem[{Z.~F. {Bostanc{\i}} {et~al.}(2015){Bostanc{\i}}, {Ak}, {Yontan}, {Bilir}, {G{\"u}ver}, {Ak}, {{\c{C}}ak{\i}rl{\i}}, {{\"O}zdarcan}, {Paunzen}, {De Cat}, {Fu}, {Zhang}, {Hou}, {Li}, {Wang}, {Zhang}, {Shi}, \& {Wu}}]{Bostanci2015}
{Bostanc{\i}}, Z.~F., {Ak}, T., {Yontan}, T., {et~al.} 2015, \bibinfo{title}{{A comprehensive study of the open cluster NGC 6866},} \mnras, 453, 1095, \dodoi{10.1093/mnras/stv1665}

% type= article
\bibitem[{Z.~F. {Bostanc{\i}} {et~al.}(2018){Bostanc{\i}}, {Yontan}, {Bilir}, {Ak}, {G{\"u}ver}, {Ak}, {Paunzen}, {Ba{\c{s}}aran}, {Vurgun}, {Akti}, {{\c{C}}elebi}, \& {{\"U}rg{\"u}p}}]{Bostanci2018}
{Bostanc{\i}}, Z.~F., {Yontan}, T., {Bilir}, S., {et~al.} 2018, \bibinfo{title}{{CCD UBV photometric study of five open clusters{\textemdash}Dolidze 36, NGC 6728, NGC 6800, NGC 7209, and Platais 1},} \apss, 363, 143, \dodoi{10.1007/s10509-018-3364-4}

% type= article
\bibitem[{J. Bovy(2015)Bovy}]{Bovy2015}
Bovy, J. 2015, \bibinfo{title}{galpy: A {p}ython Library for {G}alactic Dynamics,} The Astrophysical Journal Supplement Series, 216, 29, \dodoi{10.1088/0067-0049/216/2/29}

% type= article
\bibitem[{J. {Bovy}(2016){Bovy}}]{Bovy2016}
{Bovy}, J. 2016, \bibinfo{title}{{The Chemical Homogeneity of Open Clusters},} \apj, 817, 49, \dodoi{10.3847/0004-637X/817/1/49}

% type= article
\bibitem[{J. {Bovy} \& H.-W. {Rix}(2013){Bovy} \& {Rix}}]{Bovy2013}
{Bovy}, J., \& {Rix}, H.-W. 2013, \bibinfo{title}{{A Direct Dynamical Measurement of the Milky Way's Disk Surface Density Profile, Disk Scale Length, and Dark Matter Profile at 4 kpc <\raisebox{-0.5ex}\textasciitilde R <\raisebox{-0.5ex}\textasciitilde 9 kpc},} \apj, 779, 115, \dodoi{10.1088/0004-637X/779/2/115}

% type= article
\bibitem[{J. Bovy \& S. Tremaine(2012)Bovy \& Tremaine}]{Bovy2012}
Bovy, J., \& Tremaine, S. 2012, \bibinfo{title}{On the Local Dark Matter Density,} The Astrophysical Journal, 756, 89, \dodoi{10.1088/0004-637X/756/1/89}

% type= article
\bibitem[{A. {Bragaglia} {et~al.}(2018){Bragaglia}, {Fu}, {Mucciarelli}, {Andreuzzi}, \& {Donati}}]{Bragaglia2018}
{Bragaglia}, A., {Fu}, X., {Mucciarelli}, A., {Andreuzzi}, G., \& {Donati}, P. 2018, \bibinfo{title}{{The chemical composition of the oldest nearby open cluster Ruprecht 147},} \aap, 619, A176, \dodoi{10.1051/0004-6361/201833888}

% type= article
\bibitem[{A. {Bragaglia} {et~al.}(2008){Bragaglia}, {Sestito}, {Villanova}, {Carretta}, {Randich}, \& {Tosi}}]{Bragaglia2008}
{Bragaglia}, A., {Sestito}, P., {Villanova}, S., {et~al.} 2008, \bibinfo{title}{{Old open clusters as key tracers of Galactic chemical evolution. II. Iron and elemental abundances in NGC 2324, NGC 2477 NGC 2660, NGC 3960, and Berkeley 32},} \aap, 480, 79, \dodoi{10.1051/0004-6361:20077904}

% type= article
\bibitem[{A. {Bressan} {et~al.}(2012){Bressan}, {Marigo}, {Girardi}, {Salasnich}, {Dal Cero}, {Rubele}, \& {Nanni}}]{Bressan2012}
{Bressan}, A., {Marigo}, P., {Girardi}, L., {et~al.} 2012, \bibinfo{title}{{PARSEC: stellar tracks and isochrones with the PAdova and TRieste Stellar Evolution Code},} \mnras, 427, 127, \dodoi{10.1111/j.1365-2966.2012.21948.x}

% type= article
\bibitem[{{\L}. {Bukowiecki} {et~al.}(2011){Bukowiecki}, {Maciejewski}, {Konorski}, \& {Strobel}}]{Bukowiecki2011}
{Bukowiecki}, {\L}., {Maciejewski}, G., {Konorski}, P., \& {Strobel}, A. 2011, \bibinfo{title}{{Open Clusters in 2MASS Photometry. I. Structural and Basic Astrophysical Parameters},} \actaa, 61, 231, \dodoi{10.48550/arXiv.1107.5119}

% type= article
\bibitem[{G. {Burki}(1975){Burki}}]{Burki1975}
{Burki}, G. 1975, \bibinfo{title}{{Non-uniform extinction in open star clusters and dispersion of the photometric sequences.},} \aap, 43, 37

% type= article
\bibitem[{A. {Cabrera-Lavers} {et~al.}(2007){Cabrera-Lavers}, {Bilir}, {Ak}, {Yaz}, \& {L{\'o}pez-Corredoira}}]{Cabrera-Lavers2007}
{Cabrera-Lavers}, A., {Bilir}, S., {Ak}, S., {Yaz}, E., \& {L{\'o}pez-Corredoira}, M. 2007, \bibinfo{title}{{Estimation of Galactic model parameters in high latitudes with 2MASS},} \aap, 464, 565, \dodoi{10.1051/0004-6361:20066475}

% type= article
\bibitem[{R. {Canbay} {et~al.}(2023){Canbay}, {Bilir}, {{\"O}zd{\"o}nmez}, \& {Ak}}]{Canbay2023}
{Canbay}, R., {Bilir}, S., {{\"O}zd{\"o}nmez}, A., \& {Ak}, T. 2023, \bibinfo{title}{{Galactic Model Parameters and Spatial Density of Cataclysmic Variables in the Gaia Era: New Constraints on Population Models},} \aj, 165, 163, \dodoi{10.3847/1538-3881/acbead}

% type= article
\bibitem[{R. {Canbay} {et~al.}(2026){Canbay}, {{\c{C}}{\i}nar}, {{\c{C}}al{\i}{\textcommabelow s}kan}, {Tan{\i}k {\"O}zt{\"u}rk}, {Ta{\textcommabelow s}demir}, {Erayd{\i}n}, \& {Bilir}}]{Canbay2026b}
{Canbay}, R., {{\c{C}}{\i}nar}, D.~C., {{\c{C}}al{\i}{\textcommabelow s}kan}, {\"O}., {et~al.} 2026, \bibinfo{title}{{Chemo-dynamical Analysis of Eight UBC Open Clusters},} arXiv e-prints, arXiv:2606.31714, \dodoi{10.48550/arXiv.2606.31714}

% type= article
\bibitem[{T. {Cantat-Gaudin} {et~al.}(2020){Cantat-Gaudin}, {Anders}, {Castro-Ginard}, \& {others}}]{Cantat-Gaudin2020}
{Cantat-Gaudin}, T., {Anders}, F., {Castro-Ginard}, A., \& {others}. 2020, \bibinfo{title}{{Painting a portrait of the Galactic disc with its stellar clusters},} \aap, 640, A1, \dodoi{10.1051/0004-6361/202038192}

% type= article
\bibitem[{T. {Cantat-Gaudin} {et~al.}(2018){Cantat-Gaudin}, {Jordi}, {Vallenari}, \& {others}}]{Cantat-Gaudin2018}
{Cantat-Gaudin}, T., {Jordi}, C., {Vallenari}, A., \& {others}. 2018, \bibinfo{title}{{A Gaia DR2 view of the open cluster population in the Milky Way},} \aap, 618, A93, \dodoi{10.1051/0004-6361/201833476}

% type= article
\bibitem[{T. Cantat-Gaudin {et~al.}(2020)Cantat-Gaudin, Anders, Castro-Ginard, Jordi, Romero-G{\'o}mez, Soubiran, Casamiquela, Tarricq, Moitinho, Vallenari, Bragaglia, Krone-Martins, \& Kounkel}]{Cantat-Gaudin20}
Cantat-Gaudin, T., Anders, F., Castro-Ginard, A., {et~al.} 2020, \bibinfo{title}{Painting a portrait of the Galactic disc with its stellar clusters,} Astronomy and Astrophysics, 640, A1, \dodoi{10.1051/0004-6361/202038192}

% type= article
\bibitem[{J.~A. {Cardelli} {et~al.}(1989){Cardelli}, {Clayton}, \& {Mathis}}]{Cardelli1989}
{Cardelli}, J.~A., {Clayton}, G.~C., \& {Mathis}, J.~S. 1989, \bibinfo{title}{{The Relationship between Infrared, Optical, and Ultraviolet Extinction},} \apj, 345, 245, \dodoi{10.1086/167900}

% type= article
\bibitem[{G. {Carraro} {et~al.}(2007){Carraro}, {Geisler}, {Villanova}, {Frinchaboy}, \& {Majewski}}]{Carraro2007}
{Carraro}, G., {Geisler}, D., {Villanova}, S., {Frinchaboy}, P.~M., \& {Majewski}, S.~R. 2007, \bibinfo{title}{{Old open clusters in the outer Galactic disk},} \aap, 476, 217, \dodoi{10.1051/0004-6361:20078113}

% type= article
\bibitem[{G. {Carraro} {et~al.}(2017){Carraro}, {Turner}, {Majaess}, {Baume}, {Gamen}, \& {Molina Lera}}]{Carraro2017}
{Carraro}, G., {Turner}, D.~G., {Majaess}, D.~J., {et~al.} 2017, \bibinfo{title}{{Extinction in the Star Cluster SAI 113 and Galactic Structure in Carina},} \aj, 153, 156, \dodoi{10.3847/1538-3881/aa5c3a}

% type= article
\bibitem[{R. {Carrera} \& E. {Pancino}(2011){Carrera} \& {Pancino}}]{Carrera2011}
{Carrera}, R., \& {Pancino}, E. 2011, \bibinfo{title}{{Chemical abundance analysis of the open clusters Berkeley 32, NGC 752, Hyades, and Praesepe},} \aap, 535, A30, \dodoi{10.1051/0004-6361/201117473}

% type= article
\bibitem[{R. {Carrera} {et~al.}(2019){Carrera}, {Bragaglia}, {Cantat-Gaudin}, {Vallenari}, {Balaguer-N{\'u}{\~n}ez}, {Bossini}, {Casamiquela}, {Jordi}, {Sordo}, \& {Soubiran}}]{Carrera2019}
{Carrera}, R., {Bragaglia}, A., {Cantat-Gaudin}, T., {et~al.} 2019, \bibinfo{title}{{Open clusters in APOGEE and GALAH. Combining Gaia and ground-based spectroscopic surveys},} \aap, 623, A80, \dodoi{10.1051/0004-6361/201834546}

% type= article
\bibitem[{R. {Carrera} {et~al.}(2022){Carrera}, {Casamiquela}, {Carbajo-Hijarrubia}, {Balaguer-N{\'u}{\~n}ez}, {Jordi}, {Romero-G{\'o}mez}, {Blanco-Cuaresma}, {Cantat-Gaudin}, {Lillo-Box}, {Masana}, \& {Pancino}}]{Carrera2022}
{Carrera}, R., {Casamiquela}, L., {Carbajo-Hijarrubia}, J., {et~al.} 2022, \bibinfo{title}{{OCCASO. IV. Radial velocities and open cluster kinematics},} \aap, 658, A14, \dodoi{10.1051/0004-6361/202141832}

% type= article
\bibitem[{A. {Castro-Ginard} {et~al.}(2020){Castro-Ginard}, {Jordi}, {Luri}, {{\'A}lvarez Cid-Fuentes}, {Casamiquela}, {Anders}, {Cantat-Gaudin}, {Mongui{\'o}}, {Balaguer-N{\'u}{\~n}ez}, {Sol{\`a}}, \& {Badia}}]{Castro-Ginard2020}
{Castro-Ginard}, A., {Jordi}, C., {Luri}, X., {et~al.} 2020, \bibinfo{title}{Hunting for open clusters in {G}aia {DR}2: 582 new open clusters in the {G}alactic disc,} \aap, 635, A45, \dodoi{10.1051/0004-6361/201937386}

% type= article
\bibitem[{L. {Cavallo} {et~al.}(2024){Cavallo}, {Spina}, {Carraro}, {Magrini}, {Poggio}, {Cantat-Gaudin}, {Pasquato}, {Lucatello}, {Ortolani}, \& {Schiappacasse-Ulloa}}]{Cavallo2024}
{Cavallo}, L., {Spina}, L., {Carraro}, G., {et~al.} 2024, \bibinfo{title}{{Parameter Estimation for Open Clusters using an Artificial Neural Network with a QuadTree-based Feature Extractor},} \aj, 167, 12, \dodoi{10.3847/1538-3881/ad07e5}

% type= article
\bibitem[{H. {{\c{C}}akmak} {et~al.}(2024){{\c{C}}akmak}, {Yontan}, {Bilir}, {Banks}, {Michel}, {Soydugan}, {Ko{\c{c}}}, \& {Er{\c{c}}ay}}]{Cakmak2024}
{{\c{C}}akmak}, H., {Yontan}, T., {Bilir}, S., {et~al.} 2024, \bibinfo{title}{{Photometric and Kinematic Studies of Open Clusters Ruprecht 1 and Ruprecht 171},} Astronomische Nachrichten, 345, e20240054, \dodoi{10.1002/asna.20240054}

% type= article
\bibitem[{D.~C. {{\c{C}}{\i}nar} {et~al.}(2025){{\c{C}}{\i}nar}, {Bilir}, {{\c{S}}ahin}, \& {Plevne}}]{Cinar2025}
{{\c{C}}{\i}nar}, D.~C., {Bilir}, S., {{\c{S}}ahin}, T., \& {Plevne}, O. 2025, \bibinfo{title}{{Tracing the Galactic Origins of Selected Four G-type Stars in the Solar Neighborhood},} \aj, 170, 13, \dodoi{10.3847/1538-3881/add343}

% type= article
\bibitem[{D.~C. {{\c{C}}{\i}nar} {et~al.}(2026{\natexlab{a}}){{\c{C}}{\i}nar}, {Bisht}, {Bilir}, {Qin}, \& {Saker}}]{Cinar2026b}
{{\c{C}}{\i}nar}, D.~C., {Bisht}, D., {Bilir}, S., {Qin}, S., \& {Saker}, L. 2026{\natexlab{a}}, \bibinfo{title}{{Hot Degenerate Components in Blue Stragglers: A Multiwavelength Spectral Energy Distribution Analysis of Nine Open Clusters with Swift/UVOT},} \apj, 1006, 109, \dodoi{10.3847/1538-4357/ae7a33}

% type= article
\bibitem[{D.~C. {{\c{C}}{\i}nar} {et~al.}(2026{\natexlab{b}}){{\c{C}}{\i}nar}, {Bisht}, {Jiang}, {Elsanhoury}, {Belwal}, \& {Bilir}}]{Cinar2026c}
{{\c{C}}{\i}nar}, D.~C., {Bisht}, D., {Jiang}, I.-G., {et~al.} 2026{\natexlab{b}}, \bibinfo{title}{{Berkeley 32: A Metal-poor and Dynamically Evolved Open Cluster with Evidence of Radial Migration},} arXiv e-prints, arXiv:2607.15131, \dodoi{10.48550/arXiv.2607.15131}

% type= article
\bibitem[{D.~C. {{\c{C}}{\i}nar} {et~al.}(2026{\natexlab{c}}){{\c{C}}{\i}nar}, {Elsanhoury}, \& {Haroon}}]{Cinar2026}
{{\c{C}}{\i}nar}, D.~C., {Elsanhoury}, W.~H., \& {Haroon}, A.~A. 2026{\natexlab{c}}, \bibinfo{title}{{Exploring the near galactic centre: A comprehensive study of bulge OCs HSC 25, HSC 37, HSC 2878 utilising Gaia DR3 data},} \na, 125, 102532, \dodoi{10.1016/j.newast.2026.102532}

% type= article
\bibitem[{D.~C. {{\c{C}}{\i}nar} {et~al.}(2024){{\c{C}}{\i}nar}, {{Ta{\c{s}}demir}}, {Ko{\c{c}}}, \& {Iyer}}]{Cinar2024}
{{\c{C}}{\i}nar}, D.~C., {{Ta{\c{s}}demir}}, S., {Ko{\c{c}}}, S., \& {Iyer}, S. 2024, \bibinfo{title}{{SED Analysis of the Old Open Cluster NGC 188},} Physics and Astronomy Reports, 2, 1, \dodoi{10.26650/PAR.2024.00002}

% type= article
\bibitem[{K.~C. {Chambers} {et~al.}(2016){Chambers}, {Magnier}, {Metcalfe}, {Flewelling}, {Huber}, {Waters}, {Denneau}, {Draper}, {Farrow}, {Finkbeiner}, {Holmberg}, {Koppenhoefer}, {Price}, {Rest}, {Saglia}, {Schlafly}, {Smartt}, {Sweeney}, {Wainscoat}, {Burgett}, {Chastel}, {Grav}, {Heasley}, {Hodapp}, {Jedicke}, {Kaiser}, {Kudritzki}, {Luppino}, {Lupton}, {Monet}, {Morgan}, {Onaka}, {Shiao}, {Stubbs}, {Tonry}, {White}, {Ba{\~n}ados}, {Bell}, {Bender}, {Bernard}, {Boegner}, {Boffi}, {Botticella}, {Calamida}, {Casertano}, {Chen}, {Chen}, {Cole}, {Deacon}, {Frenk}, {Fitzsimmons}, {Gezari}, {Gibbs}, {Goessl}, {Goggia}, {Gourgue}, {Goldman}, {Grant}, {Grebel}, {Hambly}, {Hasinger}, {Heavens}, {Heckman}, {Henderson}, {Henning}, {Holman}, {Hopp}, {Ip}, {Isani}, {Jackson}, {Keyes}, {Koekemoer}, {Kotak}, {Le}, {Liska}, {Long}, {Lucey}, {Liu}, {Martin}, {Masci}, {McLean}, {Mindel}, {Misra}, {Morganson}, {Murphy}, {Obaika}, {Narayan}, {Nieto-Santisteban}, {Norberg}, {Peacock}, {Pier}, {Postman}, {Primak}, {Rae},
  {Rai}, {Riess}, {Riffeser}, {Rix}, {R{\"o}ser}, {Russel}, {Rutz}, {Schilbach}, {Schultz}, {Scolnic}, {Strolger}, {Szalay}, {Seitz}, {Small}, {Smith}, {Soderblom}, {Taylor}, {Thomson}, {Taylor}, {Thakar}, {Thiel}, {Thilker}, {Unger}, {Urata}, {Valenti}, {Wagner}, {Walder}, {Walter}, {Watters}, {Werner}, {Wood-Vasey}, \& {Wyse}}]{Chambers2016}
{Chambers}, K.~C., {Magnier}, E.~A., {Metcalfe}, N., {et~al.} 2016, \bibinfo{title}{{The Pan-STARRS1 Surveys},} arXiv e-prints, arXiv:1612.05560, \dodoi{10.48550/arXiv.1612.05560}

% type= article
\bibitem[{Y.~Q. {Chen} \& G. {Zhao}(2020){Chen} \& {Zhao}}]{Chen2020}
{Chen}, Y.~Q., \& {Zhao}, G. 2020, \bibinfo{title}{{Open clusters as tracers on radial migration of the galactic disc},} \mnras, 495, 2673, \dodoi{10.1093/mnras/staa1079}

% type= article
\bibitem[{C. {Chiappini} {et~al.}(1997){Chiappini}, {Matteucci}, \& {Gratton}}]{Chiappini1997}
{Chiappini}, C., {Matteucci}, F., \& {Gratton}, R. 1997, \bibinfo{title}{{The Chemical Evolution of the Galaxy: The Two-Infall Model},} \apj, 477, 765, \dodoi{10.1086/303726}

% type= article
\bibitem[{J. {Choi} {et~al.}(2016){Choi}, {Dotter}, {Conroy}, {Cantiello}, {Paxton}, \& {Johnson}}]{Choi2016}
{Choi}, J., {Dotter}, A., {Conroy}, C., {et~al.} 2016, \bibinfo{title}{{Mesa Isochrones and Stellar Tracks (MIST). I. Solar-scaled Models},} \apj, 823, 102, \dodoi{10.3847/0004-637X/823/2/102}

% type= article
\bibitem[{{\v{Z}}. {Chrob{\'a}kov{\'a}} {et~al.}(2022){Chrob{\'a}kov{\'a}}, {Nagy}, \& {L{\'o}pez-Corredoira}}]{Chrobakova2022}
{Chrob{\'a}kov{\'a}}, {\v{Z}}., {Nagy}, R., \& {L{\'o}pez-Corredoira}, M. 2022, \bibinfo{title}{{Warp and flare of the Galactic disc revealed with supergiants by Gaia EDR3},} \aap, 664, A58, \dodoi{10.1051/0004-6361/202243296}

% type= article
\bibitem[{B. Co{\c{s}}kuno{\v{g}}lu {et~al.}(2011)Co{\c{s}}kuno{\v{g}}lu, Ak, Bilir, Karaali, Yaz, Gilmore, Seabroke, Bienaym{\'e}, Bland-Hawthorn, Campbell, Freeman, Gibson, Grebel, Munari, Navarro, Parker, Siebert, Siviero, Steinmetz, Watson, Wyse, \& Zwitter}]{Coskunoglu2011}
Co{\c{s}}kuno{\v{g}}lu, B., Ak, S., Bilir, S., {et~al.} 2011, \bibinfo{title}{{Local stellar kinematics from {RAVE} data - {I.} Local standard of rest},} Monthly Notices of the Royal Astronomical Society, 412, 1237, \dodoi{10.1111/j.1365-2966.2010.17983.x}

% type= article
\bibitem[{T. Cover \& P. Hart(1967)Cover \& Hart}]{cover1967nearest}
Cover, T., \& Hart, P. 1967, \bibinfo{title}{Nearest neighbor pattern classification,} IEEE transactions on information theory, 13, 21

% type= article
\bibitem[{O. {Demircan} {et~al.}(2006){Demircan}, {Eker}, {Karata{\textcommabelow s}}, \& {Bilir}}]{Demircan2006}
{Demircan}, O., {Eker}, Z., {Karata{\textcommabelow s}}, Y., \& {Bilir}, S. 2006, \bibinfo{title}{{Mass loss and orbital period decrease in detached chromospherically active binaries},} \mnras, 366, 1511, \dodoi{10.1111/j.1365-2966.2005.09948.x}

% type= article
\bibitem[{A.~P. Dempster {et~al.}(1977)Dempster, Laird, \& Rubin}]{dempster1977maximum}
Dempster, A.~P., Laird, N.~M., \& Rubin, D.~B. 1977, \bibinfo{title}{Maximum likelihood from incomplete data via the EM algorithm,} Journal of the royal statistical society: series B (methodological), 39, 1

% type= inproceedings
\bibitem[{S. {Derriere} \& A.~C. {Robin}(2001){Derriere} \& {Robin}}]{Derriere2001}
{Derriere}, S., \& {Robin}, A.~C. 2001, \bibinfo{title}{{Near-infrared Surveys and the Shape of the Galactic Disc.},} in Astronomical Society of the Pacific Conference Series, Vol. 232, The New Era of Wide Field Astronomy, ed. R.~{Clowes}, A.~{Adamson}, \& G.~{Bromage}, 229

% type= article
\bibitem[{W.~S. {Dias} {et~al.}(2002){Dias}, {Alessi}, {Moitinho}, \& {L{\'e}pine}}]{Dias2002}
{Dias}, W.~S., {Alessi}, B.~S., {Moitinho}, A., \& {L{\'e}pine}, J.~R.~D. 2002, \bibinfo{title}{{New catalogue of optically visible open clusters and candidates},} \aap, 389, 871, \dodoi{10.1051/0004-6361:20020668}

% type= article
\bibitem[{W.~S. {Dias} {et~al.}(2014){Dias}, {Monteiro}, {Caetano}, {L{\'e}pine}, {Assafin}, \& {Oliveira}}]{Dias2014}
{Dias}, W.~S., {Monteiro}, H., {Caetano}, T.~C., {et~al.} 2014, \bibinfo{title}{{Proper motions of the optically visible open clusters based on the UCAC4 catalog},} \aap, 564, A79, \dodoi{10.1051/0004-6361/201323226}

% type= article
\bibitem[{W.~S. {Dias} {et~al.}(2021){Dias}, {Monteiro}, {Moitinho}, {L{\'e}pine}, {Carraro}, {Paunzen}, {Alessi}, \& {Villela}}]{Dias2021}
{Dias}, W.~S., {Monteiro}, H., {Moitinho}, A., {et~al.} 2021, \bibinfo{title}{{Updated parameters of 1743 open clusters based on Gaia DR2},} \mnras, 504, 356, \dodoi{10.1093/mnras/stab770}

% type= article
\bibitem[{S. {Dib} {et~al.}(2018){Dib}, {Schmeja}, \& {Parker}}]{Dib2018}
{Dib}, S., {Schmeja}, S., \& {Parker}, R.~J. 2018, \bibinfo{title}{{Structure and mass segregation in Galactic stellar clusters},} \mnras, 473, 849, \dodoi{10.1093/mnras/stx2413}

% type= article
\bibitem[{J. {Donada} {et~al.}(2023){Donada}, {Anders}, {Jordi}, {Masana}, {Gieles}, {Perren}, {Balaguer-N{\'u}{\~n}ez}, {Castro-Ginard}, {Cantat-Gaudin}, \& {Casamiquela}}]{Donada2023}
{Donada}, J., {Anders}, F., {Jordi}, C., {et~al.} 2023, \bibinfo{title}{{The multiplicity fraction in 202 open clusters from Gaia},} \aap, 675, A89, \dodoi{10.1051/0004-6361/202245219}

% type= article
\bibitem[{S. {D{\"o}ner} {et~al.}(2023){D{\"o}ner}, {Ak}, {{\"O}nal Ta{\textcommabelow s}}, \& {Plevne}}]{Doner2023}
{D{\"o}ner}, S., {Ak}, S., {{\"O}nal Ta{\textcommabelow s}}, {\"O}., \& {Plevne}, O. 2023, \bibinfo{title}{{The Age-Metallicity Relation in the Solar Neighbourhood},} Physics and Astronomy Reports, 1, 11, \dodoi{10.26650/PAR.2023.00002}

% type= article
\bibitem[{J. {Donor} {et~al.}(2020){Donor}, {Frinchaboy}, {Cunha}, \& {others}}]{Donor2020}
{Donor}, J., {Frinchaboy}, P.~M., {Cunha}, K., \& {others}. 2020, \bibinfo{title}{{The Open Cluster Chemical Abundances and Mapping Survey. III. APOGEE-derived Elemental Abundances for 128 Open Clusters},} \aj, 159, 199, \dodoi{10.3847/1538-3881/ab77bc}

% type= article
\bibitem[{A. {Dotter}(2016){Dotter}}]{Dotter2016}
{Dotter}, A. 2016, \bibinfo{title}{{MESA Isochrones and Stellar Tracks (MIST) 0: Methods for the Construction of Stellar Isochrones},} \apjs, 222, 8, \dodoi{10.3847/0067-0049/222/1/8}

% type= article
\bibitem[{Z. {Eker} {et~al.}(2014){Eker}, {Bilir}, {Soydugan}, {G{\"o}k{\c{c}}e}, {Soydugan}, {T{\"u}ys{\"u}z}, {{\c{S}}eny{\"u}z}, \& {Demircan}}]{Eker2014}
{Eker}, Z., {Bilir}, S., {Soydugan}, F., {et~al.} 2014, \bibinfo{title}{{The Catalogue of Stellar Parameters from the Detached Double-Lined Eclipsing Binaries in the Milky Way},} \pasa, 31, e024, \dodoi{10.1017/pasa.2014.17}

% type= article
\bibitem[{Z. {Eker} {et~al.}(2006){Eker}, {Demircan}, {Bilir}, \& {Karata{\textcommabelow s}}}]{Eker2006}
{Eker}, Z., {Demircan}, O., {Bilir}, S., \& {Karata{\textcommabelow s}}, Y. 2006, \bibinfo{title}{{Dynamical evolution of active detached binaries on the logJ$_{o}$-logM diagram and contact binary formation},} \mnras, 373, 1483, \dodoi{10.1111/j.1365-2966.2006.11073.x}

% type= article
\bibitem[{Z. {Eker} {et~al.}(2024){Eker}, {Soydugan}, \& {Bilir}}]{Eker2024}
{Eker}, Z., {Soydugan}, F., \& {Bilir}, S. 2024, \bibinfo{title}{{Fundamentals of Stars: A Critical Look at Mass-Luminosity Relations and Beyond},} Physics and Astronomy Reports, 2, 41, \dodoi{10.26650/PAR.2024.00001}

% type= article
\bibitem[{Z. {Eker} {et~al.}(2015){Eker}, {Soydugan}, {Soydugan}, {Bilir}, {Yaz G{\"o}k{\c{c}}e}, {Steer}, {T{\"u}ys{\"u}z}, {{\c{S}}eny{\"u}z}, \& {Demircan}}]{Eker2015}
{Eker}, Z., {Soydugan}, F., {Soydugan}, E., {et~al.} 2015, \bibinfo{title}{{Main-Sequence Effective Temperatures from a Revised Mass-Luminosity Relation Based on Accurate Properties},} \aj, 149, 131, \dodoi{10.1088/0004-6256/149/4/131}

% type= article
\bibitem[{Z. {Eker} {et~al.}(2018){Eker}, {Bak{\i}{\textcommabelow s}}, {Bilir}, {Soydugan}, {Steer}, {Soydugan}, {Bak{\i}{\textcommabelow s}}, {Ali{\c{c}}avu{\textcommabelow s}}, {Aslan}, \& {Alpsoy}}]{Eker2018}
{Eker}, Z., {Bak{\i}{\textcommabelow s}}, V., {Bilir}, S., {et~al.} 2018, \bibinfo{title}{{Interrelated main-sequence mass-luminosity, mass-radius, and mass-effective temperature relations},} \mnras, 479, 5491, \dodoi{10.1093/mnras/sty1834}

% type= article
\bibitem[{W.~H. {Elsanhoury} {et~al.}(2025){Elsanhoury}, {Haroon}, {Elkholy}, \& {{\c{C}}{\i}nar}}]{Elsanhoury2025}
{Elsanhoury}, W.~H., {Haroon}, A.~A., {Elkholy}, E.~A., \& {{\c{C}}{\i}nar}, D.~C. 2025, \bibinfo{title}{{Deeply comprehensive astrometric, photometric, and kinematic studies of the three OCSN open clusters with Gaia DR3},} Journal of Astrophysics and Astronomy, 46, 21, \dodoi{10.1007/s12036-025-10044-0}

% type= article
\bibitem[{W.~H. {Elsanhoury} {et~al.}(2026){Elsanhoury}, {Ta{\textcommabelow s}demir}, {{\c{C}}{\i}nar}, {Canbay}, {Haroon}, \& {Ahmed}}]{Elsanhoury2026}
{Elsanhoury}, W.~H., {Ta{\textcommabelow s}demir}, S., {{\c{C}}{\i}nar}, D.~C., {et~al.} 2026, \bibinfo{title}{{Exploring the Structure and Evolution of Four Young Open Clusters near the Galactic Mid-plane via Gaia DR3},} Research in Astronomy and Astrophysics, 26, 035020, \dodoi{10.1088/1674-4527/ae3280}

% type= article
\bibitem[{F.~R. {Ferraro} {et~al.}(2026){Ferraro}, {Lanzoni}, {Vesperini}, {Dalessandro}, {Cadelano}, {Pallanca}, {Beccari}, {Nardiello}, {Libralato}, \& {Piotto}}]{Ferraro2026}
{Ferraro}, F.~R., {Lanzoni}, B., {Vesperini}, E., {et~al.} 2026, \bibinfo{title}{{A binary-related origin mediated by environmental conditions for blue straggler stars},} Nature Communications, 17, 768, \dodoi{10.1038/s41467-025-68159-5}

% type= article
\bibitem[{D. {Foreman-Mackey} {et~al.}(2013){Foreman-Mackey}, {Hogg}, {Lang}, \& {Goodman}}]{emcee}
{Foreman-Mackey}, D., {Hogg}, D.~W., {Lang}, D., \& {Goodman}, J. 2013, \bibinfo{title}{{emcee: The MCMC Hammer},} \pasp, 125, 306, \dodoi{10.1086/670067}

% type= article
\bibitem[{N. {Frankel} {et~al.}(2018){Frankel}, {Rix}, {Ting}, {Ness}, \& {Hogg}}]{Frankel_2018}
{Frankel}, N., {Rix}, H.-W., {Ting}, Y.-S., {Ness}, M., \& {Hogg}, D.~W. 2018, \bibinfo{title}{{Measuring Radial Orbit Migration in the Galactic Disk},} \apj, 865, 96, \dodoi{10.3847/1538-4357/aadba5}

% type= article
\bibitem[{N. {Frankel} {et~al.}(2020){Frankel}, {Sanders}, {Ting}, \& {Rix}}]{Frankel2020}
{Frankel}, N., {Sanders}, J., {Ting}, Y.-S., \& {Rix}, H.-W. 2020, \bibinfo{title}{{Keeping It Cool: Much Orbit Migration, yet Little Heating, in the Galactic Disk},} \apj, 896, 15, \dodoi{10.3847/1538-4357/ab910c}

% type= article
\bibitem[{E.~D. {Friel}(1995){Friel}}]{Friel1995}
{Friel}, E.~D. 1995, \bibinfo{title}{{The Old Open Clusters of the Milky Way},} \araa, 33, 381, \dodoi{10.1146/annurev.aa.33.090195.002121}

% type= article
\bibitem[{E.~D. {Friel} {et~al.}(2002){Friel}, {Janes}, {Tavarez}, {Scott}, {Katsanis}, {Lotz}, {Hong}, \& {Miller}}]{Friel2002}
{Friel}, E.~D., {Janes}, K.~A., {Tavarez}, M., {et~al.} 2002, \bibinfo{title}{{Metallicities of Old Open Clusters},} \aj, 124, 2693, \dodoi{10.1086/344161}

% type= article
\bibitem[{ {Gaia Collaboration} {et~al.}(2016){Gaia Collaboration}, {Prusti}, {de Bruijne}, \& {others}}]{GaiaCollaboration2016}
{Gaia Collaboration}, {Prusti}, T., {de Bruijne}, J.~H.~J., \& {others}. 2016, \bibinfo{title}{{The Gaia mission},} \aap, 595, A1, \dodoi{10.1051/0004-6361/201629272}

% type= article
\bibitem[{ {Gaia Collaboration} {et~al.}(2018){Gaia Collaboration}, {Brown}, {Vallenari}, {Prusti}, {de Bruijne}, {Babusiaux}, {Bailer-Jones}, {Biermann}, {Evans}, {Eyer}, {Jansen}, {Jordi}, {Klioner}, {Lammers}, {Lindegren}, {Luri}, {Mignard}, {Panem}, {Pourbaix}, {Randich}, {Sartoretti}, {Siddiqui}, {Soubiran}, {van Leeuwen}, {Walton}, {Arenou}, {Bastian}, {Cropper}, {Drimmel}, {Katz}, {Lattanzi}, {Bakker}, {Cacciari}, {Casta{\~n}eda}, {Chaoul}, {Cheek}, {De Angeli}, {Fabricius}, {Guerra}, {Holl}, {Masana}, {Messineo}, {Mowlavi}, {Nienartowicz}, {Panuzzo}, {Portell}, {Riello}, {Seabroke}, {Tanga}, {Th{\'e}venin}, {Gracia-Abril}, {Comoretto}, {Garcia-Reinaldos}, {Teyssier}, {Altmann}, {Andrae}, {Audard}, {Bellas-Velidis}, {Benson}, {Berthier}, {Blomme}, {Burgess}, {Busso}, {Carry}, {Cellino}, {Clementini}, {Clotet}, {Creevey}, {Davidson}, {De Ridder}, {Delchambre}, {Dell'Oro}, {Ducourant}, {Fern{\'a}ndez-Hern{\'a}ndez}, {Fouesneau}, {Fr{\'e}mat}, {Galluccio}, {Garc{\'\i}a-Torres},
  {Gonz{\'a}lez-N{\'u}{\~n}ez}, {Gonz{\'a}lez-Vidal}, {Gosset}, {Guy}, {Halbwachs}, {Hambly}, {Harrison}, {Hern{\'a}ndez}, {Hestroffer}, {Hodgkin}, {Hutton}, {Jasniewicz}, {Jean-Antoine-Piccolo}, {Jordan}, {Korn}, {Krone-Martins}, {Lanzafame}, {Lebzelter}, {L{\"o}ffler}, {Manteiga}, {Marrese}, {Mart{\'\i}n-Fleitas}, {Moitinho}, {Mora}, {Muinonen}, {Osinde}, {Pancino}, {Pauwels}, {Petit}, {Recio-Blanco}, {Richards}, {Rimoldini}, {Robin}, {Sarro}, {Siopis}, {Smith}, {Sozzetti}, {S{\"u}veges}, {Torra}, {van Reeven}, {Abbas}, {Abreu Aramburu}, {Accart}, {Aerts}, {Altavilla}, {{\'A}lvarez}, {Alvarez}, {Alves}, {Anderson}, {Andrei}, {Anglada Varela}, {Antiche}, {Antoja}, {Arcay}, {Astraatmadja}, {Bach}, {Baker}, {Balaguer-N{\'u}{\~n}ez}, {Balm}, {Barache}, {Barata}, {Barbato}, {Barblan}, {Barklem}, {Barrado}, {Barros}, {Barstow}, {Bartholom{\'e} Mu{\~n}oz}, {Bassilana}, {Becciani}, {Bellazzini}, {Berihuete}, {Bertone}, {Bianchi}, {Bienaym{\'e}}, {Blanco-Cuaresma}, {Boch}, {Boeche}, {Bombrun}, {Borrachero},
  {Bossini}, {Bouquillon}, {Bourda}, {Bragaglia}, {Bramante}, {Breddels}, {Bressan}, {Brouillet}, {Br{\"u}semeister}, {Brugaletta}, {Bucciarelli}, {Burlacu}, {Busonero}, {Butkevich}, {Buzzi}, {Caffau}, {Cancelliere}, {Cannizzaro}, {Cantat-Gaudin}, {Carballo}, {Carlucci}, {Carrasco}, {Casamiquela}, {Castellani}, {Castro-Ginard}, {Charlot}, {Chemin}, {Chiavassa}, {Cocozza}, {Costigan}, {Cowell}, {Crifo}, {Crosta}, {Crowley}, {Cuypers}, {Dafonte}, {Damerdji}, {Dapergolas}, {David}, {David}, {de Laverny}, \& {De Luise}}]{GaiaCollaboration2018}
{Gaia Collaboration}, {Brown}, A.~G.~A., {Vallenari}, A., {et~al.} 2018, \bibinfo{title}{{Gaia Data Release 2. Summary of the contents and survey properties},} \aap, 616, A1, \dodoi{10.1051/0004-6361/201833051}

% type= article
\bibitem[{ {Gaia Collaboration} {et~al.}(2021){Gaia Collaboration}, {Brown}, {Vallenari}, {Prusti}, {de Bruijne}, {Babusiaux}, {Biermann}, {Creevey}, {Evans}, {Eyer}, {Hutton}, {Jansen}, {Jordi}, {Klioner}, {Lammers}, {Lindegren}, {Luri}, {Mignard}, {Panem}, {Pourbaix}, {Randich}, {Sartoretti}, {Soubiran}, {Walton}, {Arenou}, {Bailer-Jones}, {Bastian}, {Cropper}, {Drimmel}, {Katz}, {Lattanzi}, {van Leeuwen}, {Bakker}, {Cacciari}, {Casta{\~n}eda}, {De Angeli}, {Ducourant}, {Fabricius}, {Fouesneau}, {Fr{\'e}mat}, {Guerra}, {Guerrier}, {Guiraud}, {Jean-Antoine Piccolo}, {Masana}, {Messineo}, {Mowlavi}, {Nicolas}, {Nienartowicz}, {Pailler}, {Panuzzo}, {Riclet}, {Roux}, {Seabroke}, {Sordo}, {Th{\'e}venin}, {Gracia-Abril}, {Portell}, {Teyssier}, {Altmann}, {Andrae}, {Bellas-Velidis}, {Benson}, {Berthier}, {Blomme}, {Brugaletta}, {Burgess}, {Busso}, {Carry}, {Cellino}, {Cheek}, {Clementini}, {Damerdji}, {Davidson}, {Delchambre}, {Dell'Oro}, {Fern{\'a}ndez-Hern{\'a}ndez}, {Galluccio}, {Garc{\'\i}a-Lario},
  {Garc{\'\i}a-Reinaldos}, {Gonz{\'a}lez-N{\'u}{\~n}ez}, {Gosset}, {Haigron}, {Halbwachs}, {Hambly}, {Harrison}, {Hatzidimitriou}, {Heiter}, {Hern{\'a}ndez}, {Hestroffer}, {Hodgkin}, {Holl}, {Jan{\ss}en}, {Jevardat de Fombelle}, {Jordan}, {Krone-Martins}, {Lanzafame}, {L{\"o}ffler}, {Lorca}, {L{\"u}st}, {McMillan}, {Marchal}, {Marrese}, {Moitinho}, {Mora}, {Muinonen}, {Osborne}, {Pancino}, {Pauwels}, {Petit}, {Recio-Blanco}, {Richards}, {Riello}, {Rimoldini}, {Robin}, {Roegiers}, {Rybizki}, {Sarro}, {Siopis}, {Smith}, {Sozzetti}, {Utrilla}, {van Leeuwen}, {van Reeven}, {Abbas}, {Abreu Aramburu}, {Accart}, {Aerts}, {Aguado}, {Ajaj}, {Altavilla}, {{\'A}lvarez}, {{\'A}lvarez Cid-Fuentes}, {Alves}, {Anderson}, {Anglada Varela}, {Antoja}, {Audard}, {Baines}, {Baker}, {Balaguer-N{\'u}{\~n}ez}, {Balbinot}, {Balog}, {Barache}, {Barbato}, {Barros}, {Barstow}, {Bartolom{\'e}}, {Bassilana}, {Bauchet}, {Baudesson-Stella}, {Becciani}, {Bellazzini}, {Bernet}, {Bertone}, {Bianchi}, {Blanco-Cuaresma}, {Boch}, {Bombrun},
  {Bossini}, {Bouquillon}, {Bragaglia}, {Bramante}, {Breddels}, {Bressan}, {Brouillet}, {Br{\"u}semeister}, {Bucciarelli}, {Busonero}, {Butkevich}, {Buzzi}, {Caffau}, {Calabr{\`o}}, {Calvo}, {Cano}, {Cantat-Gaudin}, {Capello}, {Carballo}, {Carlucci}, {Carnerero}, {Carrasco}, {Casamiquela}, {Castellani}, {Castro-Ginard}, {Castro Sampol}, {Chaoul}, {Charlot}, {Chemin}, {Chiavassa}, {Chornay}, {Clotet}, {Cocozza}, {Collins}, {Comer{\'o}n}, {Contursi}, {Cooper}, {Cornez}, {Cowell}, {Crifo}, {Crosta}, {Crowley}, {Dafonte}, {Dapergolas}, {David}, {David}, {de Laverny}, {De Luise}, {De March}, {De Ridder}, {de Souza}, {de Teodoro}, {de Torres}, {del Peloso}, {del Pozo}, {Delbo}, {Delgado}, {Delgado}, {Delisle}, {Di Matteo}, {Diakite}, {Diener}, {Distefano}, {Dolding}, {Eappachen}, {Edvardsson}, {Enke}, {Esquej}, {Fabre}, {Fabrizio}, {Faigler}, {Fedorets}, {Fernique}, {Fienga}, {Figueras}, {Filippi}, {Findeisen}, {Fonti}, {Fraile}, {Fraser}, {Fr{\'e}zouls}, {Gai}, {Galleti}, {Gallego}, {Galli}, {G{\'a}lvez-Ortiz},
  {Gante}, {Garc{\'\i}a-Guirado}, {Garc{\'\i}a-Torres}, {Garofalo}, {Gavel}, {Gavras}, {Gerlach}, {Geyer}, {Giacobbe}, {Gilmore}, {Girona}, {Giuffrida}, {Gomel}, {Gomez}, {Gonz{\'a}lez-Marcos}, {Gonz{\'a}lez-Santamar{\'\i}a}, {Gonz{\'a}lez-Vidal}, {Granvik}, {Guti{\'e}rrez-S{\'a}nchez}, {Guy}, {Hauser}, {Haywood}, {Helmi}, {Hidalgo}, {Hilger}, {H{\l}adczuk}, {Hobbs}, {Holland}, {Huckle}, {Jardine}, {Jasniewicz}, {Jorissen}, {Juaristi Campillo}, {Julbe}, {Karbevska}, {Kervella}, {Khanna}, {Kontizas}, {Kordopatis}, {Korn}, {K{\'o}sp{\'a}l}, {Kostrzewa-Rutkowska}, {Kruszy{\'n}ska}, {Kun}, {Laizeau}, {Lanza}, {Lasne}, {Le Campion}, {Le Fustec}, {Lebreton}, {Lebzelter}, {Leccia}, {Leclerc}, {Lecoeur-Taibi}, {Liao}, {Licata}, {Lindstr{\o}m}, {Lister}, {Livanou}, {Lobel}, {Madrero Pardo}, {Managau}, {Mann}, {Marconi}, {Marcos Santos}, {Mar{\'\i}n Pina}, {Marinoni}, {Marocco}, {Marshall}, {Martin Polo}, {Mart{\'\i}n-Fleitas}, {Marton}, {Mary}, {Massari}, {Mastroserio}, {Mateu}, {Mat{\'{\i}}as}, {Mazure}, {Messina},
  {Michalik}, {Millar}, {Mints}, {Molina}, {Molinaro}, {Moln{\'a}r}, {Monari}, {Mongui{\'o}}, {Montegriffo}, {Montero}, {Mor}, {Mora}, {Morbidelli}, {Morel}, {Morris}, {Muraveva}, {Murphy}, {Musella}, {Nagy}, {Noval}, {Oca{\~n}a}, {Ogden}, {Ordenovic}, {Osinde}, {Pagani}, {Pagano}, {Palaversa}, {Palicio}, {Panahi}, {Pawlak}, {Pe{\~n}alosa Esteller}, {Penttil{\"a}}, {Piersimoni}, {Pineau}, {Plachy}, {Plum}, {Poggio}, {Poretti}, {Poujoulet}, {Pr{\v{s}}a}, {Pulone}, {Racero}, {Ragaini}, {Rainer}, {Raiteri}, {Rambaux}, {Ramos}, {Ramos-Lerate}, {Re Fiorentin}, {Regibo}, {Reyl{\'e}}, {Ripepi}, {Riva}, {Rix}, {Rixon}, {Robichon}, {Robin}, {Roelens}, {Rohrbasser}, {Romero-G{\'o}mez}, {Rowell}, {Royer}, {Ruz Mieres}, {Rybicki}, {Sadowski}, {S{\'a}ez N{\'u}{\~n}ez}, {Sagrist{\`a} Sell{\'e}s}, {Sahlmann}, {Salguero}, {Samaras}, {Sanchez Gimenez}, {Sanna}, {Santove{\~n}a}, {Sarri}, {Schartmann}, {Schultheis}, {Sciacca}, {Segol}, {Segovia}, {S{\'e}gransan}, {Semeux}, {Shahaf}, {Siddiqui}, {Siebert}, {Siltala}, {Silvelo},
  {Slezak}, {Slezak}, {Smart}, {Snaith}, {Solano}, {Solitro}, {Souami}, {Souchay}, {Spagna}, {Spoto}, {Steele}, {Steidelm{\"u}ller}, {Stephenson}, {St{\`e}r}, {S{\"u}veges}, {Szabados}, {Szegedi-Elek}, {Taris}, {Taylor}, {Teixeira}, {Teodorescu}, {Terman}, {Thiel}, {Tolomei}, {Tonello}, {Torra}, {Torra}, {Torralba Elipe}, {Trabucchi}, {Tsounis}, {Turon}, {Ulla}, {Unger}, {Vaillant}, {van Dillen}, {van Reeven}, {Vanderspek}, {Vasiliev}, {Vaughn}, {Vera-Ciro}, {Viala}, {Vicente}, {Voutsinas}, {Weiler}, {Wevers}, {Wyrzykowski}, {Yoldas}, {Yvard}, {Zhao}, {Zorec}, {Zucker}, {Zurbach}, \& {Zwitter}}]{GaiaCollaboration2021}
{Gaia Collaboration}, {Brown}, A. G.~A., {Vallenari}, A., {et~al.} 2021, \bibinfo{title}{{Gaia Early Data Release 3. Summary of the contents and survey properties},} \aap, 649, A1, \dodoi{10.1051/0004-6361/202039657}

% type= article
\bibitem[{ {Gaia Collaboration} {et~al.}(2023){Gaia Collaboration}, {Vallenari}, {Brown}, {Prusti}, {de Bruijne}, {Arenou}, {Babusiaux}, {Biermann}, {Creevey}, {Ducourant}, {Evans}, {Eyer}, {Guerra}, {Hutton}, {Jordi}, {Klioner}, {Lammers}, {Lindegren}, {Luri}, {Mignard}, {Panem}, {Pourbaix}, {Randich}, {Sartoretti}, {Soubiran}, {Tanga}, {Walton}, {Bailer-Jones}, {Bastian}, {Drimmel}, {Jansen}, {Katz}, {Lattanzi}, {van Leeuwen}, {Bakker}, {Cacciari}, {Casta{\~n}eda}, {De Angeli}, {Fabricius}, {Fouesneau}, {Fr{\'e}mat}, {Galluccio}, {Guerrier}, {Heiter}, {Masana}, {Messineo}, {Mowlavi}, {Nicolas}, {Nienartowicz}, {Pailler}, {Panuzzo}, {Riclet}, {Roux}, {Seabroke}, {Sordo}, {Th{\'e}venin}, {Gracia-Abril}, {Portell}, {Teyssier}, {Altmann}, {Andrae}, {Audard}, {Bellas-Velidis}, {Benson}, {Berthier}, {Blomme}, {Burgess}, {Busonero}, {Busso}, {C{\'a}novas}, {Carry}, {Cellino}, {Cheek}, {Clementini}, {Damerdji}, {Davidson}, {de Teodoro}, {Nu{\~n}ez Campos}, {Delchambre}, {Dell'Oro}, {Esquej},
  {Fern{\'a}ndez-Hern{\'a}ndez}, {Fraile}, {Garabato}, {Garc{\'\i}a-Lario}, {Gavras}, {Gosset}, {Haigron}, {Halbwachs}, {Hambly}, {Harrison}, {Hern{\'a}ndez}, {Hestroffer}, {Hodgkin}, {Holl}, {Jan{\ss}en}, {Jevardat de Fombelle}, {Jordan}, {Krone-Martins}, {Lanzafame}, {L{\"o}ffler}, {Lorca}, {L{\"u}st}, {Marchal}, {Marrese}, {Moitinho}, {Muinonen}, {Osborne}, {Pancino}, {Pauwels}, {Petit}, {Recio-Blanco}, {Richards}, {Riello}, {Rimoldini}, {Robin}, {Roegiers}, {Rybizki}, {Sarro}, {Siopis}, {Smith}, {Sozzetti}, {Utrilla}, {van Leeuwen}, {Abbas}, {{\`A}brah{\'a}m}, {Abreu Aramburu}, {Aerts}, {Aguado}, {Ajaj}, {Aldea-Montero}, {Altavilla}, {{\'A}lvarez}, {Alves}, {Anders}, {Anderson}, {Anglada Varela}, {Antoja}, {Baines}, {Baker}, {Balaguer-N{\'u}{\~n}ez}, {Balbinot}, {Balog}, {Barache}, {Barbato}, {Barros}, {Barstow}, {Bartolom{\'e}}, {Bassilana}, {Bauchet}, {Becciani}, {Bellazzini}, {Berihuete}, {Bernet}, {Bertone}, {Bianchi}, {Binnenfeld}, {Blanco-Cuaresma}, {Blazere}, {Boch}, {Bombrun}, {Bossini},
  {Bouquillon}, {Bragaglia}, {Bramante}, {Breddels}, {Bressan}, {Brouillet}, {Brugaletta}, {Bucciarelli}, {Burlacu}, {Butkevich}, {Buzzi}, {Caffau}, {Calabr{\`o}}, {Caputo}, {Carballo}, {Carlucci}, {Carnerero}, {Carrasco}, {Casamiquela}, {Castellani}, {Castro-Ginard}, {Chaoul}, {Charlot}, {Chemin}, {Chiaramida}, {Chiavassa}, {Chornay}, {Comoretto}, {Contursi}, {Cooper}, {Cornez}, {Cowell}, {Crifo}, {Cropper}, {Crosta}, {Crowley}, {Dafonte}, {Dapergolas}, {David}, {David}, {de Laverny}, {De Luise}, {De March}, {De Ridder}, {de Souza}, {de Torres}, {del Peloso}, {del Pozo}, {Delbo}, {Delgado}, {Delgado}, {Delisle}, {D{\'{\i}}az-Garc{\'{\i}}a}, {Diakite}, {Diener}, {Distefano}, {Dolding}, {Edvardsson}, {Enke}, {Fabre}, {Fabrizio}, {Faigler}, {Fedorets}, {Fernique}, {Fienga}, {Figueras}, {Filippi}, {Findeisen}, {Fonti}, {Frelijj}, {Fuente}, {Fukal}, {Fustes}, {Gallego}, {G{\'a}lvez-Ortiz}, {Gante}, {Garabello}, {Garc{\'\i}a-Guirado}, {Garc{\'\i}a-Reinaldos}, {Garc{\'\i}a-Torres}, {Garofalo}, {Gavel}, {Gerlach},
  {Geyer}, {Giacobbe}, {Gilmore}, {Girona}, {Giuffrida}, {Gomel}, {Gomez}, {Gonz{\'a}lez-N{\'u}{\~n}ez}, {Gonz{\'a}lez-Santamar{\'\i}a}, {Gonz{\'a}lez-Vidal}, {Granvik}, {Guillout}, {Guiraud}, {Guti{\'e}rrez-S{\'a}nchez}, {Guy}, {Hatzidimitriou}, {Hauser}, {Haywood}, {Helmer}, {Helmi}, {Sarmiento}, {Hidalgo}, {Hilger}, {H{\l}adczuk}, {Hobbs}, {Holland}, {Huckle}, {Jardine}, {Jasniewicz}, {Jean-Antoine Piccolo}, {Jim{\'e}nez-Arranz}, {Jorissen}, {Juaristi Campillo}, {Julbe}, {Karbevska}, {Kervella}, {Khanna}, {Kontizas}, {Kordopatis}, {Korn}, {K{\'o}sp{\'a}l}, {Kostrzewa-Rutkowska}, {Kruszy{\'n}ska}, {Kun}, {Laizeau}, {Lambert}, {Lanza}, {Lasne}, {Le Campion}, {Le Fustec}, {Lebreton}, {Lebzelter}, {Leccia}, {Leclerc}, {Lecoeur-Taibi}, {Liao}, {Licata}, {Lindstr{\o}m}, {Lister}, {Livanou}, {Lobel}, {Madrero Pardo}, {Managau}, {Mann}, {Marchant}, {Marconi}, {Marcos}, {Marcos Santos}, {Mar{\'{\i}}n Pina}, {Marinoni}, {Marocco}, {Marshall}, {Martin Polo}, {Mart{\'\i}n-Fleitas}, {Marton}, {Mary}, {Massari},
  {Mastroserio}, {Mateu}, {Mat{\'{\i}}as}, {Mazure}, {McMillan}, {Messina}, {Michalik}, {Millar}, {Mints}, {Molina}, {Molinaro}, {Moln{\'a}r}, {Monari}, {Mongui{\'o}}, {Montegriffo}, {Montero}, {Mor}, {Mora}, {Morbidelli}, {Morel}, {Morris}, {Muraveva}, {Murphy}, {Musella}, {Nagy}, {Noval}, {Oca{\~n}a}, {Ogden}, {Ordenovic}, {Osinde}, {Pagani}, {Pagano}, {Palaversa}, {Palicio}, {Pallas-Quintana}, {Panahi}, {Payne-Wardenaar}, {Pe{\~n}alosa Esteller}, {Penttil{\"a}}, {Pichon}, {Piersimoni}, {Pineau}, {Plachy}, {Plum}, {Poggio}, {Pr{\v{s}}a}, {Pulone}, {Racero}, {Ragaini}, {Rainer}, {Raiteri}, {Rambaux}, {Ramos}, {Ramos-Lerate}, {Re Fiorentin}, {Regibo}, {Reyl{\'e}}, {Ripepi}, {Riva}, {Rix}, {Rixon}, {Robichon}, {Robin}, {Roelens}, {Rohrbasser}, {Romero-G{\'o}mez}, {Rowell}, {Royer}, {Ruz Mieres}, {Rybicki}, {Sadowski}, {S{\'a}ez N{\'u}{\~n}ez}, {Sagrist{\`a} Sell{\'e}s}, {Sahlmann}, {Salguero}, {Samaras}, {Sanchez Gimenez}, {Sanna}, {Santove{\~n}a}, {Sarri}, {Schultheis}, {Sciacca}, {Segol}, {Segovia},
  {S{\'e}gransan}, {Semeux}, {Shahaf}, {Siddiqui}, {Siebert}, {Siltala}, {Silvelo}, {Slezak}, {Slezak}, {Smart}, {Snaith}, {Solano}, {Solitro}, {Souami}, {Souchay}, {Spagna}, {Spina}, {Spoto}, {Steele}, {Steidelm{\"u}ller}, {Stephenson}, {S{\"u}veges}, {Szabados}, {Szegedi-Elek}, {Taris}, {Taylor}, {Teixeira}, {Tolomei}, {Tonello}, {Torra}, {Torra}, {Torralba Elipe}, {Trabucchi}, {Tsounis}, {Turon}, {Ulla}, {Unglaub}, {Vaillant}, {van Dillen}, {van Reeven}, {Vanderspek}, {Varadi}, {Vasiliev}, {Vaughn}, {Vera-Ciro}, {Viala}, {Vicente}, {Voutsinas}, {Weiler}, {Wevers}, {Wyrzykowski}, {Yoldas}, {Yvard}, {Zhao}, {Zorec}, {Zucker}, {Zurbach}, \& {Zwitter}}]{GaiaCollaboration2023}
{Gaia Collaboration}, {Vallenari}, A., {Brown}, A. G.~A., {et~al.} 2023, \bibinfo{title}{{Gaia Data Release 3. Summary of the content and survey properties},} \aap, 674, A1, \dodoi{10.1051/0004-6361/202243940}

% type= article
\bibitem[{A.~M. {Geller} \& R.~D. {Mathieu}(2011){Geller} \& {Mathieu}}]{Geller2011}
{Geller}, A.~M., \& {Mathieu}, R.~D. 2011, \bibinfo{title}{{A mass transfer origin for blue stragglers in NGC 188 as revealed by half-solar-mass companions},} \nat, 478, 356, \dodoi{10.1038/nature10512}

% type= article
\bibitem[{A.~M. {Geller} \& R.~D. {Mathieu}(2012){Geller} \& {Mathieu}}]{Geller2012}
{Geller}, A.~M., \& {Mathieu}, R.~D. 2012, \bibinfo{title}{{WIYN Open Cluster Study. XLVIII. The Hard-binary Population of NGC 188},} \aj, 144, 54, \dodoi{10.1088/0004-6256/144/2/54}

% type= article
\bibitem[{A. {Gelman} \& D.~B. {Rubin}(1992){Gelman} \& {Rubin}}]{gelman1992}
{Gelman}, A., \& {Rubin}, D.~B. 1992, \bibinfo{title}{{Inference from Iterative Simulation Using Multiple Sequences},} Statistical Science, 7, 457, \dodoi{10.1214/ss/1177011136}

% type=
\bibitem[{G. Gilmore \& S. Randich(2015)Gilmore \& Randich}]{ESO_GES}
Gilmore, G., \& Randich, S. 2015, Gaia-ESO spectroscopic survey, European Southern Observatory (ESO), \dodoi{10.18727/archive/25}

% type= article
\bibitem[{G. {Gilmore} {et~al.}(2022){Gilmore}, {Randich}, {Worley}, \& {others}}]{Gilmore2022}
{Gilmore}, G., {Randich}, S., {Worley}, C.~C., \& {others}. 2022, \bibinfo{title}{{The Gaia-ESO Public Spectroscopic Survey: Implementation, data products, open cluster survey, science, and legacy},} \aap, 666, A120, \dodoi{10.1051/0004-6361/202243134}

% type= article
\bibitem[{G. {Gilmore} {et~al.}(2012){Gilmore}, {Randich}, {Asplund}, {Binney}, {Bonifacio}, {Drew}, {Feltzing}, {Ferguson}, {Jeffries}, {Micela}, {Negueruela}, {Prusti}, {Rix}, {Vallenari}, {Alfaro}, {Allende-Prieto}, {Babusiaux}, {Bensby}, {Blomme}, {Bragaglia}, {Flaccomio}, {Fran{\c{c}}ois}, {Irwin}, {Koposov}, {Korn}, {Lanzafame}, {Pancino}, {Paunzen}, {Recio-Blanco}, {Sacco}, {Smiljanic}, {Van Eck}, {Walton}, {Aden}, {Aerts}, {Affer}, {Alcala}, {Altavilla}, {Alves}, {Antoja}, {Arenou}, {Argiroffi}, {Asensio Ramos}, {Bailer-Jones}, {Balaguer-Nunez}, {Bayo}, {Barbuy}, {Barisevicius}, {Barrado y Navascues}, {Battistini}, {Bellas Velidis}, {Bellazzini}, {Belokurov}, {Bergemann}, {Bertelli}, {Biazzo}, {Bienayme}, {Bland-Hawthorn}, {Boeche}, {Bonito}, {Boudreault}, {Bouvier}, {Brandao}, {Brown}, {de Bruijne}, {Burleigh}, {Caballero}, {Caffau}, {Calura}, {Capuzzo-Dolcetta}, {Caramazza}, {Carraro}, {Casagrande}, {Casewell}, {Chapman}, {Chiappini}, {Chorniy}, {Christlieb}, {Cignoni}, {Cocozza}, {Colless},
  {Collet}, {Collins}, {Correnti}, {Covino}, {Crnojevic}, {Cropper}, {Cunha}, {Damiani}, {David}, {Delgado}, {Duffau}, {Edvardsson}, {Eldridge}, {Enke}, {Eriksson}, {Evans}, {Eyer}, {Famaey}, {Fellhauer}, {Ferreras}, {Figueras}, {Fiorentino}, {Flynn}, {Folha}, {Franciosini}, {Frasca}, {Freeman}, {Fremat}, {Friel}, {Gaensicke}, {Gameiro}, {Garzon}, {Geier}, {Geisler}, {Gerhard}, {Gibson}, {Gomboc}, {Gomez}, {Gonzalez-Fernandez}, {Gonzalez Hernandez}, {Gosset}, {Grebel}, {Greimel}, {Groenewegen}, {Grundahl}, {Guarcello}, {Gustafsson}, {Hadrava}, {Hatzidimitriou}, {Hambly}, {Hammersley}, {Hansen}, {Haywood}, {Heber}, {Heiter}, {Held}, {Helmi}, {Hensler}, {Herrero}, {Hill}, {Hodgkin}, {Huelamo}, {Huxor}, {Ibata}, {Jackson}, {de Jong}, {Jonker}, {Jordan}, {Jordi}, {Jorissen}, {Katz}, {Kawata}, {Keller}, {Kharchenko}, {Klement}, {Klutsch}, {Knude}, {Koch}, {Kochukhov}, {Kontizas}, {Koubsky}, {Lallement}, {de Laverny}, {van Leeuwen}, {Lemasle}, {Lewis}, {Lind}, {Lindstrom}, {Lobel}, {Lopez Santiago}, {Lucas},
  {Ludwig}, {Lueftinger}, {Magrini}, {Maiz Apellaniz}, {Maldonado}, {Marconi}, {Marino}, {Martayan}, {Martinez-Valpuesta}, {Matijevic}, {McMahon}, {Messina}, {Meyer}, {Miglio}, {Mikolaitis}, {Minchev}, {Minniti}, {Moitinho}, {Momany}, {Monaco}, {Montalto}, {Monteiro}, {Monier}, {Montes}, {Mora}, {Moraux}, {Morel}, \& {Mowlavi}}]{GES2012}
{Gilmore}, G., {Randich}, S., {Asplund}, M., {et~al.} 2012, \bibinfo{title}{{The Gaia-ESO Public Spectroscopic Survey},} The Messenger, 147, 25

% type= article
\bibitem[{S. {Gokmen} {et~al.}(2023){Gokmen}, {Eker}, {Yontan}, {Bilir}, {Ak}, {Ak}, {Banks}, \& {Sarajedini}}]{Gokmen2023}
{Gokmen}, S., {Eker}, Z., {Yontan}, T., {et~al.} 2023, \bibinfo{title}{{CCD UBV and Gaia DR3 Analyses of the Open Clusters King 6 and NGC 1605},} \aj, 166, 263, \dodoi{10.3847/1538-3881/ad08b0}

% type= article
\bibitem[{G.~M. {Green} {et~al.}(2019){Green}, {Schlafly}, {Zucker}, {Speagle}, \& {Finkbeiner}}]{Green2019}
{Green}, G.~M., {Schlafly}, E., {Zucker}, C., {Speagle}, J.~S., \& {Finkbeiner}, D. 2019, \bibinfo{title}{{A 3D Dust Map Based on Gaia, Pan-STARRS 1, and 2MASS},} \apj, 887, 93, \dodoi{10.3847/1538-4357/ab5362}

% type= article
\bibitem[{C.~R. {Harris} {et~al.}(2020){Harris}, {Millman}, {van der Walt}, {Gommers}, {Virtanen}, {Cournapeau}, {Wieser}, {Taylor}, {Berg}, {Smith}, {Kern}, {Picus}, {Hoyer}, {van Kerkwijk}, {Brett}, {Haldane}, {del R{\'\i}o}, {Wiebe}, {Peterson}, {G{\'e}rard-Marchant}, {Sheppard}, {Reddy}, {Weckesser}, {Abbasi}, {Gohlke}, \& {Oliphant}}]{Harris20}
{Harris}, C.~R., {Millman}, K.~J., {van der Walt}, S.~J., {et~al.} 2020, \bibinfo{title}{{Array programming with NumPy},} \nat, 585, 357, \dodoi{10.1038/s41586-020-2649-2}

% type= article
\bibitem[{T. {Hasegawa} {et~al.}(2004){Hasegawa}, {Malasan}, {Kawakita}, {Obayashi}, {Kurabayashi}, {Nakai}, {Hyakkai}, \& {Arimoto}}]{Hasegawa2004}
{Hasegawa}, T., {Malasan}, H.~L., {Kawakita}, H., {et~al.} 2004, \bibinfo{title}{{New Photometric Data of Old Open Clusters in the Anti-Galactic Center Region},} \pasj, 56, 295, \dodoi{10.1093/pasj/56.2.295}

% type= article
\bibitem[{A. {Hourihane} {et~al.}(2023){Hourihane}, {Fran{\c{c}}ois}, {Worley}, {Magrini}, {Gonneau}, {Casey}, {Gilmore}, {Randich}, {Sacco}, {Recio-Blanco}, {Korn}, {Allende Prieto}, {Smiljanic}, {Blomme}, {Bragaglia}, {Walton}, {Van Eck}, {Bensby}, {Lanzafame}, {Frasca}, {Franciosini}, {Damiani}, {Lind}, {Bergemann}, {Bonifacio}, {Hill}, {Lobel}, {Montes}, {Feuillet}, {Tautvai{\v{s}}ien{\.{e}}}, {Guiglion}, {Tabernero}, {Gonz{\'a}lez Hern{\'a}ndez}, {Gebran}, {Van der Swaelmen}, {Mikolaitis}, {Daflon}, {Merle}, {Morel}, {Lewis}, {Gonz{\'a}lez Solares}, {Murphy}, {Jeffries}, {Jackson}, {Feltzing}, {Prusti}, {Carraro}, {Biazzo}, {Prisinzano}, {Jofr{\'e}}, {Zaggia}, {Drazdauskas}, {Stonkut{\'e}}, {Marfil}, {Jim{\'e}nez-Esteban}, {Mahy}, {Guti{\'e}rrez Albarr{\'a}n}, {Berlanas}, {Santos}, {Morbidelli}, {Spina}, \& {Minkevi{\v{c}}i{\={u}}t{\.{e}}}}]{Hourihane2023}
{Hourihane}, A., {Fran{\c{c}}ois}, P., {Worley}, C.~C., {et~al.} 2023, \bibinfo{title}{{The Gaia-ESO Survey: Homogenisation of stellar parameters and elemental abundances},} \aap, 676, A129, \dodoi{10.1051/0004-6361/202345910}

% type= article
\bibitem[{E.~L. {Hunt} \& S. {Reffert}(2024){Hunt} \& {Reffert}}]{Hunt2024}
{Hunt}, E.~L., \& {Reffert}, S. 2024, \bibinfo{title}{{Improving the open cluster census. III. Using cluster masses, radii, and dynamics to create a cleaned open cluster catalogue},} \aap, 686, A42, \dodoi{10.1051/0004-6361/202348662}

% type= article
\bibitem[{J.~D. Hunter(2007)Hunter}]{Hunter07}
Hunter, J.~D. 2007, \bibinfo{title}{Matplotlib: A 2D Graphics Environment,} CSE, 9, 90

% type= article
\bibitem[{A. {Hypki} \& M. {Giersz}(2017){Hypki} \& {Giersz}}]{Hypki2017}
{Hypki}, A., \& {Giersz}, M. 2017, \bibinfo{title}{{mocca code for star cluster simulations - VI. Bimodal spatial distribution of blue stragglers},} \mnras, 471, 2537, \dodoi{10.1093/mnras/stx1718}

% type= article
\bibitem[{C. {Ibano{\v{g}}lu} {et~al.}(2006){Ibano{\v{g}}lu}, {Soydugan}, {Soydugan}, \& {Dervi{\textcommabelow s}o{\v{g}}lu}}]{Ibanoglu2006}
{Ibano{\v{g}}lu}, C., {Soydugan}, F., {Soydugan}, E., \& {Dervi{\textcommabelow s}o{\v{g}}lu}, A. 2006, \bibinfo{title}{{Angular momentum evolution of Algol binaries},} \mnras, 373, 435, \dodoi{10.1111/j.1365-2966.2006.11052.x}

% type= article
\bibitem[{A. {Irrgang} {et~al.}(2013){Irrgang}, {Wilcox}, {Tucker}, \& {Schiefelbein}}]{Irrgang2013}
{Irrgang}, A., {Wilcox}, B., {Tucker}, E., \& {Schiefelbein}, L. 2013, \bibinfo{title}{{Milky Way mass models for orbit calculations},} \aap, 549, A137, \dodoi{10.1051/0004-6361/201220540}

% type= article
\bibitem[{H.~R. {Jacobson} {et~al.}(2011){Jacobson}, {Friel}, \& {Pilachowski}}]{Jacobson2011}
{Jacobson}, H.~R., {Friel}, E.~D., \& {Pilachowski}, C.~A. 2011, \bibinfo{title}{{Abundances in the Old Open Cluster NGC~2141},} \aj, 141, 58, \dodoi{10.1088/0004-6256/141/2/58}

% type= article
\bibitem[{V.~V. {Jadhav} \& A. {Subramaniam}(2021){Jadhav} \& {Subramaniam}}]{Jadhav2021bss}
{Jadhav}, V.~V., \& {Subramaniam}, A. 2021, \bibinfo{title}{{Blue straggler stars in open clusters using Gaia: dependence on cluster parameters and possible formation pathways},} \mnras, 507, 1699, \dodoi{10.1093/mnras/stab2264}

% type= article
\bibitem[{Y.~C. {Joshi} {et~al.}(2024){Joshi}, {Deepak}, \& {Malhotra}}]{Joshi2024}
{Joshi}, Y.~C., {Deepak}, \& {Malhotra}, S. 2024, \bibinfo{title}{{A study on the metallicity gradients in the galactic disk using open clusters},} Frontiers in Astronomy and Space Sciences, 11, 1348321, \dodoi{10.3389/fspas.2024.1348321}

% type= article
\bibitem[{H. {Karag{\"o}z} {et~al.}(2025){Karag{\"o}z}, {Yontan}, {Bilir}, {Plevne}, {Ak}, {Ak}, {Canbay}, \& {Banks}}]{Karagoz25}
{Karag{\"o}z}, H., {Yontan}, T., {Bilir}, S., {et~al.} 2025, \bibinfo{title}{{A Multidata Approach to Open Clusters: Roslund 3 and Ruprecht 174 in CCD UBV and Gaia DR3 Context},} \aj, 170, 149, \dodoi{10.3847/1538-3881/adef16}

% type= article
\bibitem[{N.~V. {Kharchenko} {et~al.}(2013){Kharchenko}, {Piskunov}, {Schilbach}, {R{\"o}ser}, \& {Scholz}}]{Kharchenko2013}
{Kharchenko}, N.~V., {Piskunov}, A.~E., {Schilbach}, E., {R{\"o}ser}, S., \& {Scholz}, R.-D. 2013, \bibinfo{title}{{Global survey of star clusters in the Milky Way. II. The catalogue of basic parameters},} \aap, 558, A53, \dodoi{10.1051/0004-6361/201322302}

% type= article
\bibitem[{I. King(1962)King}]{King62}
King, I. 1962, \bibinfo{title}{The structure of star clusters. {I.} an empirical density law,} Astronomical Journal, 67, 471, \dodoi{10.1086/108756}

% type= article
\bibitem[{C. {Kobayashi} {et~al.}(2020){Kobayashi}, {Karakas}, \& {Lugaro}}]{Kobayashi2020}
{Kobayashi}, C., {Karakas}, A.~I., \& {Lugaro}, M. 2020, \bibinfo{title}{{The Origin of Elements from Carbon to Uranium},} \apj, 900, 179, \dodoi{10.3847/1538-4357/abae65}

% type= article
\bibitem[{C. {Kobayashi} {et~al.}(2006){Kobayashi}, {Umeda}, {Nomoto}, {Tominaga}, \& {Ohkubo}}]{Kobayashi2006}
{Kobayashi}, C., {Umeda}, H., {Nomoto}, K., {Tominaga}, N., \& {Ohkubo}, T. 2006, \bibinfo{title}{{Galactic Chemical Evolution: Carbon through Zinc},} \apj, 653, 1145, \dodoi{10.1086/508914}

% type= article
\bibitem[{S. {Ko{\c{c}}} {et~al.}(2022){Ko{\c{c}}}, {Yontan}, {Bilir}, {Canbay}, {Ak}, {Banks}, {Ak}, \& {Paunzen}}]{Koc2022}
{Ko{\c{c}}}, S., {Yontan}, T., {Bilir}, S., {et~al.} 2022, \bibinfo{title}{{A Photometric and Astrometric Study of the Open Clusters NGC 1664 and NGC 6939},} \aj, 163, 191, \dodoi{10.3847/1538-3881/ac58a0}

% type= article
\bibitem[{M. {Kounkel} {et~al.}(2020){Kounkel}, {Covey}, \& {Stassun}}]{Kounkel2020}
{Kounkel}, M., {Covey}, K., \& {Stassun}, K.~G. 2020, \bibinfo{title}{{Untangling the Galaxy. II. Structure within 3 kpc},} \aj, 160, 279, \dodoi{10.3847/1538-3881/abc0e6}

% type= article
\bibitem[{J.~D. Kruijssen(2014)Kruijssen}]{kruijssen2014}
Kruijssen, J.~D. 2014, \bibinfo{title}{Globular cluster formation in the context of galaxy formation and evolution,} Class. Quantum Grav., 31, 244006

% type= article
\bibitem[{C.~J. Lada \& E.~A. Lada(2003)Lada \& Lada}]{lada2003}
Lada, C.~J., \& Lada, E.~A. 2003, \bibinfo{title}{Embedded clusters in molecular clouds,} Annu. Rev. Astron. Astrophys., 41, 57

% type= article
\bibitem[{S.~K. {Leggett}(1992){Leggett}}]{Leggett1992}
{Leggett}, S.~K. 1992, \bibinfo{title}{{Infrared Colors of Low-Mass Stars},} \apjs, 82, 351, \dodoi{10.1086/191720}

% type= article
\bibitem[{P.~J.~T. {Leonard}(1989){Leonard}}]{Leonard1989}
{Leonard}, P. J.~T. 1989, \bibinfo{title}{{Stellar Collisions in Globular Clusters and the Blue Straggler Problem},} \aj, 98, 217, \dodoi{10.1086/115138}

% type= article
\bibitem[{E. {Linck} \& R.~D. {Mathieu}(2026){Linck} \& {Mathieu}}]{Linck2026}
{Linck}, E., \& {Mathieu}, R.~D. 2026, \bibinfo{title}{{The Distribution of Blue Straggler Stars in the Color-Magnitude Diagrams of Old Open Clusters},} arXiv e-prints, arXiv:2605.14187.
\newblock \doarXiv{2605.14187}

% type= article
\bibitem[{L. {Lindegren} {et~al.}(2021){Lindegren}, {Klioner}, {Hern{\'a}ndez}, {Bombrun}, {Ramos-Lerate}, {Steidelm{\"u}ller}, {Bastian}, {Biermann}, {de Torres}, {Gerlach}, {Geyer}, {Hilger}, {Hobbs}, {Lammers}, {McMillan}, {Stephenson}, {Casta{\~n}eda}, {Davidson}, {Fabricius}, {Gracia-Abril}, {Portell}, {Abbas}, {Altmann}, {Anglada Varela}, {Balaguer-N{\'u}{\~n}ez}, {Balog}, {Barache}, {Becciani}, {Bertone}, {Bianchi}, {Bouquillon}, {Brown}, {Bucciarelli}, {Busonero}, {Butkevich}, {Buzzi}, {Cancelliere}, {Carlucci}, {Charlot}, {Cheek}, {Clementini}, {Contursi}, {Cooper}, {Crifo}, {Crosta}, {Dafonte}, {Damerdji}, {David}, {de Laverny}, {Delchambre}, {Dell'Oro}, {D{\'{\i}}az-Garc{\'{\i}}a}, {Distefano}, {Dolding}, {Edvardsson}, {Enke}, {Esquej}, {Fabre}, {Fabrizio}, {Faigler}, {Fedorets}, {Fernique}, {Fienga}, {Figueras}, {Fouron}, {Fouesneau}, {Fr{\'e}mat}, {Galluccio}, {Garc{\'\i}a-Torres}, {Gavras}, {Gomel}, {Gonz{\'a}lez-N{\'u}{\~n}ez}, {Gonz{\'a}lez-Vidal}, {Granvik}, {Guerrier}, {Guillout}, {Guiraud},
  {Guti{\'e}rrez-S{\'a}nchez}, {Guy}, {Hauser}, {Haywood}, {Heiter}, {Hestroffer}, {Holl}, {Huckle}, {Jardine}, {Jasniewicz}, {Jean-Antoine Piccolo}, {Jorissen}, {Juaristi Campillo}, {Julbe}, {Karampelas}, {Kervella}, {Khanna}, {Korn}, {Kostrzewa-Rutkowska}, {Kruszy{\'n}ska}, {Lambert}, {Lanza}, {Lasne}, {Le Campion}, {Le Fustec}, {Lebreton}, {Lebzelter}, {Leccia}, {Leclerc}, {Lecoeur-Taibi}, {Liao}, {Licata}, {Lister}, {Livanou}, {Lobel}, {L{\"o}ffler}, {Managau}, {Mann}, {Marchal}, {Marconi}, {Marinoni}, {Marocco}, {Marshall}, {Mart{\'\i}n-Fleitas}, {Masana}, {Mastroserio}, {Mazure}, {Messineo}, {Michalik}, {Millar}, {Mints}, {Molina}, {Molinaro}, {Moln{\'a}r}, {Mora}, {Morbidelli}, {Morel}, {Morris}, {Muinonen}, {Osinde}, {Pagani}, {Pagano}, {Palaversa}, {Panahi}, {Panuzzo}, {Pauwels}, {Petit}, {Pichon}, {Piersimoni}, {Pineau}, {Plum}, {Poggio}, {Poretti}, {Pr{\v{s}}a}, {Pulone}, {Racero}, {Ragaini}, {Rainer}, {Raiteri}, {Rambaux}, {Ramos}, {Re Fiorentin}, {Regibo}, {Reyl{\'e}}, {Riclet}, {Riello},
  {Rimoldini}, {Robin}, {Roegiers}, {Rohrbasser}, {Rowell}, {Royer}, {Rybizki}, {Sadowski}, {Sagrist{\`a} Sell{\'e}s}, {Sahlmann}, {Salguero}, {Samaras}, {Sanchez Gimenez}, {Sanna}, {Sartoretti}, {Sciacca}, {Segol}, {Segovia}, {S{\'e}gransan}, {Semeux}, {Shahaf}, {Siddiqui}, {Siebert}, {Siltala}, {Slezak}, {Smart}, {Solano}, {Solitro}, {Souami}, {Souchay}, {Spagna}, {Spoto}, {Steele}, {Tariso}, {Taris}, {Teixeira}, {Thiel}, {Thuillot}, {Tonello}, {Torra}, {Torra}, {Turon}, {Unger}, {Vaillant}, {van Dillen}, {van Reeven}, {Vasiliev}, {Vera-Ciro}, {Viala}, {Vicente}, {Voutsinas}, {Weiler}, {Wyrzykowski}, {Yoldas}, {Yvard}, {Zhao}, {Zorec}, {Zucker}, {Zurbach}, \& {Zwitter}}]{Lindegren2021}
{Lindegren}, L., {Klioner}, S.~A., {Hern{\'a}ndez}, J., {et~al.} 2021, \bibinfo{title}{{Gaia Early Data Release 3. The astrometric solution},} \aap, 649, A2, \dodoi{10.1051/0004-6361/202039709}

% type= article
\bibitem[{L. Liu \& X. Pang(2019)Liu \& Pang}]{Liu2019}
Liu, L., \& Pang, X. 2019, \bibinfo{title}{{A Catalog of Newly Identified Star Clusters in {G}aia {DR}2},} \apjs, 245, 32, \dodoi{10.3847/1538-4365/ab530a}

% type= article
\bibitem[{M. {L{\'o}pez-Corredoira} {et~al.}(2002){L{\'o}pez-Corredoira}, {Cabrera-Lavers}, {Garz{\'o}n}, \& {Hammersley}}]{Lopez2002}
{L{\'o}pez-Corredoira}, M., {Cabrera-Lavers}, A., {Garz{\'o}n}, F., \& {Hammersley}, P.~L. 2002, \bibinfo{title}{{Old stellar Galactic disc in near-plane regions according to 2MASS: Scales, cut-off, flare and warp},} \aap, 394, 883, \dodoi{10.1051/0004-6361:20021175}

% type= article
\bibitem[{M. {L{\'o}pez-Corredoira} \& J. {Molg{\'o}}(2014){L{\'o}pez-Corredoira} \& {Molg{\'o}}}]{Lopez2014}
{L{\'o}pez-Corredoira}, M., \& {Molg{\'o}}, J. 2014, \bibinfo{title}{{Flare in the Galactic stellar outer disc detected in SDSS-SEGUE data},} \aap, 567, A106, \dodoi{10.1051/0004-6361/201423706}

% type= article
\bibitem[{Y.~L. {Lu} {et~al.}(2024){Lu}, {Minchev}, {Buck}, {Khoperskov}, {Steinmetz}, {Libeskind}, {Cescutti}, {Freeman}, \& {Ratcliffe}}]{Lu2024}
{Lu}, Y.~L., {Minchev}, I., {Buck}, T., {et~al.} 2024, \bibinfo{title}{{There is no place like home - finding birth radii of stars in the Milky Way},} \mnras, 535, 392, \dodoi{10.1093/mnras/stae2364}

% type= article
\bibitem[{L. {Magrini} {et~al.}(2009){Magrini}, {Sestito}, {Randich}, \& {Galli}}]{Magrini2009}
{Magrini}, L., {Sestito}, P., {Randich}, S., \& {Galli}, D. 2009, \bibinfo{title}{{The evolution of the Galactic metallicity gradient from high-resolution spectroscopy of open clusters},} \aap, 494, 95, \dodoi{10.1051/0004-6361:200810634}

% type= article
\bibitem[{L. {Magrini} {et~al.}(2017){Magrini}, {Sestito}, {Randich}, {Galli}, \& {others}}]{Magrini2017}
{Magrini}, L., {Sestito}, P., {Randich}, S., {Galli}, D., \& {others}. 2017, \bibinfo{title}{{The Gaia-ESO Survey: radial metallicity gradients and age-metallicity relation of stars in the Milky Way disk},} \aap, 603, A2, \dodoi{10.1051/0004-6361/201630099}

% type= article
\bibitem[{M. {Mapelli} {et~al.}(2004){Mapelli}, {Sigurdsson}, {Colpi}, {Ferraro}, {Possenti}, {Rood}, {Sills}, \& {Beccari}}]{Mapelli2004}
{Mapelli}, M., {Sigurdsson}, S., {Colpi}, M., {et~al.} 2004, \bibinfo{title}{{The Contribution of Primordial Binaries to the Blue Straggler Population in 47 Tucanae},} \apjl, 605, L29, \dodoi{10.1086/386370}

% type= article
\bibitem[{M. {Mapelli} {et~al.}(2006){Mapelli}, {Sigurdsson}, {Ferraro}, {Colpi}, {Possenti}, \& {Lanzoni}}]{Mapelli2006}
{Mapelli}, M., {Sigurdsson}, S., {Ferraro}, F.~R., {et~al.} 2006, \bibinfo{title}{{The radial distribution of blue straggler stars and the nature of their progenitors},} \mnras, 373, 361, \dodoi{10.1111/j.1365-2966.2006.11038.x}

% type= article
\bibitem[{R.~D. {Mathieu} \& A.~M. {Geller}(2009){Mathieu} \& {Geller}}]{Mathieu2009}
{Mathieu}, R.~D., \& {Geller}, A.~M. 2009, \bibinfo{title}{{A binary star fraction of 76 per cent and unusual orbit parameters for the blue stragglers of NGC 188},} \nat, 462, 1032, \dodoi{10.1038/nature08568}

% type= article
\bibitem[{W.~H. {McCrea}(1964){McCrea}}]{McCrea1964}
{McCrea}, W.~H. 1964, \bibinfo{title}{{Extended main-sequence of some stellar clusters},} \mnras, 128, 147, \dodoi{10.1093/mnras/128.2.147}

% type= article
\bibitem[{G. McLachlan(2000)McLachlan}]{mclachlan2000finite}
McLachlan, G. 2000, \bibinfo{title}{Finite mixture models,} A wiley-interscience publication

% type= article
\bibitem[{P.~J. {McMillan}(2017){McMillan}}]{McMillan2017}
{McMillan}, P.~J. 2017, \bibinfo{title}{{The mass distribution and gravitational potential of the Milky Way},} \mnras, 465, 76, \dodoi{10.1093/mnras/stw2759}

% type= article
\bibitem[{A. {McWilliam}(1997){McWilliam}}]{McWilliam1997}
{McWilliam}, A. 1997, \bibinfo{title}{{Abundance Ratios and Galactic Chemical Evolution},} \araa, 35, 503, \dodoi{10.1146/annurev.astro.35.1.503}

% type= article
\bibitem[{A.~P. {Milone} {et~al.}(2012){Milone}, {Piotto}, {Bedin}, {Aparicio}, {Anderson}, {Sarajedini}, {Marino}, {Moretti}, {Davies}, {Chaboyer}, {Dotter}, {Hempel}, {Mar{\'\i}n-Franch}, {Majewski}, {Paust}, {Reid}, {Rosenberg}, \& {Siegel}}]{Milone2012}
{Milone}, A.~P., {Piotto}, G., {Bedin}, L.~R., {et~al.} 2012, \bibinfo{title}{{The ACS survey of Galactic globular clusters. XII. Photometric binaries along the main sequence},} \aap, 540, A16, \dodoi{10.1051/0004-6361/201016384}

% type= article
\bibitem[{I. {Minchev} {et~al.}(2018){Minchev}, {Anders}, {Recio-Blanco}, \& {others}}]{Minchev2018}
{Minchev}, I., {Anders}, F., {Recio-Blanco}, A., \& {others}. 2018, \bibinfo{title}{{New constraints on radial migration in the Milky Way disk: multi-element abundance patterns},} \mnras, 481, 1645, \dodoi{10.1093/mnras/sty2033}

% type= article
\bibitem[{I. {Minchev} {et~al.}(2014){Minchev}, {Chiappini}, \& {Martig}}]{Minchev2014}
{Minchev}, I., {Chiappini}, C., \& {Martig}, M. 2014, \bibinfo{title}{{Chemodynamical Evolution of the Milky Way Disk. II. Variations with Galactic Radius and Height above the Midplane},} \apj, 804, L9, \dodoi{10.1088/2041-8205/804/1/L9}

% type= article
\bibitem[{I. {Minchev} {et~al.}(2011){Minchev}, {Famaey}, {Combes}, {Di Matteo}, {Mouhcine}, \& {Wozniak}}]{Minchev2011}
{Minchev}, I., {Famaey}, B., {Combes}, F., {et~al.} 2011, \bibinfo{title}{{Radial migration in galactic disks due to bars and transient spiral structures},} \aap, 527, A147, \dodoi{10.1051/0004-6361/201015139}

% type= article
\bibitem[{M. Miyamoto \& R. Nagai(1975)Miyamoto \& Nagai}]{Miyamoto1975}
Miyamoto, M., \& Nagai, R. 1975, \bibinfo{title}{Three-dimensional models for the distribution of mass in galaxies,} Publications of the Astronomical Society of Japan, 27, 533

% type= article
\bibitem[{Y. {Momany} {et~al.}(2006){Momany}, {Zaggia}, {Gilmore}, {Piotto}, {Carraro}, {Bedin}, \& {de Angeli}}]{Momany2006}
{Momany}, Y., {Zaggia}, S., {Gilmore}, G., {et~al.} 2006, \bibinfo{title}{{Outer structure of the Galactic warp and flare: explaining the Canis Major over-density},} \aap, 451, 515, \dodoi{10.1051/0004-6361:20054081}

% type= article
\bibitem[{J.~F. {Navarro} {et~al.}(1996){Navarro}, {Frenk}, \& {White}}]{Navarro1996}
{Navarro}, J.~F., {Frenk}, C.~S., \& {White}, S. D.~M. 1996, \bibinfo{title}{{The Structure of Cold Dark Matter Halos},} Astrophysical Journal, 462, 563, \dodoi{10.1086/177173}

% type= article
\bibitem[{M. {Netopil} {et~al.}(2022){Netopil}, {Oralhan}, {{\c{C}}akmak}, {Michel}, \& {Karata{\textcommabelow s}}}]{Netopil2022}
{Netopil}, M., {Oralhan}, {\.I}.~A., {{\c{C}}akmak}, H., {Michel}, R., \& {Karata{\textcommabelow s}}, Y. 2022, \bibinfo{title}{{The Galactic metallicity gradient shown by open clusters in the light of radial migration},} \mnras, 509, 421, \dodoi{10.1093/mnras/stab2961}

% type= article
\bibitem[{M. {Netopil} {et~al.}(2016){Netopil}, {Paunzen}, {Heiter}, \& {Soubiran}}]{Netopil2016}
{Netopil}, M., {Paunzen}, E., {Heiter}, U., \& {Soubiran}, C. 2016, \bibinfo{title}{{On the metallicity of open clusters. III. Homogenised sample},} \aap, 585, A150, \dodoi{10.1051/0004-6361/201526370}

% type= inproceedings
\bibitem[{P.~E. {Nissen}(2004){Nissen}}]{Nissen2004}
{Nissen}, P.~E. 2004, \bibinfo{title}{{Thin and Thick Galactic Disks},} in Origin and Evolution of the Elements, ed. A.~{McWilliam} \& M.~{Rauch}, 154, \dodoi{10.48550/arXiv.astro-ph/0310326}

% type= article
\bibitem[{K. {Nomoto} {et~al.}(2013){Nomoto}, {Kobayashi}, \& {Tominaga}}]{Nomoto2013}
{Nomoto}, K., {Kobayashi}, C., \& {Tominaga}, N. 2013, \bibinfo{title}{{Nucleosynthesis in Stars and the Chemical Enrichment of Galaxies},} \araa, 51, 457, \dodoi{10.1146/annurev-astro-082812-140956}

% type= article
\bibitem[{J.~M. {Otto} {et~al.}(2026){Otto}, {Frinchaboy}, {Myers}, {Johnson}, {Donor}, {Hossain}, {M{\'e}sz{\'a}ros}, {Wallace}, {Cunha}, {Bhattarai}, {Sinha}, {Zasowski}, {Loebman}, {Wiggins}, {Price-Whelan}, {Spoo}, {Souto}, {Bizyaev}, {Pan}, {Saydjari}, \& {Song}}]{Otto2026}
{Otto}, J.~M., {Frinchaboy}, P.~M., {Myers}, N.~R., {et~al.} 2026, \bibinfo{title}{{The Open Cluster Chemical Abundances and Mapping Survey. VIII. Galactic Chemical Gradient and Azimuthal Analysis from SDSS/MWM DR19},} \aj, 171, 91, \dodoi{10.3847/1538-3881/ae28d8}

% type= article
\bibitem[{L. {Pasquini} {et~al.}(2002){Pasquini}, {Avila}, {Blecha}, {Cacciari}, {Cayatte}, {Colless}, {Damiani}, {de Propris}, {Dekker}, {di Marcantonio}, {Farrell}, {Gillingham}, {Guinouard}, {Hammer}, {Kaufer}, {Hill}, {Marteaud}, {Modigliani}, {Mulas}, {North}, {Popovic}, {Rossetti}, {Royer}, {Santin}, {Schmutzer}, {Simond}, {Vola}, {Waller}, \& {Zoccali}}]{Pasquini2002}
{Pasquini}, L., {Avila}, G., {Blecha}, A., {et~al.} 2002, \bibinfo{title}{{Installation and commissioning of FLAMES, the VLT Multifibre Facility},} The Messenger, 110, 1

% type= article
\bibitem[{M.~J. {Pecaut} \& E.~E. {Mamajek}(2013){Pecaut} \& {Mamajek}}]{Pecaut2013}
{Pecaut}, M.~J., \& {Mamajek}, E.~E. 2013, \bibinfo{title}{{Intrinsic Colors, Temperatures, and Bolometric Corrections of Pre-main-sequence Stars},} \apjs, 208, 9, \dodoi{10.1088/0067-0049/208/1/9}

% type= article
\bibitem[{F. Pedregosa {et~al.}(2011)Pedregosa, Varoquaux, Gramfort, Michel, Thirion, Grisel, Blondel, Prettenhofer, Weiss, Dubourg, {et~al.}}]{scikit-learn}
Pedregosa, F., Varoquaux, G., Gramfort, A., {et~al.} 2011, \bibinfo{title}{Scikit-learn: Machine learning in Python,} the Journal of machine Learning research, 12, 2825

% type= article
\bibitem[{O. {Plevne} \& F. {Akbaba}(2026){Plevne} \& {Akbaba}}]{Plevne2026}
{Plevne}, O., \& {Akbaba}, F. 2026, \bibinfo{title}{{Astrophysical Parameters of 5056 Open Star Clusters from Bayesian Nested Sampling with PARSEC Isochrones},} arXiv e-prints, arXiv:2605.23802, \dodoi{10.48550/arXiv.2605.23802}

% type= article
\bibitem[{O. {Plevne} {et~al.}(2020){Plevne}, {{\"O}nal Ta{\textcommabelow s}}, {Bilir}, \& {Seabroke}}]{Plevne2020}
{Plevne}, O., {{\"O}nal Ta{\textcommabelow s}}, {\"O}., {Bilir}, S., \& {Seabroke}, G.~M. 2020, \bibinfo{title}{{Multiwavelength Absolute Magnitudes and Colors of Red Clump Stars in the Gaia Era},} \apj, 893, 108, \dodoi{10.3847/1538-4357/ab80bb}

% type= article
\bibitem[{S.~F. {Portegies Zwart} {et~al.}(2010){Portegies Zwart}, {McMillan}, \& {Gieles}}]{PortegiesZwart2010}
{Portegies Zwart}, S.~F., {McMillan}, S. L.~W., \& {Gieles}, M. 2010, \bibinfo{title}{{Young Massive Star Clusters},} \araa, 48, 431, \dodoi{10.1146/annurev-astro-081309-130834}

% type= article
\bibitem[{S. {Randich} {et~al.}(2022){Randich}, {Gilmore}, {Magrini}, \& {others}}]{Randich2022}
{Randich}, S., {Gilmore}, G., {Magrini}, L., \& {others}. 2022, \bibinfo{title}{{The Gaia-ESO Public Spectroscopic Survey: Motivation, implementation, GIRAFFE data processing, analysis, and final data products},} \aap, 666, A121, \dodoi{10.1051/0004-6361/202243141}

% type= article
\bibitem[{B. {Ratcliffe} {et~al.}(2025){Ratcliffe}, {Khoperskov}, {Minchev}, {Lee}, {Buck}, {Marques}, {Bernaldez}, {Lu}, \& {Steinmetz}}]{Ratcliffe2025}
{Ratcliffe}, B., {Khoperskov}, S., {Minchev}, I., {et~al.} 2025, \bibinfo{title}{{Evolution of the radial interstellar medium metallicity gradient in the Milky Way disk since redshift {\ensuremath{\approx}}3},} \aap, 698, A267, \dodoi{10.1051/0004-6361/202452658}

% type= article
\bibitem[{B. {Ratcliffe} {et~al.}(2026){Ratcliffe}, {Khoperskov}, {Lee}, {Minchev}, {Di Matteo}, {van de Ven}, {Haywood}, {Marques}, {Bernaldez}, {Krajnovi{\'c}}, \& {Steinmetz}}]{Ratcliffe2026}
{Ratcliffe}, B., {Khoperskov}, S., {Lee}, N., {et~al.} 2026, \bibinfo{title}{{Rediscovering the Milky Way with an orbit superposition approach and APOGEE data: IV. The disc growth and history of star formation},} \aap, 706, A103, \dodoi{10.1051/0004-6361/202557057}

% type= article
\bibitem[{C. {Reyl{\'e}} {et~al.}(2009){Reyl{\'e}}, {Marshall}, {Robin}, \& {Schultheis}}]{Reyle2009}
{Reyl{\'e}}, C., {Marshall}, D.~J., {Robin}, A.~C., \& {Schultheis}, M. 2009, \bibinfo{title}{{The Milky Way's external disc constrained by 2MASS star counts},} \aap, 495, 819, \dodoi{10.1051/0004-6361/200811341}

% type= article
\bibitem[{G.~R. Ricker {et~al.}(2015)Ricker, Winn, Vanderspek, Latham, Bakos, Bean, Berta-Thompson, Brown, Buchhave, Butler, {et~al.}}]{ricker2015transiting}
Ricker, G.~R., Winn, J.~N., Vanderspek, R., {et~al.} 2015, \bibinfo{title}{Transiting Exoplanet Survey Satellite (TESS),} Journal of Astronomical Telescopes, Instruments, and Systems, 1, 014003, \dodoi{10.1117/1.JATIS.1.1.014003}

% type= article
\bibitem[{G.~R. {Ricker} {et~al.}(2015){Ricker}, {Winn}, {Vanderspek}, {Latham}, {Bakos}, {Bean}, {Berta-Thompson}, {Brown}, {Buchhave}, {Butler}, {Butler}, {Chaplin}, {Charbonneau}, {Christensen-Dalsgaard}, {Clampin}, {Deming}, {Doty}, {De Lee}, {Dressing}, {Dunham}, {Endl}, {Fressin}, {Ge}, {Henning}, {Holman}, {Howard}, {Ida}, {Jenkins}, {Jernigan}, {Johnson}, {Kaltenegger}, {Kawai}, {Kjeldsen}, {Laughlin}, {Levine}, {Lin}, {Lissauer}, {MacQueen}, {Marcy}, {McCullough}, {Morton}, {Narita}, {Paegert}, {Palle}, {Pepe}, {Pepper}, {Quirrenbach}, {Rinehart}, {Sasselov}, {Sato}, {Seager}, {Sozzetti}, {Stassun}, {Sullivan}, {Szentgyorgyi}, {Torres}, {Udry}, \& {Villasenor}}]{Ricker2015}
{Ricker}, G.~R., {Winn}, J.~N., {Vanderspek}, R., {et~al.} 2015, \bibinfo{title}{{Transiting Exoplanet Survey Satellite (TESS)},} Journal of Astronomical Telescopes, Instruments, and Systems, 1, 014003, \dodoi{10.1117/1.JATIS.1.1.014003}

% type= article
\bibitem[{M. {Riello} {et~al.}(2021){Riello}, {De Angeli}, {Evans}, {Montegriffo}, {Carrasco}, {Busso}, {Palaversa}, {Burgess}, {Diener}, {Davidson}, {Rowell}, {Fabricius}, {Jordi}, {Bellazzini}, {Pancino}, {Harrison}, {Cacciari}, {van Leeuwen}, {Hambly}, {Hodgkin}, {Osborne}, {Altavilla}, {Barstow}, {Brown}, {Castellani}, {Cowell}, {De Luise}, {Gilmore}, {Giuffrida}, {Hidalgo}, {Holland}, {Marinoni}, {Pagani}, {Piersimoni}, {Pulone}, {Ragaini}, {Rainer}, {Richards}, {Sanna}, {Walton}, {Weiler}, \& {Yoldas}}]{Riello2021}
{Riello}, M., {De Angeli}, F., {Evans}, D.~W., {et~al.} 2021, \bibinfo{title}{{Gaia Early Data Release 3. Photometric content and validation},} \aap, 649, A3, \dodoi{10.1051/0004-6361/202039587}

% type= article
\bibitem[{A.~C. {Robin} {et~al.}(2012){Robin}, {Luri}, {Reyl{\'e}}, {Isasi}, {Grux}, {Blanco-Cuaresma}, {Arenou}, {Babusiaux}, {Belcheva}, {Drimmel}, {Jordi}, {Krone-Martins}, {Masana}, {Mauduit}, {Mignard}, {Mowlavi}, {Rocca-Volmerange}, {Sartoretti}, {Slezak}, \& {Sozzetti}}]{Robin12}
{Robin}, A.~C., {Luri}, X., {Reyl{\'e}}, C., {et~al.} 2012, \bibinfo{title}{{Gaia Universe model snapshot. A statistical analysis of the expected contents of the Gaia catalogue},} \aap, 543, A100, \dodoi{10.1051/0004-6361/201118646}

% type= article
\bibitem[{R. {Ro{\v s}kar} {et~al.}(2008){Ro{\v s}kar}, {Debattista}, {Quinn}, {Stinson}, \& {Wadsley}}]{Roskar2008}
{Ro{\v s}kar}, R., {Debattista}, V.~P., {Quinn}, T.~R., {Stinson}, G.~S., \& {Wadsley}, J. 2008, \bibinfo{title}{{Riding the Spiral Waves: Implications of Stellar Migration for the Properties of Galactic Disks},} \apj, 684, L79, \dodoi{10.1086/592231}

% type= article
\bibitem[{E.~F. {Schlafly} \& D.~P. {Finkbeiner}(2011){Schlafly} \& {Finkbeiner}}]{Schlafly2011}
{Schlafly}, E.~F., \& {Finkbeiner}, D.~P. 2011, \bibinfo{title}{{Measuring Reddening with Sloan Digital Sky Survey Stellar Spectra and Recalibrating SFD},} \apj, 737, 103, \dodoi{10.1088/0004-637X/737/2/103}

% type= article
\bibitem[{R. {Sch{\"o}nrich} \& J. {Binney}(2009){Sch{\"o}nrich} \& {Binney}}]{Schoenrich2009}
{Sch{\"o}nrich}, R., \& {Binney}, J. 2009, \bibinfo{title}{{Chemical evolution with radial mixing},} \mnras, 396, 203, \dodoi{10.1111/j.1365-2966.2009.14750.x}

% type= article
\bibitem[{J.~A. {Sellwood} \& J.~J. {Binney}(2002){Sellwood} \& {Binney}}]{Sellwood2002}
{Sellwood}, J.~A., \& {Binney}, J.~J. 2002, \bibinfo{title}{{Radial mixing in galactic discs},} \mnras, 336, 785, \dodoi{10.1046/j.1365-8711.2002.05806.x}

% type= article
\bibitem[{A. {Sinha} {et~al.}(2024){Sinha}, {Zasowski}, {Frinchaboy}, {Cunha}, {Souto}, {Tayar}, \& {Stassun}}]{Sinha2024}
{Sinha}, A., {Zasowski}, G., {Frinchaboy}, P., {et~al.} 2024, \bibinfo{title}{{A Comprehensive Study of Open Cluster Chemical Homogeneity Using APOGEE and Milky Way Mapper Abundances},} \apj, 975, 89, \dodoi{10.3847/1538-4357/ad78e1}

% type= article
\bibitem[{M.~F. Skrutskie {et~al.}(2006)Skrutskie, Cutri, Stiening, Weinberg, Schneider, Carpenter, Beichman, Capps, Chester, Elias, {et~al.}}]{skrutskie2006two}
Skrutskie, M.~F., Cutri, R., Stiening, R., {et~al.} 2006, \bibinfo{title}{The two micron all sky survey (2MASS),} The Astronomical Journal, 131, 1163

% type= article
\bibitem[{A. {Sollima} {et~al.}(2007){Sollima}, {Beccari}, {Ferraro}, {Fusi Pecci}, \& {Sarajedini}}]{Sollima2007}
{Sollima}, A., {Beccari}, G., {Ferraro}, F.~R., {Fusi Pecci}, F., \& {Sarajedini}, A. 2007, \bibinfo{title}{{The fraction of binary systems in the core of 13 low-density Galactic globular clusters},} \mnras, 380, 781, \dodoi{10.1111/j.1365-2966.2007.12116.x}

% type= article
\bibitem[{C. {Soubiran} {et~al.}(2018){Soubiran}, {Cantat-Gaudin}, {Romero-G{\'o}mez}, {Casamiquela}, {Jordi}, {Vallenari}, {Antoja}, {Balaguer-N{\'u}{\~n}ez}, {Bossini}, {Bragaglia}, {Carrera}, {Castro-Ginard}, {Figueras}, {Heiter}, {Katz}, {Krone-Martins}, {Le Campion}, {Moitinho}, \& {Sordo}}]{Soubiran2018}
{Soubiran}, C., {Cantat-Gaudin}, T., {Romero-G{\'o}mez}, M., {et~al.} 2018, \bibinfo{title}{{Open cluster kinematics with Gaia DR2},} \aap, 619, A155, \dodoi{10.1051/0004-6361/201834020}

% type= article
\bibitem[{L. {Spina} {et~al.}(2022){Spina}, {Ness}, {Nordlander}, \& {others}}]{Spina2022}
{Spina}, L., {Ness}, M., {Nordlander}, T., \& {others}. 2022, \bibinfo{title}{{The GALAH Survey: chemical tagging and tracing the birth of open clusters},} \mnras, 511, 1217, \dodoi{10.1093/mnras/stab3671}

% type= article
\bibitem[{L. {Spina} {et~al.}(2021){Spina}, {Ting}, {De Silva}, {Frankel}, {Sharma}, {Cantat-Gaudin}, {Joyce}, {Stello}, {Karakas}, {Asplund}, {Nordlander}, {Casagrande}, {D'Orazi}, {Casey}, {Cottrell}, {Tepper-Garc{\'\i}a}, {Baratella}, {Kos}, {{\v{C}}otar}, {Bland-Hawthorn}, {Buder}, {Freeman}, {Hayden}, {Lewis}, {Lin}, {Lind}, {Martell}, {Schlesinger}, {Simpson}, {Zucker}, \& {Zwitter}}]{Spina2021}
{Spina}, L., {Ting}, Y.-S., {De Silva}, G.~M., {et~al.} 2021, \bibinfo{title}{{The GALAH survey: tracing the Galactic disc with open clusters},} \mnras, 503, 3279, \dodoi{10.1093/mnras/stab471}

% type= article
\bibitem[{S. {Ta{\c{s}}demir} {et~al.}(2025){Ta{\c{s}}demir}, {{\c{C}}{\i}nar}, {Canbay}, {Ta{\c{s}}tan}, {Elsanhoury}, \& {Haroon}}]{Tasdemir2025}
{Ta{\c{s}}demir}, S., {{\c{C}}{\i}nar}, D.~C., {Canbay}, R., {et~al.} 2025, \bibinfo{title}{{Comprehensive Analysis of Middle-Aged Open Cluster NGC 6793 in Vulpecula via Gaia DR3 Data},} Physics and Astronomy Reports, 3, 1, \dodoi{10.26650/PAR.2025.00003}

% type= article
\bibitem[{S. {Ta{\c{s}}demir} \& T. {Yontan}(2023){Ta{\c{s}}demir} \& {Yontan}}]{Tasdemir23}
{Ta{\c{s}}demir}, S., \& {Yontan}, T. 2023, \bibinfo{title}{{Analysis of the Young Open Cluster Trumpler 2 Using Gaia DR3 Data},} Physics and Astronomy Reports, 1, 1, \dodoi{10.26650/PAR.2023.00001}

% type= article
\bibitem[{A.~L. {Tadross}(2011){Tadross}}]{Tadross2011}
{Tadross}, A.~L. 2011, \bibinfo{title}{{A Catalog of 120 NGC Open Star Clusters},} Journal of Korean Astronomical Society, 44, 1, \dodoi{10.5303/JKAS.2011.44.1.001}

% type= article
\bibitem[{B. {Tan{\i}k {\"O}zt{\"u}rk} {et~al.}(2025){Tan{\i}k {\"O}zt{\"u}rk}, {Bilir}, {Yontan}, {Plevne}, {Ak}, {Ak}, {Canbay}, \& {Banks}}]{Tanik2025}
{Tan{\i}k {\"O}zt{\"u}rk}, B., {Bilir}, S., {Yontan}, T., {et~al.} 2025, \bibinfo{title}{{A Comprehensive Study of Czernik 41 and NGC 1342 Using CCD UBV and Gaia DR3 Data},} \aj, 170, 164, \dodoi{10.3847/1538-3881/adefe0}

% type= article
\bibitem[{Y. {Tarricq} {et~al.}(2021){Tarricq}, {Soubiran}, {Casamiquela}, {Cantat-Gaudin}, {Chemin}, {Anders}, {Antoja}, {Romero-G{\'o}mez}, {Figueras}, {Jordi}, {Bragaglia}, {Balaguer-N{\'u}{\~n}ez}, {Carrera}, {Castro-Ginard}, {Moitinho}, {Ramos}, \& {Bossini}}]{Tarricq2021}
{Tarricq}, Y., {Soubiran}, C., {Casamiquela}, L., {et~al.} 2021, \bibinfo{title}{{3D kinematics and age distribution of the open cluster population},} \aap, 647, A19, \dodoi{10.1051/0004-6361/202039388}

% type= article
\bibitem[{S. {Ta{\textcommabelow s}demir} {et~al.}(2026){Ta{\textcommabelow s}demir}, {Elsanhoury}, {{\c{C}}{\i}nar}, {Haroon}, \& {Bilir}}]{Tasedemir2026}
{Ta{\textcommabelow s}demir}, S., {Elsanhoury}, W.~H., {{\c{C}}{\i}nar}, D.~C., {Haroon}, A., \& {Bilir}, S. 2026, \bibinfo{title}{{Comprehensive Gaia DR3-based Astrometric, Photometric, and Kinematic Studies of the Binary Open Cluster h and {\ensuremath{\chi}} Persei},} Research in Astronomy and Astrophysics, 26, 045016, \dodoi{10.1088/1674-4527/ae3d11}

% type= article
\bibitem[{S. {Tun{\c{c}}el G{\"u}{\c{c}}tekin} {et~al.}(2019){Tun{\c{c}}el G{\"u}{\c{c}}tekin}, {Bilir}, {Karaali}, {Plevne}, \& {Ak}}]{Tuncel2019}
{Tun{\c{c}}el G{\"u}{\c{c}}tekin}, S., {Bilir}, S., {Karaali}, S., {Plevne}, O., \& {Ak}, S. 2019, \bibinfo{title}{{Vertical and radial metallicity gradients in high latitude galactic fields with SDSS},} Advances in Space Research, 63, 1360, \dodoi{10.1016/j.asr.2018.10.041}

% type= article
\bibitem[{P. {Virtanen} {et~al.}(2020){Virtanen}, {Gommers}, {Oliphant}, {Haberland}, {Reddy}, {Cournapeau}, {Burovski}, {Peterson}, {Weckesser}, {Bright}, {van der Walt}, {Brett}, {Wilson}, {Millman}, {Mayorov}, {Nelson}, {Jones}, {Kern}, {Larson}, {Carey}, {Polat}, {Feng}, {Moore}, {VanderPlas}, {Laxalde}, {Perktold}, {Cimrman}, {Henriksen}, {Quintero}, {Harris}, {Archibald}, {Ribeiro}, {Pedregosa}, {van Mulbregt}, \& {SciPy 1. 0 Contributors}}]{Virtanen20}
{Virtanen}, P., {Gommers}, R., {Oliphant}, T.~E., {et~al.} 2020, \bibinfo{title}{{SciPy 1.0: fundamental algorithms for scientific computing in Python},} Nature Methods, 17, 261, \dodoi{10.1038/s41592-019-0686-2}

% type= article
\bibitem[{C. {Viscasillas V{\'a}zquez} {et~al.}(2022){Viscasillas V{\'a}zquez}, {Magrini}, {Casali}, {Tautvai{\v{s}}ien{\.{e}}}, {Spina}, {Van der Swaelmen}, {Randich}, {Bensby}, {Bragaglia}, {Friel}, {Feltzing}, {Sacco}, {Turchi}, {Jim{\'e}nez-Esteban}, {D'Orazi}, {Delgado-Mena}, {Mikolaitis}, {Drazdauskas}, {Minkevi{\v{c}}i{\={u}}t{\.{e}}}, {Stonkut{\.{e}}}, {Bagdonas}, {Montes}, {Guiglion}, {Baratella}, {Tabernero}, {Gilmore}, {Alfaro}, {Francois}, {Korn}, {Smiljanic}, {Bergemann}, {Franciosini}, {Gonneau}, {Hourihane}, {Worley}, \& {Zaggia}}]{Viscasillas2022}
{Viscasillas V{\'a}zquez}, C., {Magrini}, L., {Casali}, G., {et~al.} 2022, \bibinfo{title}{{The Gaia-ESO survey: Age-chemical-clock relations spatially resolved in the Galactic disc},} \aap, 660, A135, \dodoi{10.1051/0004-6361/202142937}

% type= article
\bibitem[{T. {von Hippel} {et~al.}(2006){von Hippel}, {Jefferys}, {Scott}, {Stein}, {Winget}, {De Gennaro}, {Dam}, \& {Jeffery}}]{vonHippel2006}
{von Hippel}, T., {Jefferys}, W.~H., {Scott}, J., {et~al.} 2006, \bibinfo{title}{{Inverting Color-Magnitude Diagrams to Access Precise Star Cluster Parameters: A Bayesian Approach},} \apj, 645, 1436, \dodoi{10.1086/504369}

% type= article
\bibitem[{T. {von Hippel} {et~al.}(2002){von Hippel}, {Steinhauer}, {Sarajedini}, \& {Deliyannis}}]{vonHippel2002}
{von Hippel}, T., {Steinhauer}, A., {Sarajedini}, A., \& {Deliyannis}, C.~P. 2002, \bibinfo{title}{{WIYN Open Cluster Study. XI. WIYN 3.5 meter Deep Photometry of M35 (NGC 2168)},} \aj, 124, 1555, \dodoi{10.1086/341951}

% type= article
\bibitem[{A.~I. {Wiggins} {et~al.}(2025){Wiggins}, {Quinn}, {Oeur}, {Loebman}, {Frinchaboy}, {Daniel}, {McCluskey}, {Otto}, {Woodward}, {D'Onghia}, {Wetzel}, {Parul}, {Bhattarai}, \& {Cozzi}}]{Wiggins2025}
{Wiggins}, A.~I., {Quinn}, J.~R., {Oeur}, M., {et~al.} 2025, \bibinfo{title}{{Understanding the Origin and Dynamical Evolution of the Unique Open Star Cluster Berkeley 20 Using FIRE Simulations},} \apjl, 995, L25, \dodoi{10.3847/2041-8213/ae21bf}

% type= article
\bibitem[{S.~E. {Woosley} \& T.~A. {Weaver}(1995){Woosley} \& {Weaver}}]{Woosley1995}
{Woosley}, S.~E., \& {Weaver}, T.~A. 1995, \bibinfo{title}{{The Evolution and Explosion of Massive Stars. II. Explosive Hydrodynamics and Nucleosynthesis},} \apjs, 101, 181, \dodoi{10.1086/192237}

% type= article
\bibitem[{Z.-Y. {Wu} {et~al.}(2009){Wu}, {Zhou}, {Ma}, \& {Du}}]{Wu2009}
{Wu}, Z.-Y., {Zhou}, X., {Ma}, J., \& {Du}, C.-H. 2009, \bibinfo{title}{{The orbits of open clusters in the Galaxy},} \mnras, 399, 2146, \dodoi{10.1111/j.1365-2966.2009.15416.x}

% type= article
\bibitem[{T. {Yontan}(2023a){Yontan}}]{Yontan2023a}
{Yontan}, T. 2023a, \bibinfo{title}{{An Investigation of Open Clusters Berkeley 68 and Stock 20 Using CCD UBV and Gaia DR3 Data},} \aj, 165, 79, \dodoi{10.3847/1538-3881/aca6f0}

% type= article
\bibitem[{T. {Yontan} {et~al.}(2023b){Yontan}, {Bilir}, {{\c{C}}akmak}, {Ra{\'u}l}, {Banks}, {Soydugan}, {Canbay}, \& {Ta{\c{s}}demir}}]{Yontan2023b}
{Yontan}, T., {Bilir}, S., {{\c{C}}akmak}, H., {et~al.} 2023b, \bibinfo{title}{{CCD UBV and Gaia DR3 based analysis of NGC 189, NGC 1758 and NGC 7762 open clusters},} Advances in Space Research, 72, 1454, \dodoi{10.1016/j.asr.2023.04.015}

% type= article
\bibitem[{T. {Yontan} \& R. {Canbay}(2023c){Yontan} \& {Canbay}}]{Yontan23c}
{Yontan}, T., \& {Canbay}, R. 2023c, \bibinfo{title}{{Comprehensive Analysis of the Open Cluster Collinder 74},} Physics and Astronomy Reports, 1, 65, \dodoi{10.26650/PAR.2023.00008}

% type= article
\bibitem[{T. {Yontan} {et~al.}(2015){Yontan}, {Bilir}, {Bostanc{\i}}, {Ak}, {Karaali}, {G{\"u}ver}, {Ak}, {Duran}, \& {Paunzen}}]{Yontan2015}
{Yontan}, T., {Bilir}, S., {Bostanc{\i}}, Z.~F., {et~al.} 2015, \bibinfo{title}{{CCD UBVRI photometry of NGC 6811},} \apss, 355, 267, \dodoi{10.1007/s10509-014-2175-5}

% type= article
\bibitem[{T. {Yontan} {et~al.}(2019){Yontan}, {Bilir}, {Bostanc{\i}}, {Ak}, {Ak}, {G{\"u}ver}, {Paunzen}, {{\"U}rg{\"u}p}, {{\c{C}}elebi}, {Akti}, \& {G{\"o}kmen}}]{Yontan2019}
{Yontan}, T., {Bilir}, S., {Bostanc{\i}}, Z.~F., {et~al.} 2019, \bibinfo{title}{{CCD UBV photometric and Gaia astrometric study of eight open clusters{\textemdash}ASCC 115, Collinder 421, NGC 6793, NGC 7031, NGC 7039, NGC 7086, Roslund 1 and Stock 21},} \apss, 364, 152, \dodoi{10.1007/s10509-019-3640-y}

% type= article
\bibitem[{T. {Yontan} {et~al.}(2022){Yontan}, {{\c{C}}akmak}, {Bilir}, {Banks}, {Ra{\'u}l}, {Canbay}, {Ko{\c{c}}}, {Ta{\c{s}}demir}, {Er{\c{c}}ay}, {Tan{\i}k Ozt{\"u}rk}, \& {Dursun}}]{Yontan2022}
{Yontan}, T., {{\c{C}}akmak}, T., {Bilir}, S., {et~al.} 2022, \bibinfo{title}{{A Study of the NGC 1193 and NGC 1798 Open Clusters Using CCD UBV Photometric and Gaia EDR3 Data},} \rmxaa, 58, 333, \dodoi{10.22201/ia.01851101p.2022.58.02.14}

% type= article
\bibitem[{T. {Yontan} {et~al.}(2026){Yontan}, {Bilir}, {{\c{C}}akmak}, {Plevne}, {Soydugan}, {Ta{\textcommabelow s}demir}, {Canbay}, {{\c{C}}{\i}nar}, {Ko{\c{c}}}, {Karag{\"o}z}, {Tan{\i}k {\"O}zt{\"u}rk}, {Akbaba}, {{\c{C}}al{\i}{\textcommabelow s}kan}, \& {Banks}}]{Yontan2026}
{Yontan}, T., {Bilir}, S., {{\c{C}}akmak}, H., {et~al.} 2026, \bibinfo{title}{{Kings of the Milky Way: A Homogeneous Gaia DR3 Analysis of King Open Clusters and the Galactic Disc Metallicity Gradient},} arXiv e-prints, arXiv:2608.08216, \dodoi{10.48550/arXiv.2608.08216}

% type= article
\bibitem[{H. Zinnecker \& H.~W. Yorke(2007)Zinnecker \& Yorke}]{zinnecker2007}
Zinnecker, H., \& Yorke, H.~W. 2007, \bibinfo{title}{Toward understanding massive star formation,} Annu. Rev. Astron. Astrophys., 45, 481

\end{thebibliography}
\bibliographystyle{aasjournalv7}
\appendix

\section{Gaia DR3 Data Retrieval and Quality Filtering}\label{app:data_quality}

Table~\ref{tab:photometric_errors} summarises the photometric precision of the Gaia DR3 catalog compiled for Berkeley~36 after the quality-selection steps described below. Sources were queried from the Gaia DR3 archive through an ADQL statement centred on the cluster coordinates, adopting a search radius of 21~arcmin and restricting the query to sources brighter than $G = 21$~mag. For clarity, only the core selection conditions (sky position, search cone, and magnitude limit) are given in the main text; the complete ADQL statement is not reproduced here for brevity.

Prior to the membership computation, we required all retrieved sources to have a full five-parameter astrometric solution, as identified by the \texttt{astrometric\_params\_solved} flag. This step guarantees that proper motions and parallaxes are simultaneously available for every star entering the subsequent membership determination. Sources flagged as having a less robust astrometric solution were not discarded outright; instead, they were kept in the working catalog, since their effect on the final membership probabilities is expected to be minor -- stars with poorly constrained astrometry tend to scatter away from the cluster locus in proper-motion/parallax space and are consequently assigned low membership probabilities by construction. Applying these criteria to the initial query results in a working sample of 27{,}016 sources in the direction of Berkeley~36, as listed in Table~\ref{tab:photometric_errors}.

We further examined how the photometric errors evolve with apparent magnitude before proceeding to the colour--magnitude diagram analysis. As anticipated, both the $G$-band and $(G_{\rm BP}-G_{\rm RP})$ uncertainties stay small at the bright end of the sample and grow steadily toward fainter stars, with a pronounced steepening past $G \sim 20$~mag. The magnitude-binned statistics in Table~\ref{tab:photometric_errors} quantify this behaviour directly. Guided by this trend, we restrict the photometric sample used for the CMD analysis to $G \leq 20.5$~mag, a choice that keeps the bulk of the well-measured stars while excluding the regime where colour errors rise sharply and could otherwise distort the cluster sequence.

\begin{table}[h]
\centering
\small
\renewcommand{\arraystretch}{1}
\caption{Mean photometric uncertainties in Gaia $G$ magnitude and $(G_{\rm BP}-G_{\rm RP})$ colour for sources toward Berkeley~36, binned by apparent magnitude.}
\label{tab:photometric_errors}
\begin{tabular}{c|ccc}
\toprule
$G$ (mag) & $N$ & $\sigma_{G}$ & $\sigma_{G_{\rm BP}-G_{\rm RP}}$ \\
\midrule
6--14   &   406 & 0.003 & 0.006 \\
14--15  &   475 & 0.003 & 0.005 \\
15--16  &   894 & 0.003 & 0.006 \\
16--17  &  1723 & 0.003 & 0.009 \\
17--18  &  2913 & 0.003 & 0.015 \\
18--19  &  4642 & 0.003 & 0.034 \\
19--20  &  6533 & 0.004 & 0.070 \\
20--21  &  8237 & 0.009 & 0.175 \\
21--23  &  1193 & 0.023 & 0.401 \\
\midrule
Total   & 27016 & 0.006 & 0.097 \\
\bottomrule
\end{tabular}
\end{table}

\section{Blue and Yellow Straggler Star Catalog} \label{sec:appendix_bss_b36}

Table~\ref{tab:appendix_bss} lists the full sample of blue straggler star (BSS) and yellow straggler star (YSS) candidates discussed in Section~\ref{sec:BSS_YSS}, together with the variable stars identified toward the cluster from the Gaia DR3 variability classification. For each object, we provide the equatorial coordinates, the $G$-band magnitude, the cluster membership probability, and the star type (BSS or YSS). The superscript $^{a}$ indicates the origin of the BSS/YSS identification: (1) \citet{Jadhav2021bss}; (2) This work.

\begin{table*}
\centering
\caption{Full list of BSS and YSS candidates identified toward the cluster, together with their membership probabilities. The superscript $^{a}$ indicates the source of the identification: (1) \citet{Jadhav2021bss}; (2) this work.} \label{tab:appendix_bss}
\renewcommand{\arraystretch}{1}
\setlength{\tabcolsep}{4pt}
\scriptsize
\begin{longtable}{l c c c c l c | l c c c c l c}
\toprule
No. & $\alpha$ & $\delta$ & $G$ & $P$ & Type & Ref.$^{a}$ &
No. & $\alpha$ & $\delta$ & $G$ & $P$ & Type & Ref.$^{a}$ \\
   & (hh:mm:ss) & (dd:mm:ss) & (mag) & &  &  &
   & (hh:mm:ss) & (dd:mm:ss) & (mag) & &  &  \\
\midrule
\endfirsthead
\multicolumn{14}{c}{\tablename\ \thetable{} -- \textit{Continued}} \\
\toprule
No. & $\alpha$ & $\delta$ & $G$ & $P$ & Type & Ref.$^{a}$ &
No. & $\alpha$ & $\delta$ & $G$ & $P$ & Ref.$^{a}$ \\
   & (hh:mm:ss) & (dd:mm:ss) & (mag) & &  &  &
   & (hh:mm:ss) & (dd:mm:ss) & (mag) & &  &  \\
\midrule
\endhead
\midrule
\endfoot
\bottomrule
\endlastfoot
 01 & 07:16:26.60 & -13:11:42.1 & 16.64 & 1.00 & BSS & 1 & 48 & 07:16:39.93 & -13:00:30.6 & 17.17 & 0.84 & BSS & 2 \\
 02 & 07:16:27.09 & -13:11:05.5 & 16.28 & 1.00 & BSS & 1 & 49 & 07:16:26.63 & -12:59:36.4 & 16.64 & 1.00 & BSS & 2 \\
 03 & 07:16:29.41 & -13:12:32.6 & 16.63 & 0.99 & BSS & 2 & 50 & 07:16:35.55 & -13:24:00.3 & 16.68 & 0.82 & BSS & 2 \\
 04 & 07:16:21.55 & -13:10:10.5 & 17.32 & 1.00 & BSS & 2 & 51 & 07:15:38.38 & -13:06:01.9 & 17.33 & 1.00 & BSS & 2 \\
 05 & 07:16:30.18 & -13:12:53.1 & 15.97 & 1.00 & BSS & 1 & 52 & 07:17:15.58 & -13:11:00.1 & 16.72 & 0.97 & BSS & 2 \\
 06 & 07:16:24.67 & -13:14:00.1 & 16.88 & 1.00 & BSS & 2 & 53 & 07:17:02.91 & -13:21:00.3 & 16.57 & 1.00 & BSS & 2 \\
 07 & 07:16:19.51 & -13:09:57.7 & 16.56 & 0.99 & BSS & 1 & 54 & 07:16:58.50 & -13:01:42.1 & 16.59 & 1.00 & BSS & 2 \\
 08 & 07:16:20.66 & -13:09:43.3 & 16.90 & 1.00 & BSS & 1 & 55 & 07:16:01.01 & -12:59:47.6 & 14.90 & 1.00 & BSS & 2 \\
 09 & 07:16:24.46 & -13:09:32.8 & 16.14 & 1.00 & BSS & 1 & 56 & 07:17:10.15 & -13:19:20.8 & 16.92 & 1.00 & BSS & 2 \\
 10 & 07:16:24.33 & -13:14:10.8 & 16.08 & 1.00 & BSS & 1 & 57 & 07:17:01.40 & -13:01:33.4 & 16.60 & 1.00 & BSS & 2 \\
 11 & 07:16:25.10 & -13:09:26.1 & 17.09 & 0.99 & BSS & 2 & 58 & 07:17:16.40 & -13:06:22.6 & 16.78 & 0.97 & BSS & 2 \\
 12 & 07:16:34.03 & -13:11:09.7 & 16.80 & 1.00 & BSS & 1 & 59 & 07:16:47.94 & -12:59:03.4 & 16.12 & 0.72 & BSS & 2 \\
 13 & 07:16:25.14 & -13:08:52.8 & 17.17 & 1.00 & BSS & 1 & 60 & 07:15:59.00 & -13:24:47.8 & 14.17 & 1.00 & BSS & 2 \\
 14 & 07:16:34.87 & -13:10:27.0 & 17.00 & 0.70 & BSS & 2 & 61 & 07:15:57.16 & -13:25:09.6 & 17.27 & 0.86 & BSS & 2 \\
 15 & 07:16:20.18 & -13:15:15.7 & 17.23 & 1.00 & BSS & 1 & 62 & 07:17:08.97 & -13:01:41.3 & 16.45 & 1.00 & BSS & 2 \\
 16 & 07:16:11.19 & -13:09:34.9 & 16.81 & 1.00 & BSS & 1 & 63 & 07:15:26.78 & -13:05:56.7 & 15.22 & 0.80 & BSS & 2 \\
 17 & 07:16:40.00 & -13:10:00.1 & 15.07 & 0.76 & BSS & 1 & 64 & 07:15:59.57 & -12:57:48.8 & 16.05 & 1.00 & BSS & 2 \\
 18 & 07:16:03.66 & -13:14:08.5 & 17.14 & 0.88 & BSS & 2 & 65 & 07:15:49.75 & -12:58:47.5 & 15.64 & 0.89 & BSS & 2 \\
 19 & 07:16:04.21 & -13:08:26.8 & 16.41 & 0.98 & BSS & 2 & 66 & 07:17:08.68 & -13:00:41.7 & 16.96 & 1.00 & BSS & 2 \\
 20 & 07:16:16.00 & -13:06:08.7 & 17.04 & 1.00 & BSS & 1 & 67 & 07:15:55.21 & -12:57:44.9 & 17.01 & 1.00 & BSS & 2 \\
 21 & 07:16:34.06 & -13:18:01.7 & 16.48 & 0.81 & BSS & 1 & 68 & 07:15:19.75 & -13:09:26.9 & 16.58 & 0.95 & BSS & 2 \\
 22 & 07:16:21.64 & -13:04:42.0 & 16.05 & 1.00 & BSS & 2 & 69 & 07:17:28.68 & -13:10:06.3 & 16.55 & 1.00 & BSS & 2 \\
 23 & 07:16:40.10 & -13:18:00.6 & 15.66 & 0.86 & BSS & 1 & 70 & 07:17:28.58 & -13:15:04.4 & 17.15 & 0.71 & BSS & 2 \\
 24 & 07:16:32.53 & -13:18:55.1 & 16.28 & 1.00 & BSS & 1 & 71 & 07:16:23.16 & -12:55:28.8 & 15.56 & 0.89 & BSS & 2 \\
 25 & 07:16:22.74 & -13:19:26.7 & 15.82 & 0.95 & BSS & 2 & 72 & 07:16:38.46 & -13:28:52.8 & 17.09 & 1.00 & BSS & 2 \\
 26 & 07:16:49.99 & -13:07:16.9 & 16.46 & 0.94 & BSS & 2 & 73 & 07:17:23.79 & -13:21:33.9 & 16.70 & 1.00 & BSS & 2 \\
 27 & 07:16:23.10 & -13:19:50.6 & 16.57 & 0.99 & BSS & 1 & 74 & 07:15:10.50 & -13:12:30.6 & 17.34 & 1.00 & BSS & 2 \\
 28 & 07:16:53.52 & -13:15:19.2 & 15.98 & 0.98 & BSS & 2 & 75 & 07:17:21.41 & -13:00:40.2 & 17.03 & 1.00 & BSS & 2 \\
 29 & 07:16:58.53 & -13:13:30.3 & 17.29 & 0.89 & BSS & 2 & 76 & 07:15:16.41 & -13:19:32.0 & 16.94 & 0.98 & BSS & 2 \\
 30 & 07:15:48.27 & -13:12:48.4 & 15.54 & 1.00 & BSS & 2 & 77 & 07:15:34.13 & -13:25:33.6 & 16.71 & 0.99 & BSS & 2 \\
 31 & 07:16:59.77 & -13:12:31.6 & 14.61 & 0.99 & BSS & 1 & 78 & 07:17:33.39 & -13:03:41.1 & 16.48 & 0.96 & BSS & 2 \\
 32 & 07:17:02.46 & -13:13:23.1 & 17.34 & 0.93 & BSS & 2 & 79 & 07:17:23.84 & -12:59:32.0 & 17.20 & 1.00 & BSS & 2 \\
 33 & 07:17:03.09 & -13:15:17.6 & 13.64 & 0.80 & BSS & 2 & 80 & 07:15:10.13 & -13:05:12.4 & 16.93 & 0.95 & BSS & 2 \\
 34 & 07:17:05.28 & -13:14:11.6 & 13.55 & 1.00 & BSS & 2 & 81 & 07:15:37.11 & -13:27:17.6 & 14.71 & 0.97 & BSS & 2 \\
 35 & 07:15:58.01 & -13:20:13.8 & 16.75 & 0.98 & BSS & 2 & 82 & 07:16:46.23 & -12:53:19.3 & 15.71 & 1.00 & BSS & 2 \\
 36 & 07:15:56.25 & -13:03:47.6 & 16.93 & 1.00 & BSS & 2 & 83 & 07:17:28.68 & -12:59:37.6 & 16.69 & 1.00 & BSS & 2 \\
 37 & 07:16:31.22 & -13:22:25.2 & 15.52 & 0.92 & BSS & 2 & 84 & 07:16:23.43 & -13:15:58.2 & 15.99 & 1.00 & YSS & 2 \\
 38 & 07:17:06.35 & -13:15:17.2 & 16.00 & 0.99 & BSS & 2 & 85 & 07:16:41.65 & -13:12:05.5 & 16.61 & 1.00 & YSS & 2 \\
 39 & 07:17:06.53 & -13:08:21.7 & 15.26 & 0.98 & BSS & 2 & 86 & 07:16:40.28 & -13:20:19.8 & 16.64 & 1.00 & YSS & 2 \\
 40 & 07:16:13.07 & -13:22:35.8 & 14.92 & 0.72 & BSS & 2 & 87 & 07:15:46.08 & -13:04:03.2 & 15.94 & 1.00 & YSS & 2 \\
 41 & 07:17:09.81 & -13:11:07.8 & 15.96 & 1.00 & BSS & 2 & 88 & 07:16:01.98 & -13:00:23.1 & 15.14 & 0.82 & YSS & 2 \\
 42 & 07:15:53.01 & -13:03:31.1 & 17.37 & 0.89 & BSS & 2 & 89 & 07:17:20.06 & -13:07:25.5 & 15.46 & 1.00 & YSS & 2 \\
 43 & 07:17:09.66 & -13:13:30.2 & 17.06 & 0.97 & BSS & 2 & 90 & 07:16:58.68 & -13:24:34.0 & 14.68 & 0.89 & YSS & 2 \\
 44 & 07:16:49.79 & -13:21:31.1 & 16.90 & 0.77 & BSS & 2 & 91 & 07:15:22.90 & -13:18:37.8 & 16.45 & 0.98 & YSS & 2 \\
 45 & 07:16:25.93 & -13:00:09.9 & 16.18 & 1.00 & BSS & 2 & 92 & 07:17:15.66 & -12:59:25.5 & 16.78 & 1.00 & YSS & 2 \\
 46 & 07:17:05.02 & -13:17:54.9 & 15.65 & 1.00 & BSS & 2 & 93 & 07:15:08.05 & -13:10:01.4 & 16.64 & 1.00 & YSS & 2 \\
 47 & 07:16:20.69 & -13:00:01.3 & 16.90 & 1.00 & BSS & 2 & & & & & & \\
\end{longtable}
\end{table*}

\end{document}